\documentclass[prb,twocolumn,aps,showpacs,superscriptaddress,epsfig]{revtex4}
\usepackage{amssymb}
\usepackage{amsbsy}
\usepackage{amsmath}
\usepackage{epsfig}
\usepackage{graphicx}
\usepackage{subfigure}

\usepackage{hyperref}
\hypersetup{
    colorlinks=true,
    linkcolor=blue,
    filecolor=cyan,
    urlcolor=magenta,
    citecolor=red,
}
\begin{document}
\title{Hinge mode dynamics of periodically driven higher-order Weyl semimetals}
\author{Somsubhra Ghosh}
\affiliation{School of Physical Sciences, Indian Association for the Cultivation of Science, Kolkata, 700032, India.}
\author{Kush Saha}
\affiliation{National Institute of Science Education and Research, Jatni, Odisha 752050, India.}
\affiliation{Homi Bhabha National Institute, Training School Complex, Anushakti Nagar, Mumbai 400094, India. }
\author{K. Sengupta}
\affiliation{School of Physical Sciences, Indian Association for the Cultivation of Science, Kolkata, 700032, India.}
\date{\today}
\begin{abstract}
We study the stroboscopic dynamics of hinge modes of a second-order
topological material modeled by a tight-binding free fermion
Hamiltonian on a cubic lattice in the intermediate drive frequency
regime for both discrete (square pulse) and continuous (cosine)
periodic drive protocols. We analyze the Floquet phases of this
system and show that its quasienergy spectrum becomes almost gapless
in the large drive amplitude regime at special drive frequencies.
Away from these frequencies, the gapped quasienergy spectrum
supports weakly dispersing Floquet hinge modes. Near them, these
hinge modes penetrate into the bulk and eventually become
indistinguishable from the bulk modes. We provide an analytic,
albeit perturbative, expression for the Floquet Hamiltonian using
Floquet perturbation theory (FPT) which explains this phenomenon and
leads to analytic expressions of these special frequencies. We also
show that in the large drive amplitude regime, the zero energy hinge
modes corresponding to the static tight-binding Hamiltonian display
qualitatively different dynamics at these special frequencies. We
discuss possible local density of state measurement using a scanning
tunneling microscope which can test our theory.
\end{abstract}
\maketitle

\section{Introduction}

Topological materials have been a subject of intense theoretical and
experimental studies in recent year \cite{toporev}. The study of
these systems began with spin-Hall systems \cite{kane05},
topological insulators \cite{toporef}, and Dirac and Weyl semimetals
\cite{diracref,weylref}. The key property of these materials which
distinguishes them from, for example, trivial insulators, is
manifestation of the bulk-boundary correspondence \cite{ashwinrev}.
In these materials, non-trivial topology of the bulk bands results
in the presence of symmetry protected surface states. Thus a
$d$-dimensional solid hosting a topological phase exhibits
topologically protected gapless states localized in its $d-1$
dimensional surface.

More recently, a new class of topological materials, dubbed as
higher-order topological materials (HOTMs), have been studied
intensively
\cite{hotmref1,hotmref2,hotmref3,hotmref4,hotmref5,hotmref6,hotmref7,hotmref8,hotmref9,hotmref10,hotmref11,hotmref12,hotmref13,hotmref14,hotmref15,hotmref16,hweylref0,hweylref1,hweylref2,hweylref3,hweylref4,hweylref5,hweylref6,hweylref7,hdiracref1,hdiracref2,hdiracref3,hdiracref4,hdiracref5,hdiracref6}.
A $n^{\rm th}$ order HOTM has non-zero $2^n$ moment in the bulk
(quadrupole for $n=2$ and octupole for $n=3$) and hosts $d-n$
dimensional topologically protected states on its edges or corners;
all other higher dimensional surface modes are gapped out. For
example in two-dimensional (2D) second-order topological materials
(SOTM), there is a non-zero quadrupole moment in the bulk and the
edge states are gapped, while topologically protected states appear
at the corner
\cite{hotmref2,hotmref4,hotmref5,hotmref6,hotmref8,hotmref9,hotmref11,hotmref12,hotmref13,hotmref14,hotmref16}.
Similarly 3D SOTMs host gapless hinge modes with gapped surface
states \cite{hotmref4,hotmref5,hotmref7,hotmref9}. A class of such
materials include the higher-order Dirac semimetals (HODS)
\cite{hdiracref1,hdiracref2,hdiracref3,hdiracref4,hdiracref5,hdiracref6}
and the more recently found higher-order Weyl semimetals (HOWS)
\cite{hweylref0,
hweylref1,hweylref2,hweylref3,hweylref4,hweylref5,hweylref6,hweylref7}.
Apart from the standard Fermi arc states of the typical Weyl
semimetals, HOWSs also host gapless hinge states with quantized
charge.

The physics of closed quantum systems driven out of equilibrium has
also been studied extensively in recent years.
\cite{rev1,rev2,rev3,rev4,rev5,rev6,rev7,rev8}. The quantum dynamics
of such systems involving periodic drive protocols are of particular
interest; they exhibit a host of phenomena which usually have no
counterpart in either equilibrium or aperiodically driven quantum
systems \cite{ap1,ap2,ap3}. Such phenomena include dynamical
freezing \cite{df1,df2,df3,df4,df5}, dynamical localization
\cite{dloc1,dloc2,dloc3,dloc4}, dynamical phase transitions
\cite{dtran1,dtran2,dtran3}, presence of time crystalline phases
\cite{tc1,tc2,tc3}, and possibility of tuning ergodic properties of
a quantum system \cite{ergoref}. More interestingly, it is realized
that such drives can be used to engineer transition between
topologically trivial and non-trivial phases of matter
\cite{topo1,topo2,topo3,topo4,topo5}.

In this work, we study a driven tight-binding hopping Hamiltonian
for free fermions which is known to host topological
Weyl-semimetallic phase in equilibrium \cite{hweylref1}. We study
this system for continuous and discrete drive protocols. The summary
of our main results and their connection to existing ones are
charted below.

\subsection{Summary of results}
\label{ressumm}
 The central
results that we obtain from our study are as follows.
\begin{itemize}
\item First, we
chart out the phase diagram of the equilibrium model and demonstrate
that it hosts second-order topological phases with a bulk gap and
zero energy hinge states. These states serve as initial states in
our dynamics studies.
\item Second, we use FPT to compute the Floquet
Hamiltonian of the driven system corresponding to both discrete
square pulse and continuous cosine drive protocols. The Floquet
phases obtained from these perturbative analytic Hamiltonians agree
remarkably well with that obtained from exact numerics in the large
drive amplitude and intermediate frequency regime where second-order
Magnus expansion fails.
\item Third, we demonstrate the existence of
special drive frequencies for which the bulk Floquet spectrum
becomes almost gapless. At these frequencies, the first order
analytic Floquet Hamiltonian leads to a gapless spectrum; thus the
contribution to the gap in the Floquet spectrum comes from higher
order terms which are small. This picture is corroborated by exact
numerical study of the system which also indicates a drastic
reduction of the Floquet spectrum gap at these special frequencies.
\item Fourth, we find that the Floquet spectrum supports weakly dispersing
hinge modes when the drive frequency is different from these special
frequencies. We provide analytic expressions of these hinge states
for the discrete protocol for a representative drive frequency
starting from the perturbative first order Floquet Hamiltonian; we
find the analytic expressions to be qualitatively similar to those
obtained from exact numerics. In contrast, the hinge modes of the
Floquet spectrum delocalizes into the bulk and becomes almost
indistinguishable from the bulk modes at these special drive
frequencies.
\item Fifth, we study the manifestation of such
a small bulk Floquet gap on the dynamics of the hinge modes.
Starting from an initial zero energy eigenstate of the equilibrium
Hamiltonian which is localized at one of the hinges, we show, by
computing the spatially resolved probability distribution of the
driven hinge state, that the dynamics keeps the state localized to
the initial hinge when the bulk Floquet gap is large. In contrast,
at the special drive frequencies where the bulk Floquet gap becomes
small, the hinge modes show coherent propagation between diagonally
opposite hinges with a fixed periodicity. We provide an analytic
estimate of this periodicity using the first order perturbative
Floquet Hamiltonian.
\item Finally, we point out that such a periodic variation would
reflect in the local density of state (LDOS) of the fermions and is
hence measurable by a scanning tunneling microscope (STM). This
allows for the possibility of verification of our theoretical
results in standard STM experiments.
\end{itemize}

\subsection{Comparison with existing works}
Most of the theoretical efforts in the direction of Floquet
engineering of HOTMs have been based on either a class of hopping
Hamiltonians on specific lattices
\cite{hotmbandcontref1,hotmbanddiscrref2,hotmbanddiscrref3,hotmbanddiscrref4,hotmbandcontref5,hotmbanddiscrref6,hotmbanddiscrref7,hotmbanddiscrref8,hotmbanddiscrref9,hotmbanddiscrref10,hotmbanddiscrref11}
or driven topological superconductors
\cite{hotmsupdiscrref1,hotmsupcontref2,hotmsupcontref3,hotmsupcontref4,hotmsupdiscrref5,hotmsupdiscrref6,hotmsupdiscrref7,
hotmsupdiscrref8}. The drive protocols followed in these studies are
either continuous arising from interaction of such systems with
light
\cite{hotmbandcontref1,hotmbandcontref5,hotmsupcontref2,hotmsupcontref3,hotmsupcontref4,hotmsupdiscrref8}
or specifically engineered discrete ones where one of the
Hamiltonian parameters are changed discontinuously with time
\cite{hotmbanddiscrref2,hotmbanddiscrref3,hotmbanddiscrref4,hotmbanddiscrref6,hotmbanddiscrref7,hotmbanddiscrref8,hotmbanddiscrref9,hotmbanddiscrref10,hotmbanddiscrref11,hotmsupdiscrref1,hotmsupdiscrref5,hotmsupdiscrref6,hotmsupdiscrref7,
hotmsupdiscrref8}. These studies clearly establish that such
periodic driving can be used to engineer higher-order topological
Floquet phases even when the ground state of the equilibrium parent
Hamiltonian do not host such a phase. The theoretical analysis
leading to this result may be classified into two distinct
categories. The first involves construction of exact Floquet
unitaries for discrete protocols followed by their numerical
analysis to unravel the existence of the higher-order Floquet phase
\cite{hotmsupdiscrref5,hotmsupdiscrref6,hotmsupdiscrref7}. The
second class of studies, carried out for both discrete and
continuous protocols, involves analytic computation of the Floquet
Hamiltonian of the system in the high-frequency regime using
perturbation techniques which uses $T$ as the  expansion parameter
\cite{hotmbandcontref1,hotmbanddiscrref2,hotmbanddiscrref3,hotmsupdiscrref1}.
The latter class provide analytic insight into the properties of the
Floquet Hamiltonian only in the high-frequency regime where such low
$T$ expansions are accurate. To the best of our knowledge, such
studies have not been extended to the intermediate frequency regime
where these perturbative methods fail. Our study, on the other hand,
uses FPT to explore this intermediate frequency regime both
analytically and numerically. In the process, we encounter features
like dispersion of hinge modes and closing of bulk band gap, which
have no analogue both in the undriven and in the high frequency
driven version. As we discuss in detail in Sec. \ref{2pipoint}, this
closing of the band gap, in addition to being an artifact of the
first order theory, doesn't lead to a change in topology, as this
isn't accompanied by a band inversion. Nevertheless, this leaves
dynamical signatures, which serve as diagnostic tools of our Floquet
phases. To the best of our knowledge, such diagnosis of Floquet
phases has not been discussed in the literature so far.

The plan of the rest of the work are as follows. In Sec.\
\ref{sec:Hamiltonian}, we define the starting Hamiltonian and chart
out its equilibrium phase diagram. This is followed by Sec.\
\ref{sec:floquet_PT} where we derive the analytic, albeit
perturbative, Floquet Hamiltonian using FPT for both discrete and
continuous drive protocols. Next, in Sec.\ \ref{flp}, we discuss the
Floquet phases and compare the FPT results with those from exact
numerics. This is followed by Sec.\ \ref{dyncor} where we discuss
the dynamics of the hinge modes. Finally we discuss our main results
and conclude in Sec.\ \ref{diss}. A derivation of the Floquet
Hamiltonian for both continuous and discrete protocol using Magnus
expansion is presented in the appendix.

\section{\label{sec:Hamiltonian} Model Hamiltonian and Equilibrium Phases}

\begin{figure}
    \vspace{-1\baselineskip}
    \includegraphics*[width=0.49\linewidth]{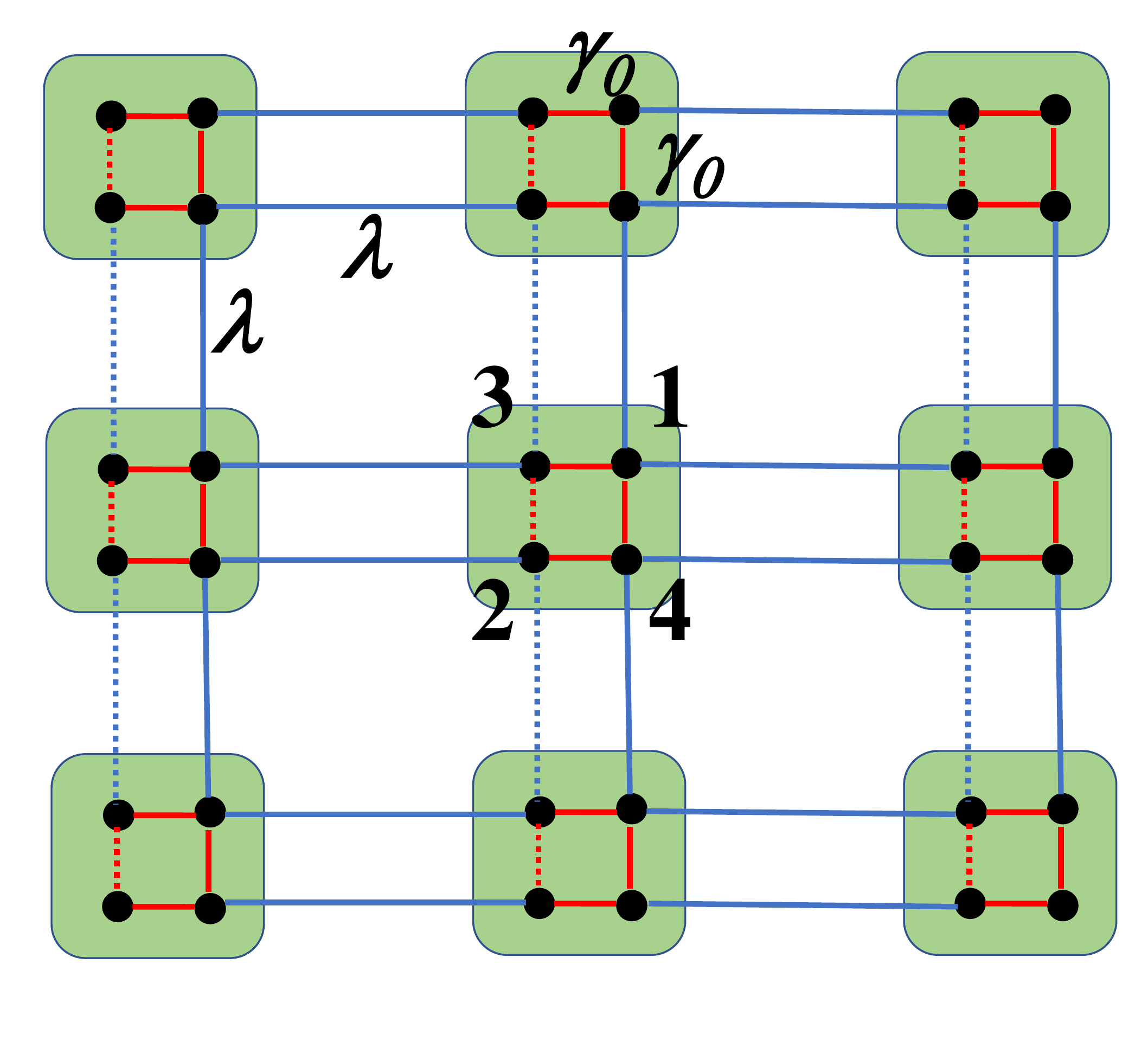}
    \includegraphics*[width=0.49\linewidth]{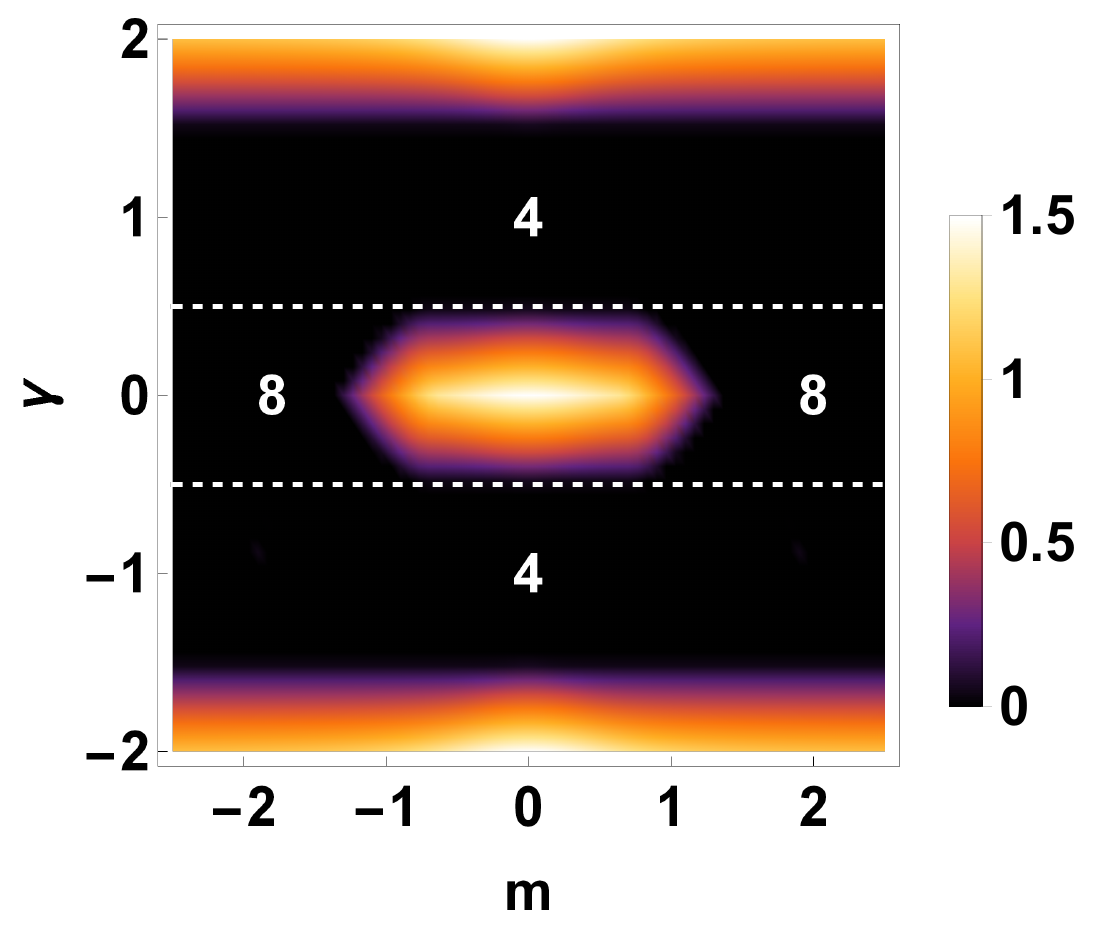}
    \includegraphics*[width=0.49\linewidth]{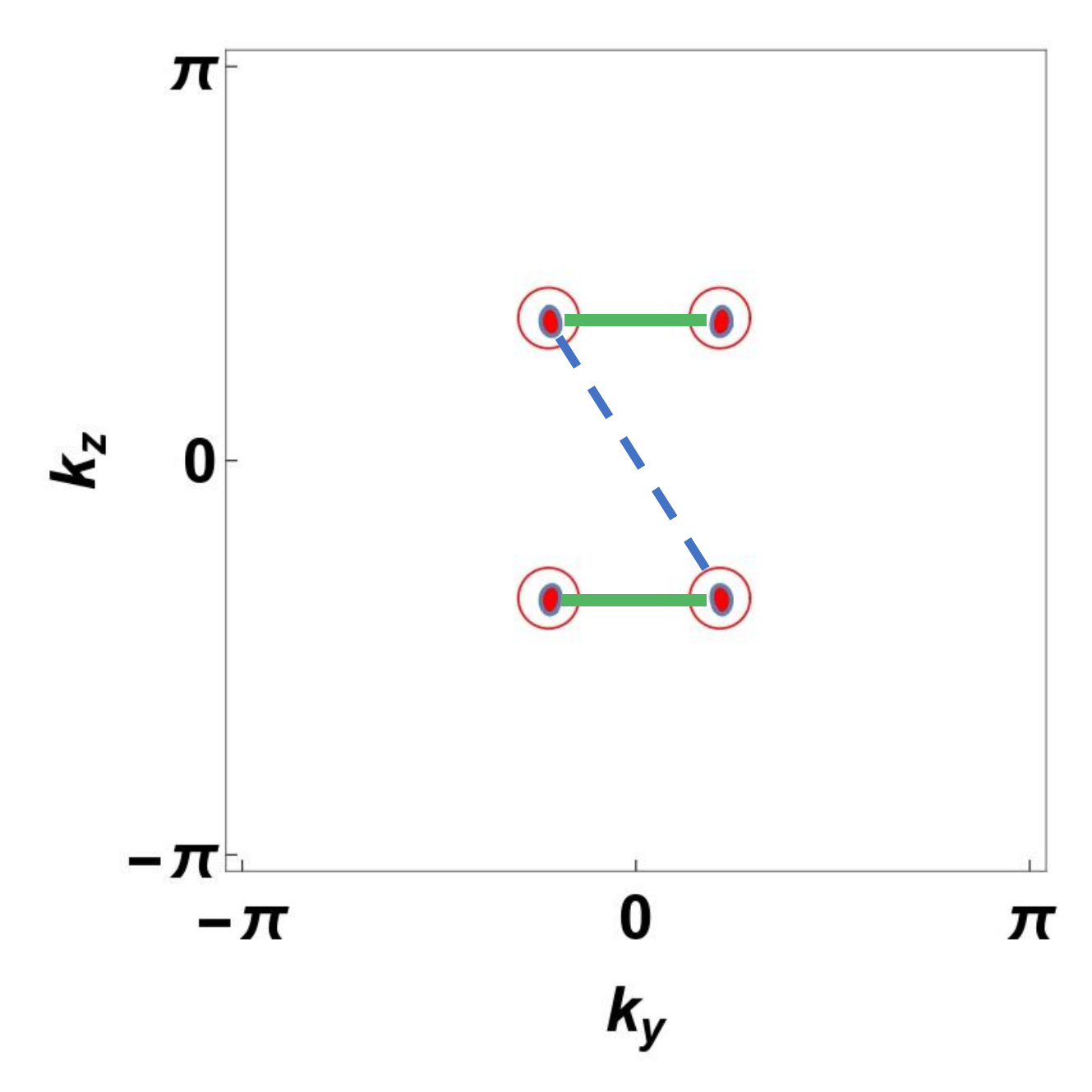}
    \includegraphics*[width=0.49\linewidth]{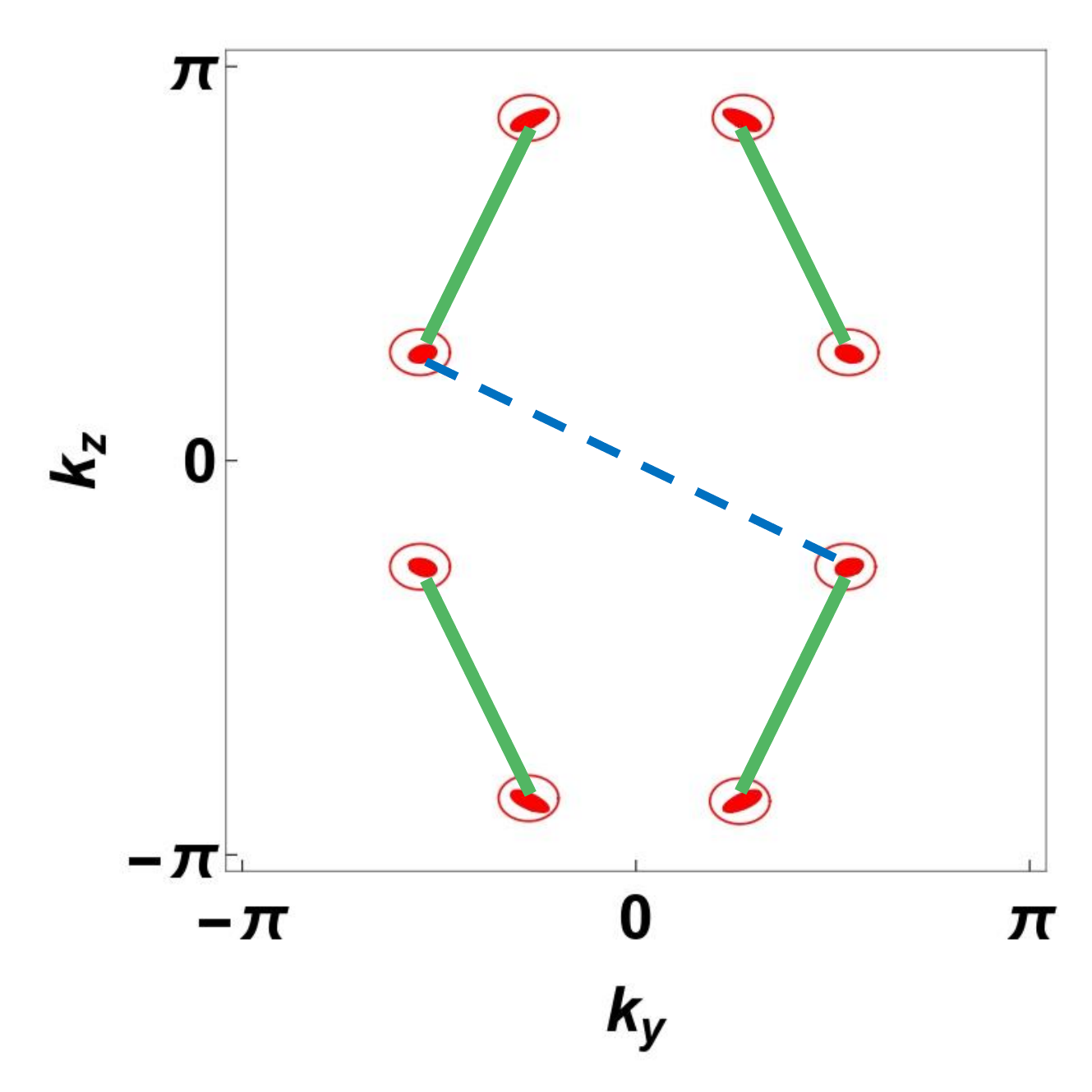}
\caption{Top Panel: Left: A lattice realization of a two-dimensional
quadrapolar insulator (QI) with four orbitals as indicated by 1-4
numbers. The 3D structure is obtained by stacking 2D QIs. The red
and blue lines denote intra-cell and inter-cell hopping,
respectively. Right: Plot of the bulk bandgap as obtained from
Eq.~(\ref{eqn:energy}) for different values of $\gamma$ and $m$. The
black regime is gapless with 4 or 8 Weyl nodes as indicated and the
rest including the central hexagon-like regime are gapped. Bottom
Panel: The location of Weyl points for $\gamma=-1.0$ (left) and
$m=0.75$ and  $\gamma=-0.2$ and $m=2.0$ (right). The green solid and
blue dashed lines are schematic representations of the surface and
hinge arcs respectively. All energies are scaled in units of
$\lambda$ and the circles for plots in the bottom panels are guide
to the eye.
    \label{fig:spectrum}}
\end{figure}

We begin with the low-energy model tight-binding Hamiltonian on a
cubic lattice involving four spinless fermions within an unit cell
hosting higher-order Weyl semimetal phases \cite{hweylref1}. A
schematic picture of the model is shown in the top left panel of
Fig.\ \ref{fig:spectrum}. In momentum-space, the Hamiltonian of this
system is given by
\begin{eqnarray}
H &=& \sum_{\vec k} \psi_{\vec k}^{\dagger} H_0(\vec k) \psi_{\vec k} \nonumber \\
 H_0(\vec k)&=& a_4\Gamma_1+ a_2 \Gamma_2
 + a_3 \Gamma_3+ a_1 \Gamma_4 + i a_5 \Gamma_2\Gamma_3, \nonumber\\
 a_{1(2)} &=& (\gamma_z + \lambda \cos k_{x(y)}), \quad \gamma_z= \gamma_0 + \frac{\lambda}{2} \cos k_z, \nonumber\\
 a_{3(4)} &=& \lambda \sin k_{x(y)}, \quad a_5= m_0 \sin k_z,
 \label{eqn:ham0}
\end{eqnarray}
where $\psi_k$ denotes a four-component annihilation operator for
fermions, the lattice spacing has been set to unity,
$\gamma_0$($\lambda)$ denotes intra-(inter-)cell hopping amplitudes
as shown in Fig.\ \ref{fig:spectrum}, $\vec k= (k_x,k_y,k_z)$
indicates crystal momenta, and the matrices $\Gamma_{\mu}$ are
given, in terms of outer product of two Pauli matrices $\tau$ and
$\sigma$, by
\begin{eqnarray}
\Gamma_{\alpha} &=& -\tau_y\otimes \sigma_{\alpha},\,\,
\Gamma_0 = \tau_z \otimes I_{\sigma}, \,\, \Gamma_4=\tau_x \otimes I_{\sigma}. \label{gammadef}
\end{eqnarray}
Here $I_{\tau}$ and $I_{\sigma}$ denote $2 \times 2$ identity
matrices and the index $\alpha$ takes values $1,2,3$. The matrices
$\Gamma_{\mu}$ satisfies the commutation relation
$\{\Gamma_{\mu},\Gamma_{\nu}\}=2\delta_{\mu\nu} I_{\tau} \otimes
I_{\sigma}$.

The model in Eq.~(\ref{eqn:ham0}) preserves inversion $\mathcal
{I}=I_{\tau} \otimes \sigma_y$ and mirror $M_y=\tau_x \otimes
\sigma_x$ symmetries, while time-reversal $T_0=\mathcal{K}$ (where
${\mathcal K}$ denotes complex conjugation), the four-fold
rotational symmetry $C_4^z$  and mirror along $x$, $M_x=\tau_x
\otimes \sigma_z$, are broken. In addition, the model preserves $M_x
T_0$. We note that for the model $M_z$ is defined through ${\mathcal
I}= M_x M_y M_z$ and that for $m_0=0$, $H_0$ preserves $C_4^z$ along
with other symmetries mentioned above and hosts a higher-order
topological semimetal phase.

The energy spectrum of Eq.~(\ref{eqn:ham0}) is given by
\begin{eqnarray}
E_{\pm,\pm} &=& \pm \sqrt{\sum_{i=1,5} a_i^2 \pm 2 |a_5|
\sqrt{a_1^2+a_4^2}}, \label{eqn:energy}
\end{eqnarray}
where $E_{+-}$ and $E_{--}$ correspond to the lowest conduction and
highest valence band respectively. For Fermi energy $\epsilon_F=0$,
the band spectrum in Eq.~(\ref{eqn:energy}) can be gapped or gapless
depending on the dimensionless parameters $\gamma=\gamma_0/\lambda$
and $m=m_0/\lambda$. It is evident from the top right panel of Fig.\
\ref{fig:spectrum} that the spectrum is gapless for
$-1.5\le\gamma\le1.5$ except the central hexagonal-like regime,
satisfying $|\gamma| \le 0.5$ and
\begin{eqnarray}
|m| & \le & \sqrt{(1-|\gamma|)+\frac{1}{2} \sqrt{3-8|\gamma|+4\gamma^2}}, \label{cond1}
\end{eqnarray}
where the spectrum is gapped.
\begin{figure}
    \vspace{-1\baselineskip}
    \includegraphics*[width=0.49\linewidth]{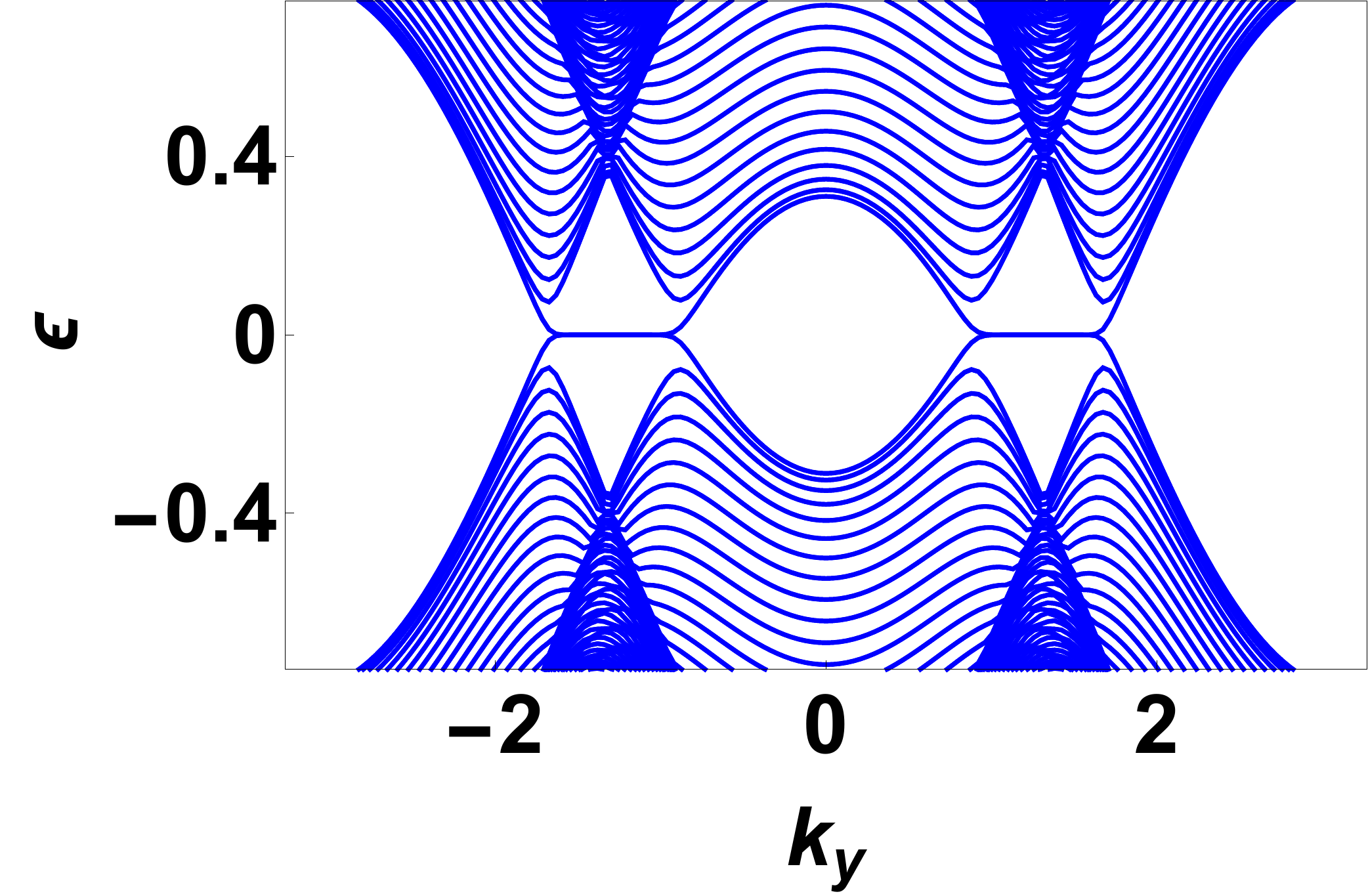}
    \includegraphics*[width=0.49\linewidth]{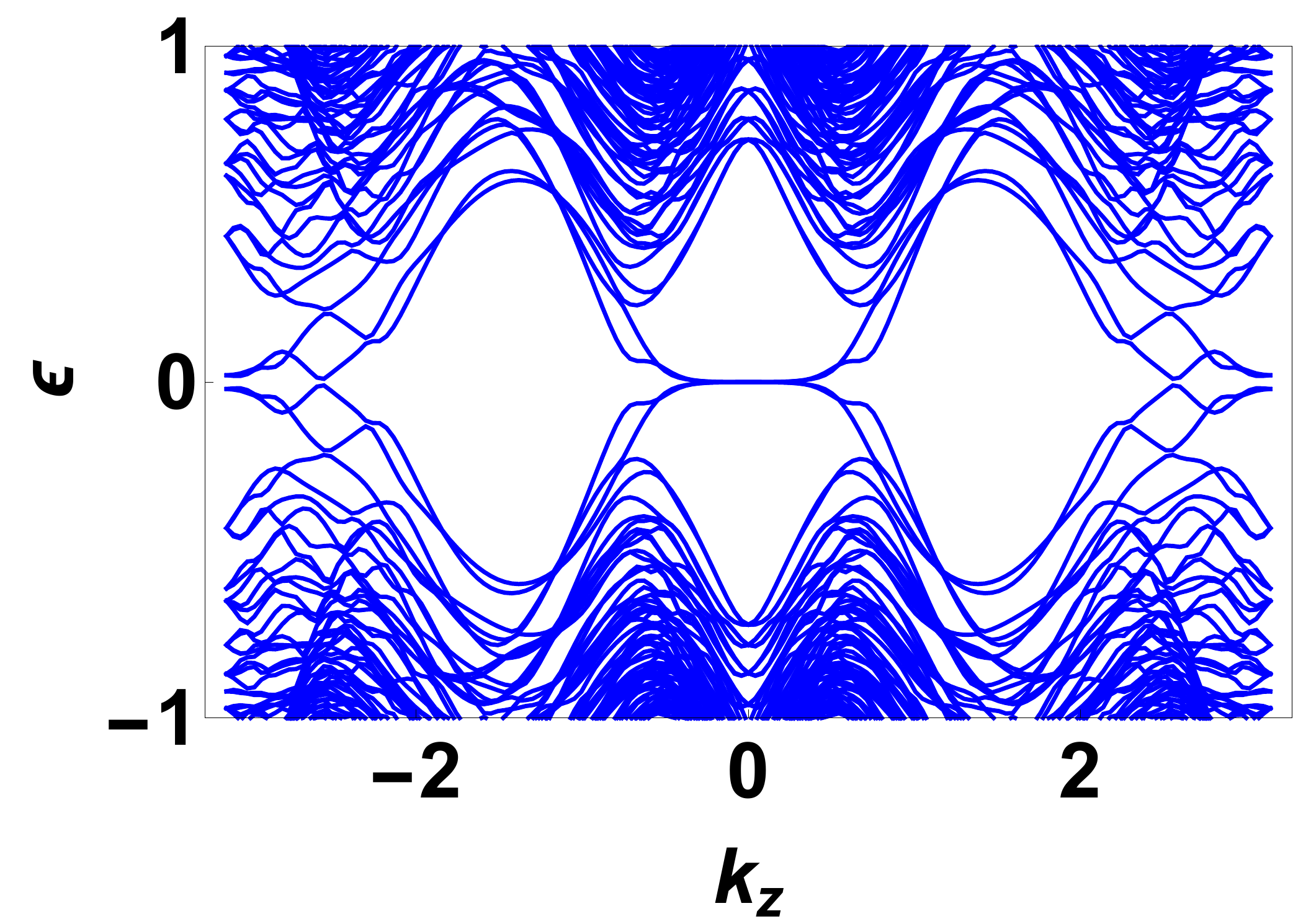}\\
        \includegraphics*[width=0.49\linewidth]{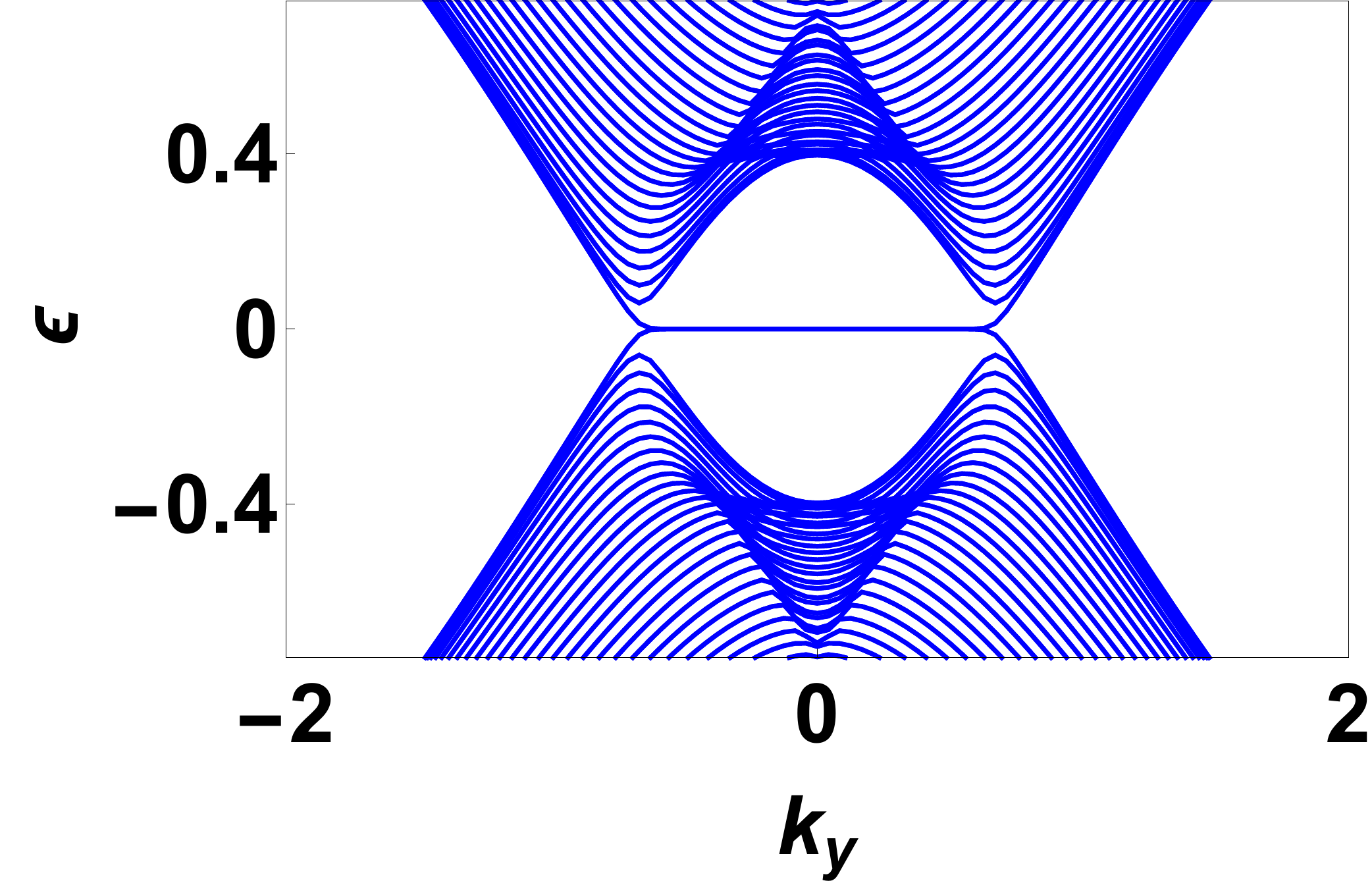}
    \includegraphics*[width=0.49\linewidth]{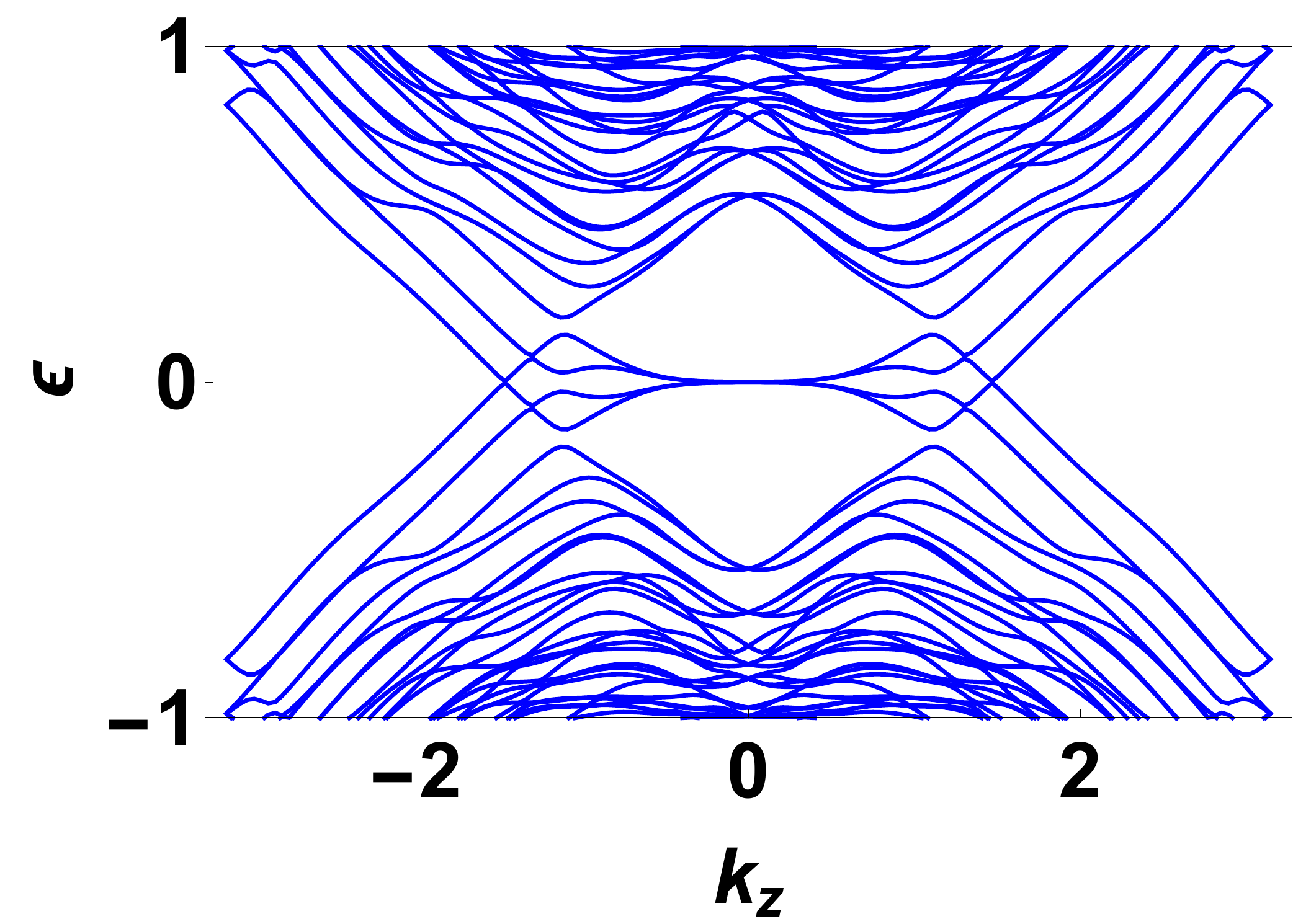}\\
    \includegraphics*[width=0.49\linewidth]{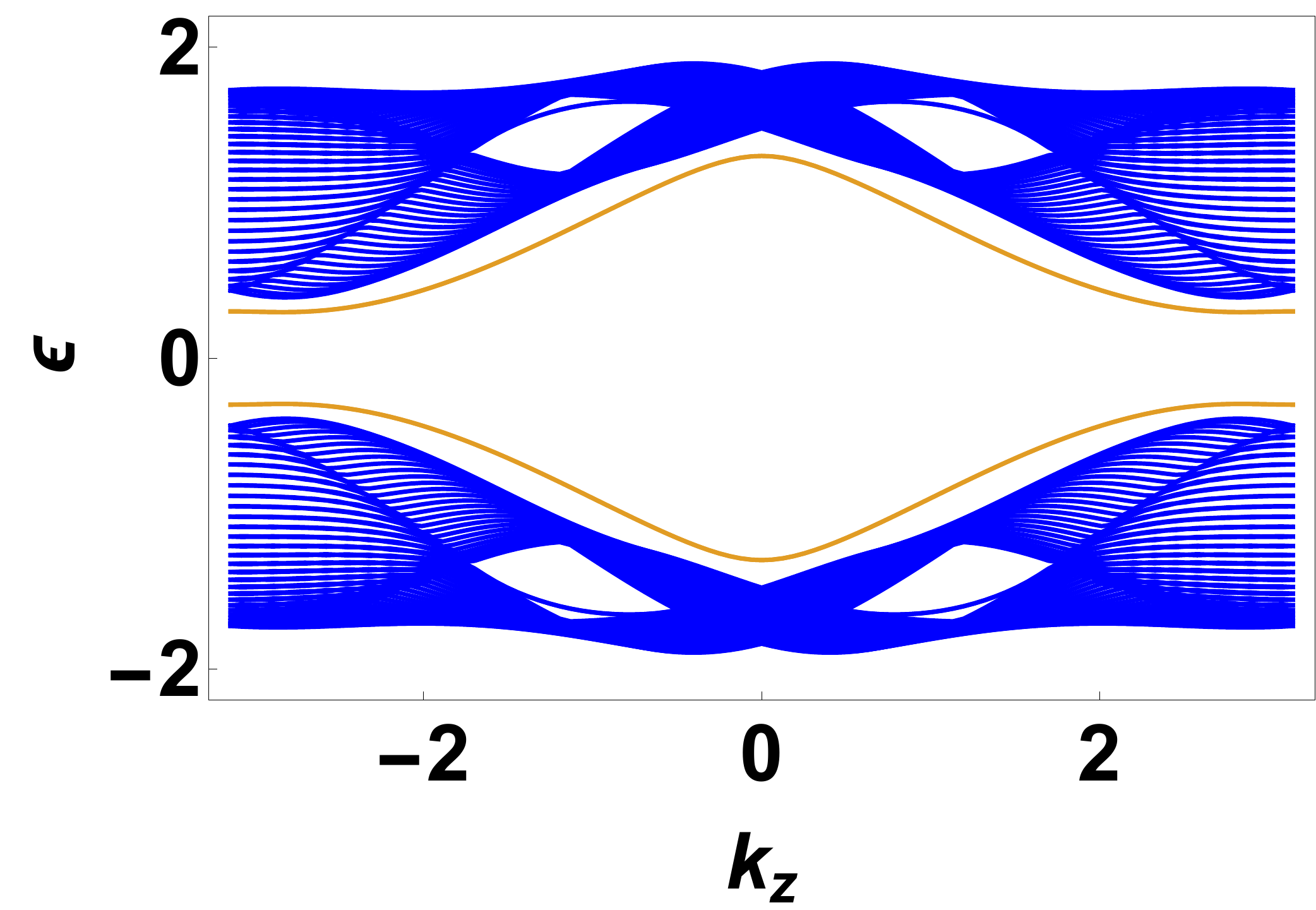}
    \includegraphics*[width=0.49\linewidth]{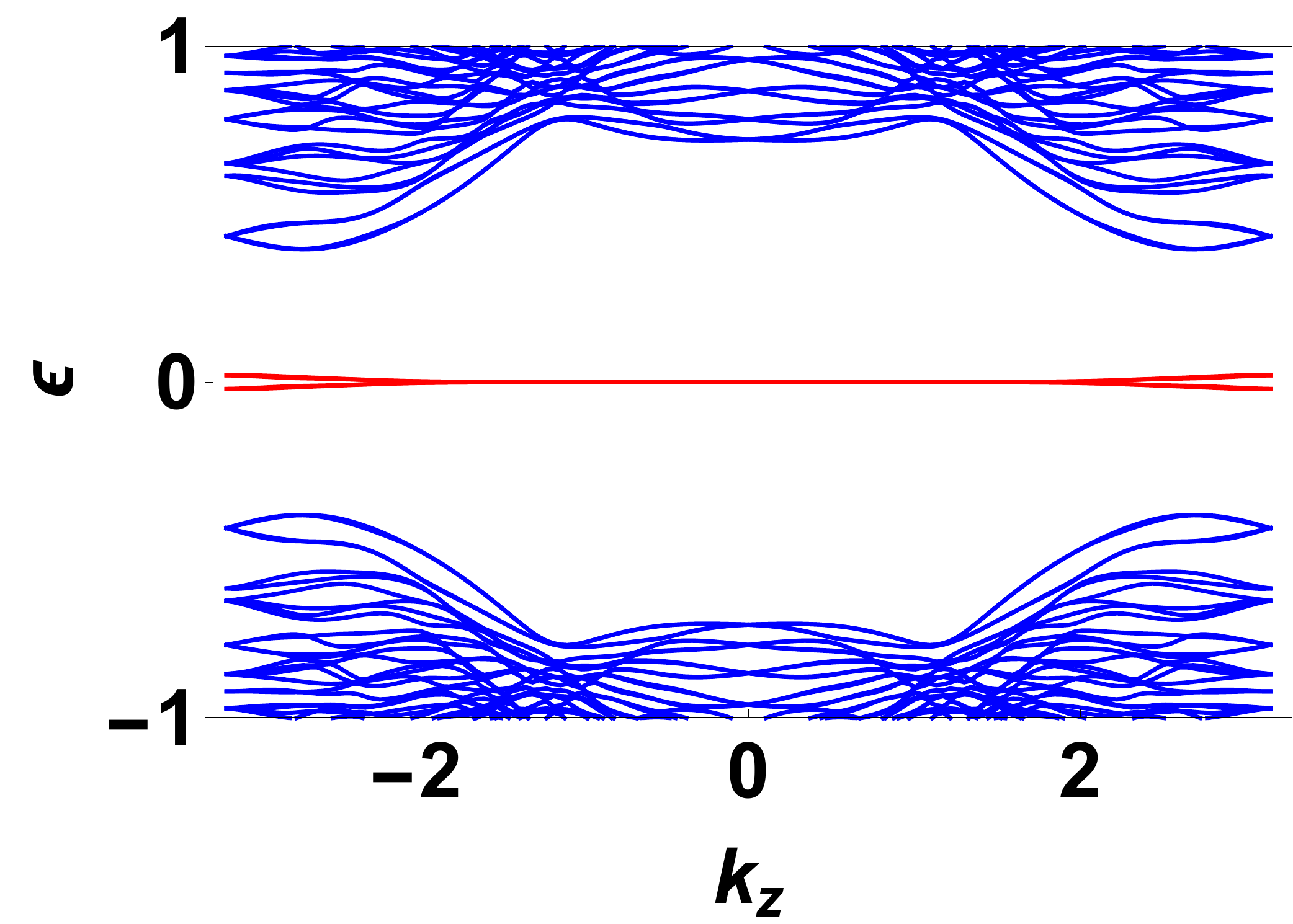}
\caption{Top Panels: Left: Plot of surface Fermi arc of undriven
Hamiltonian (Eq.~(\ref{eqn:ham0})) as a function of $k_y$ with open
boundary condition (OBC) along $z$ and $k_x=0$ for gapless bulk
regime with eight Weyl nodes ($\gamma=-0.2,\text{ }m=2.0$). Right:
The gapless hinge mode is obtained as a function of $k_z$ with OBC
along $x$ and $y$ with the same parameter set. It connects two
middle node at $k_z=0$. Middle Panels: Same as the corresponding top
panels but for gapless bulk with four Weyl nodes
($\gamma=-1.0,\text{ }m=0.75$). Bottom panels: Left: Energy spectrum
for gapped bulk ($\gamma=-0.2$, $m=0.4$) as a function of $k_z$ with
$k_x=0$ and OBC along $y$. The surface localized modes, indicated by
orange solid lines, are gapped. Right: The hinge mode, obtained with
OBC along $x$ and $y$ denoted by red line is gapless, reflecting the
typical quadrupolar insulating phase.}
    \label{fig:8wylsurfhinge}
\end{figure}

The gapless regime can further be divided into two regimes based on
the number of Weyl nodes. Note that the Weyl nodes in this
particular model lie in the $k_y-k_z$ plane (bottom panels of Fig.\
\ref{fig:spectrum}) (instead of the high symmetric lines $k_x=0$ and
$k_y=0$). The gapless regime with $|\gamma|<0.5$ exhibits eight Weyl
nodes which are connected through surface Fermi arc along the
$z$-surface, as shown in the top left panel of Fig.\
\ref{fig:8wylsurfhinge}. A further cut of this surface either along
$x$ or $y$ does not give rise to any hinge mode. Instead, the hinge
mode exists along $k_z$, connecting two Weyl nodes closest to
$k_z=0$ as depicted in the top right panel of Fig.\
\ref{fig:8wylsurfhinge}. Thus the hinge and surface modes are
perpendicular to each other in the present model . As we move away
from $|\gamma|< 0.5$ to $|\gamma|\ge 0.5$, four Weyl nodes
annihilate in pairs while the rest four retain. As before, we find
gapless surface and hinge modes connecting Weyl nodes at the center
of momenta $k_y=0$ and $k_z=0$, respectively (center left and right
panels of Fig.\ \ref{fig:8wylsurfhinge}). Finally, for the central
hexagonal-like regime with $|\gamma|<0.5$ shown in the right panel
of Fig.\ \ref{fig:spectrum}, all Weyl nodes annihilate in pairs,
resulting in gapped insulating phase. In this regime, the surface
bands denoted by orange solid line in the bottom left panel of Fig.\
\ref{fig:8wylsurfhinge} are gapped, whereas the hinge mode is found
to be gapless (see bottom right panel Fig.\
\ref{fig:8wylsurfhinge}), indicating the standard {\it quadrupolar}
topological insulating (QTI) phase.

\section{\label{sec:floquet_PT} Floquet Perturbation theory}

The studies of topological properties of a periodically driven
closed quantum system relies on computation of its Floquet
Hamiltonian $H_F$ which is related to its unitary evolution operator
$U(T,0)$ via the relation \cite{rev8}
\begin{eqnarray}
U(T,0) &=& {\mathcal T}_t \left[ e^{-i \int_0^T H(t)
dt/\hbar}\right] = e^{-i H_F T/\hbar}, \label{fldef}
\end{eqnarray}
where ${\mathcal T}_t$ denotes time ordering, $T= 2\pi/\omega_D$ is
the time period of the drive, $\omega_D$ is the drive frequency, and
$\hbar$ is Planck's constant. The knowledge of $H_F$ allows one to
compute Floquet eigenstates; it is well known that they host
non-trivial topological properties exhibiting transitions from
trivial to non-trivial topological phases as a function of the drive
frequency. It is well known that a driven system may host such
topological phases and associated transitions between them even when
the corresponding equilibrium Hamiltonian is topologically trivial
\cite{topo1,topo2,topo3,topo4,topo5}.

The exact computation of the Floquet Hamiltonian for a generic
quantum system is difficult; one therefore relies on several
perturbative method for its computation \cite{rev8,flrev}. One such
perturbative scheme is the Magnus expansion where $T$ is taken as
the perturbation parameter. However, the convergence of such an
expansion is difficult to ascertain; moreover, it almost always
fails to provide qualitatively accurate results in the intermediate
or low frequency regimes. In contrast, the properties of $H_F$ in
the intermediate drive frequency regime is known to be
well-described by the Floquet perturbation theory which uses the
inverse of the drive amplitude as the perturbation parameter
\cite{dsenref,cooperref, roopayanref}.

In this section, we shall provide an analytic computation of the
Floquet Hamiltonian of the model in the presence of a periodic drive
which is implemented by making $\gamma$ a periodic function of time:
$\gamma(t)=\gamma_0 + \gamma_1(t)$ using Floquet perturbation theory
(for a review of this method, see Ref.\ \onlinecite{rev8}). The
precise time dependence of $\gamma_1(t)$ depends on the protocol. In
this section, we shall study two such protocols. The first, which
constitutes a square pulse, leads to
\begin{align}
\gamma_1(t)=\begin{cases} -\gamma_1, & 0\leq t< \frac{T}{2}\\ \gamma_1, &\frac{T}{2}\leq t\leq T, \end{cases} \label{sqp}
\end{align}
where $\gamma_1$ is the amplitude of the pulse and $T$ denotes its
time period. The second constitutes a continuous protocol for which
\begin{eqnarray}
\gamma_1(t) = \gamma_1 \cos \omega_D t.\label{cp}
\end{eqnarray}
For implementing the FPT, we write $H(t)=H_0 +H'(t)$, where \begin{eqnarray}
H'(t)=\gamma_1(t)(\Gamma_2+\Gamma_4).
\end{eqnarray}
Note that $H'(t)$ does not break any of the additional symmetries discussed earlier.

We focus on the regime of large drive amplitude, i.e., $\gamma_1\gg
\gamma_0,\lambda, m_0$. In this case, we can treat $H'(t)$ exactly
and $H_0$ perturbatively to find the Floquet unitary and hence
$H_F$. The first term in such an expansion constitutes the unitary
evolution operator $U_0(t,0)$ given by
\begin{eqnarray}
U_0(t,0) &=&  {\mathcal T}_t \exp\left[\frac{-i}{\hbar} \int_0^t dt' H'(t') \right].
\label{zerothorder}
\end{eqnarray}
For the square pulse protocol given by Eq.\ \ref{sqp}, this yields
\begin{eqnarray}
U_0^{(s)}(t,0) &=& \exp[i \gamma_1 t (\Gamma_2+\Gamma_4)/\hbar] \quad 0\le t\le T/2, \label{zerosq} \\
&=& \exp[i \gamma_1 (T-t) (\Gamma_2+\Gamma_4)/\hbar] \quad T/2 \le
t\le T, \nonumber
\end{eqnarray}
which yields $U_0^{(s)}(T,0)=I$ and $H_F^{(0;s)}=0$. For the continuous drive protocol given by Eq.\ \ref{cp}, one gets
\begin{eqnarray}
U_0^{(c)} &=& \exp \left[-\frac{i \gamma_1 \sin \omega_D t}{\omega_D
\hbar}(\Gamma_2 +\Gamma_4) \right]\label{zeroc}
\end{eqnarray}
which also yields $H_F^{(0;c)}=0$.

Next we consider the first order term in the Floquet perturbation
theory which is given by
\begin{eqnarray}
U_1(T,0) &=&  \frac{-i}{\hbar} \int_0^T U_0^{\dagger}(t,0) H_0 U_0(t,0), \nonumber\\
H_F^{(1)} &=& \frac{i\hbar}{T} U_1(T,0).
\end{eqnarray}
A straightforward calculation using Eqs.\ \ref{eqn:ham0} and
\ref{zerosq} yields for the square pulse protocol
\begin{widetext}
\begin{align}
H_{F}^{(1);s}=&\nonumber
\frac{1}{2}\left((a_1+a_2)(\Gamma_4+\Gamma_2)+i a_5(\Gamma_2\Gamma_3+\Gamma_3\Gamma_4)\right)\nonumber
+\frac{\hbar\sin{(\sqrt{2}\gamma_1 T/\hbar)}}{2\sqrt{2}\gamma_1 T} \left((a_1-a_2)(\Gamma_4-\Gamma_2)+i a_5(\Gamma_2\Gamma_3-\Gamma_3\Gamma_4)\right)\nonumber\\
&+i \frac{\hbar(\cos{(\sqrt{2}\gamma_1 T/\hbar)}-1)}{2\gamma_1
T}\left((a_1-a_2)\Gamma_2\Gamma_4+(\Gamma_2+\Gamma_4)(a_3\Gamma_3+a_4\Gamma_1)+ia_5\Gamma_3
\right) +\frac{\hbar\sin{(\sqrt{2}\gamma_1
T/\hbar)}}{\sqrt{2}\gamma_1 T}(a_3\Gamma_3+a_4\Gamma_1).
\label{eqn:floham1s}
\end{align}
\end{widetext}
Note that the symmetries of the undriven Hamiltonian is retained in
the effective Floquet Hamiltonian $H_F^{(1);s}$. A similar
calculation for the continuous protocol using  Eqs.\ \ref{eqn:ham0}
and \ref{zeroc} yields
\begin{widetext}
\begin{eqnarray}
    H_F^{(1);c} &=& \frac{1}{2}\left[(a_1+a_2)(\Gamma_2+\Gamma_4)+ia_5(\Gamma_2\Gamma_3+\Gamma_3\Gamma_4)\right]+\mathcal{J}_0\left(\frac{\sqrt{2}\gamma_1 T}{\hbar\pi} \right) \Big[a_3\Gamma_3+a_4\Gamma_1+\frac{1}{2}((a_1-a_2)(\Gamma_4-\Gamma_2)
    \nonumber\\
    &&  +ia_5(\Gamma_2\Gamma_3-\Gamma_3\Gamma_4))\Big], \label{eqn:floham1c}
\end{eqnarray}
\end{widetext}
where $J_0$ denotes zeroth order Bessel function of the first kind
and we have used the identity $\exp[i \alpha \sin x]= \sum_n
J_n(\alpha) \exp[i n x]$.

We note that for the square pulse protocol $H_F^{(1);s}$ takes a
particularly simple form around $\sqrt{2} \gamma_1 T/\hbar= 2 n \pi$
(for $n\ne 0 \in Z$) where only the first term in Eq.\
\ref{eqn:floham1s} survives. We shall see in Sec.\ \ref{2pipoint}
that this leads to presence of ring of Weyl nodes in the spectrum of
$H_F^{(1);s}$. A similar simplification occurs for the continuous
drive protocol at $\sqrt{2} \gamma_1 T = \pi \alpha_n$, where
$\alpha_n$ denotes the position of the $n^{\rm th}$ zero of the
Bessel function. These features, for either square pulse or
continuous protocols, are difficult to obtain within a Magnus
expansion as shown via explicit calculation in the appendix; in
fact, it can be shown that our results in Eqs.\ \ref{eqn:floham1s}
and \ref{eqn:floham1c} constitute an infinite re-summation of a
class of terms in the Magnus expansion \cite{ergoref}.

Next, we compute the second order terms in the perturbative
expansion. The expression for such terms can be written, in terms of
$H_0(t)= U_0^{\dagger} (t,0) H_0 U_0(t,0)$ as
\begin{eqnarray}
U_2(T,0) &=& \left(\frac{-i}{\hbar}\right)^2 \int_0^T dt_1 H_0(t_1) \int_0^{t_1} dt_2 H_0(t_2), \nonumber\\
H_F^{(2)} &=& \frac{i \hbar}{T} \left(U_2(T,0)-\frac{1}{2}
U_1^2(T,0)\right). \label{order2}
\end{eqnarray}
For the square pulse protocol given by Eq. \ref{sqp}, it turns out
that $U_0(t,0)= U_0(T-t,T/2)$ for all $t\le T/2$. It can be shown
that this symmetry allows one to write
\begin{equation}
U_2^{s}(T,0)=2 \left(\frac{-i}{\hbar}\right)^2 \int_0^{T/2}dt_1
H_0(t_1)\int_0^{T/2}dt_2  H_0(t_2),
    \label{eqn:U2}
\end{equation}
thereby implying that $U_2^{s}(T,0)=\frac{1}{2}[U_1^{s}(T,0)]^2$.
Thus the second order correction, $H_F^{(2);s}$, vanishes
identically. So for the square pulse protocol, $H_F^{(1);s}$
provides the Floquet Hamiltonian up to third order in perturbation
theory.

The symmetry property mentioned above does not hold for the
continuous protocol given in Eq.\ \ref{cp}. In this case one obtains
a finite contribution to the second order Floquet Hamiltonian. The
computation is straightforward, though somewhat cumbersome. The
final result is
\begin{widetext}
\begin{eqnarray}
H_F^{(2);c} &=& -\frac{1}{\sqrt{2}} \sum_{n=-\infty}^{\infty}
\frac{\mathcal{J}_{2n+1}\left(\frac{\sqrt{2} \gamma_1 T}{\hbar \pi} \right)}
{(2n+1)\hbar \omega_D} \Big[ \left((a_1^2-a_2^2+a_5^2)
(\Gamma_2-\Gamma_4)-2(a_1+a_2)(a_3\Gamma_3+a_4\Gamma_1)
-2 i a_1a_5(\Gamma_2+\Gamma_4)\Gamma_3 \right.\nonumber\\
&& \left. +2i a_3a_5\Gamma_2\Gamma_4\right) +
\mathcal{J}_0\left(\frac{\sqrt{2} \gamma_1 T}{\hbar \pi }
\right)\left( \left((a_1-a_2)^2+a_5^2+2(a_3^2+a_4^2)
\right)(\Gamma_2+\Gamma_4) + 2 i a_1 a_5
(\Gamma_2-\Gamma_4)\Gamma_3\right)\Big]. \label{floham2c}
\end{eqnarray}
\end{widetext}
We note that for $\hbar \omega_D \sim {\rm O}(\gamma_1)$, the second
order Floquet terms are suppressed by a factor of ${\rm
O}(1/\gamma_1)$; thus in this regime, we expect the first order term
to be reasonably accurate.

Before closing this section, we note that the presence of such
gapless first order Floquet Hamiltonian implies that at least at the
high-frequency regime, the exact Floquet Hamiltonian will at most
have a tiny gap in its spectrum. This is due to the fact that such a
gap can only originate from higher order terms which are small in
the high-frequency regime. We shall discuss this issue and its
implication for the hinge modes of the model in more details in
Secs.\ \ref{flp} and \ref{dyncor}.

\section{Floquet Phases}\label{flp}
In this section, we analyze the spectrum of $H_F^{(1)}$ in the
intermediate and high-frequency regime and compare the result with
those obtained from numerical computation of exact $H_F$. For the
sake of concreteness, we shall mainly focus on the regime where
$H(t=0)$ hosts a quadrupolar insulating ground state. In Sec.\
\ref{bpsec}, we discuss the Floquet phases of the model. This is
followed by analysis of the properties of $H_F$ for some special
drive frequencies in Secs.\ \ref{pipoint} and \ref{2pipoint}.

\subsection{The bulk spectrum}
\label{bpsec}

The computation of the exact Floquet Hamiltonian, which shall be
used for obtaining the Floquet phases, can be carried out as
follows. For the square pulse protocol, we write the Hamiltonian
$H[\gamma=\gamma_0\pm \gamma_1] = H_{\pm}$. In terms of $H_{\pm}$,
one can write the evolution operator as
\begin{eqnarray}
U_s^{\rm ex}(T,0) &=& e^{-i H_+ T/(2\hbar)} e^{-i H_-T/(2\hbar)}.
\label{unitevol1}
\end{eqnarray}
To evaluate $U_s^{\rm ex}$, we first obtain the eigenvalues and
eigenvectors of $H_{\pm}$. This can be done analytically for
periodic boundary condition, but needs to be done numerically for
open boundary condition. We use the latter here for extracting the
properties of the Floquet phases. We define the corresponding
eigenvalues and eigenvectors as $\epsilon_n^{\pm}$ and
$|n^{\pm}\rangle$. In the basis of these eigenstates $U_s^{\rm
ex}(T,0)$ can be written as
\begin{eqnarray}
U_s^{\rm ex}(T,0) &=& \sum_{n_+,m_-} e^{-i (\epsilon_n^+
+\epsilon_m^-)T/(2\hbar)} c^{-+}_{mn} |m^-\rangle\langle n^+|.
\label{urep}
\end{eqnarray}
The diagonalization of $U_s^{\rm ex}$ in this basis leads to the
eigenvalues $\lambda_p= e^{i \theta_p}$ and the corresponding
eigenvectors $|p\rangle$. The exact Floquet eigenvalues and
eigenvectors can then be found as
\begin{eqnarray}
H_F^{\rm ex} &=& \sum_p \epsilon_{p}^{F;\rm ex} |p\rangle\langle p|,
\,\, \epsilon_{p}^{F;\rm ex} = \frac{\hbar}{T} \arccos\{ {\rm
Re}[\lambda_p] \}. \label{floex}
\end{eqnarray}
The ground state of $H_F^{\rm ex}$ is then used to distinguish
between the different Floquet phases. This is typically done by
examining the bulk gap and surface modes in the Floquet spectrum and
also by determining the presence of the hinge modes. A similar
analysis is done using the perturbative Floquet Hamiltonian
$H_{F}^{(1);s}$.

For the continuous protocol, the computation of the Floquet
Hamiltonian turns to be more challenging. The procedure involves
division of the evolution operator into $N$ trotter steps; the width
of these steps $\delta t= T/N$ are chosen such that $H[t_j] \simeq
H[t_j+\delta t]$ for any Trotter slice $t_j$. For our purpose,
numerically we find that $N=50$ is enough to satisfy this criteria;
all data corresponding to $N>50$ coincides with their $N\simeq 50$
counterparts for all frequencies studied in this work. Writing the
eigenvalues and eigenfunctions of $H(t_j)$ as $\epsilon_n^j$ and
$|n^j\rangle$ respectively, we express the evolution operator as
\begin{eqnarray}
U_c^{\rm ex} &=&  \prod_{j=1,N} \sum_n e^{-i \epsilon_n^j T/\hbar}
|n^j\rangle \langle n^j|. \label{evocon}
\end{eqnarray}
This is then diagonalized to find the corresponding eigenvalues and
eigenvectors. The rest of the analysis follows the same steps as
detailed for the discrete case. The results obtained from the exact
Floquet Hamiltonian are compared to those obtained from perturbative
result $H_F^c = H_F^{(1)c}+ H_F^{(2)c}$.

\begin{figure}
    \vspace{-1\baselineskip}
    \includegraphics*[width=0.49\linewidth]{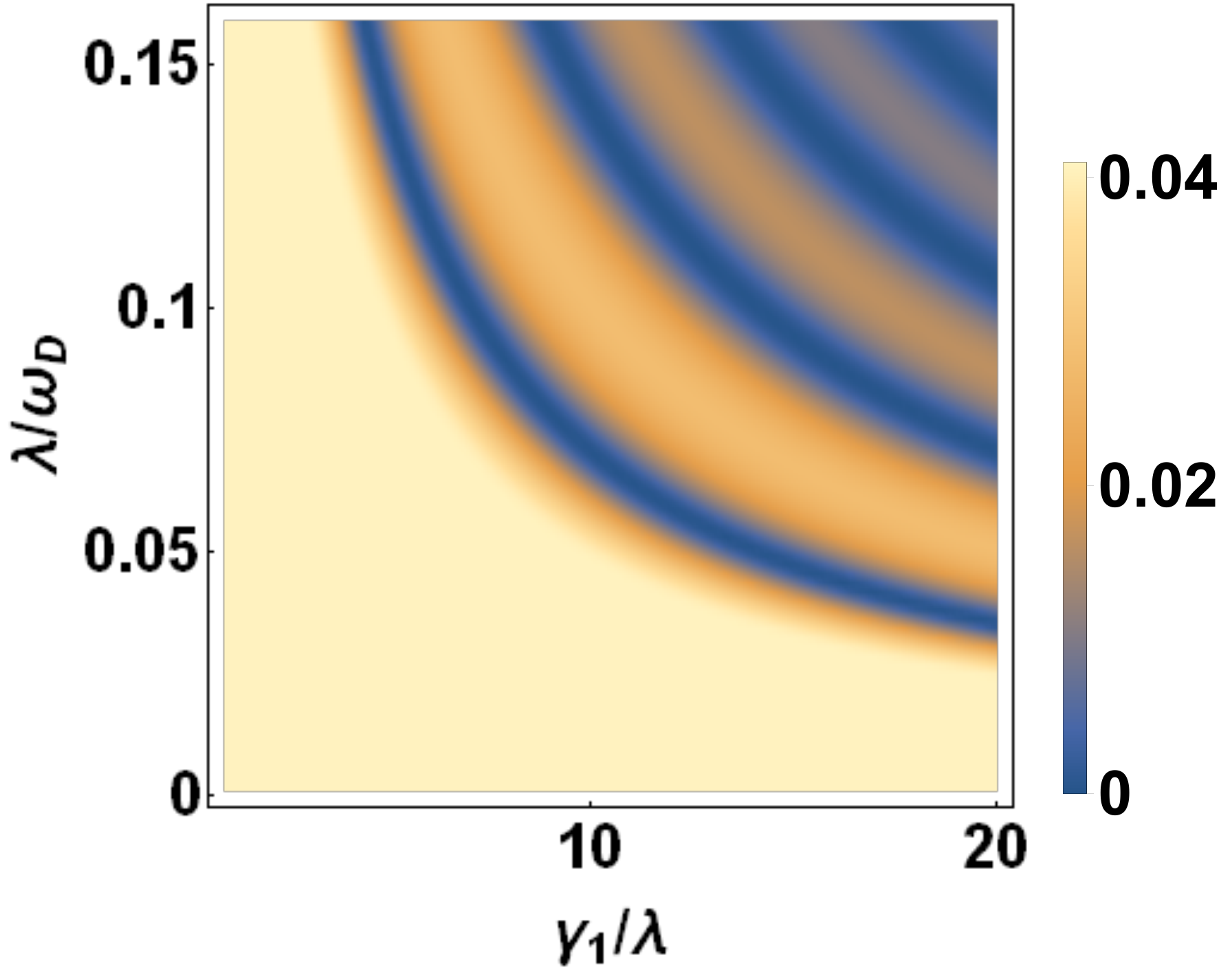}
    \includegraphics*[width=0.49\linewidth]{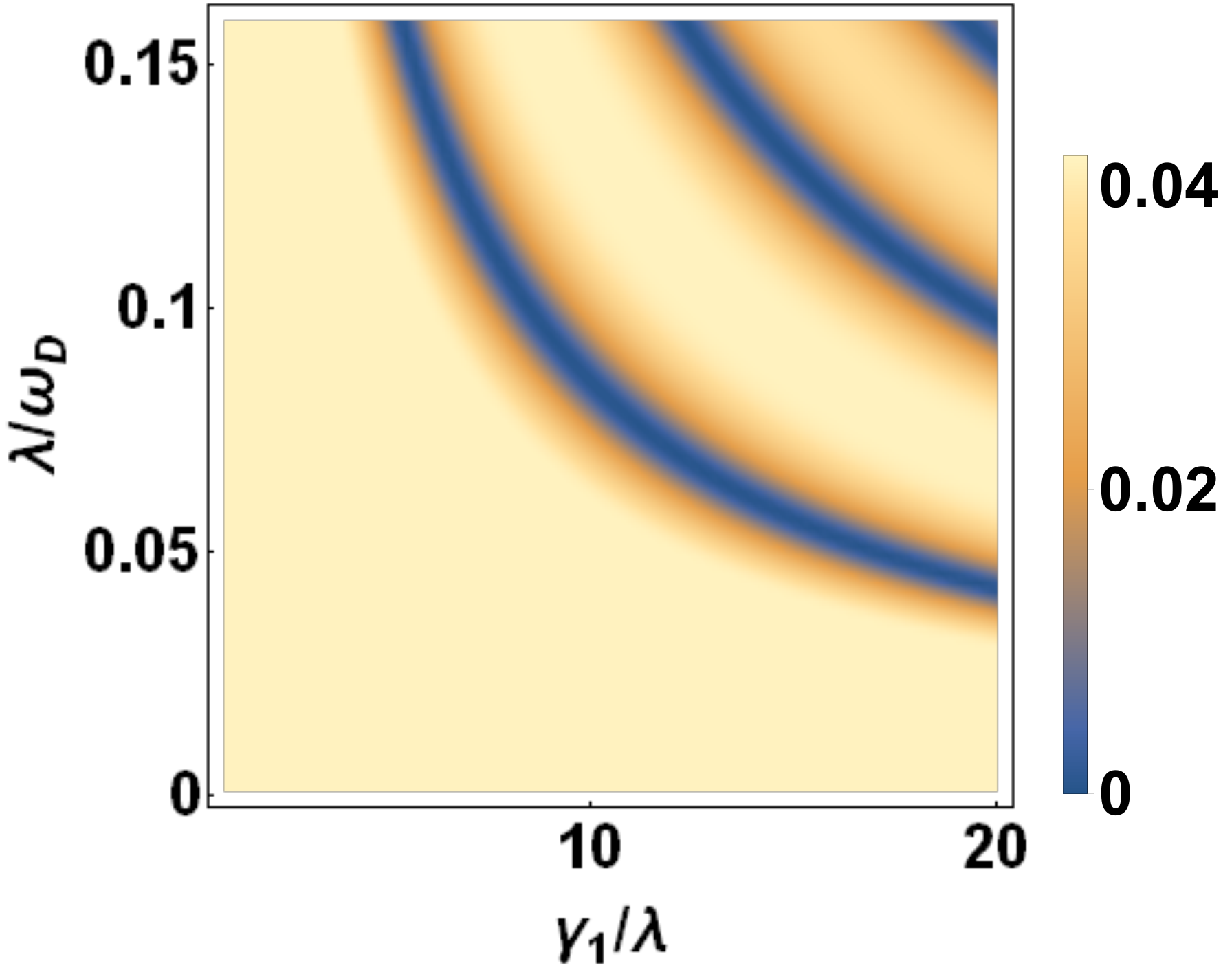}
    \includegraphics*[width=0.49\linewidth]{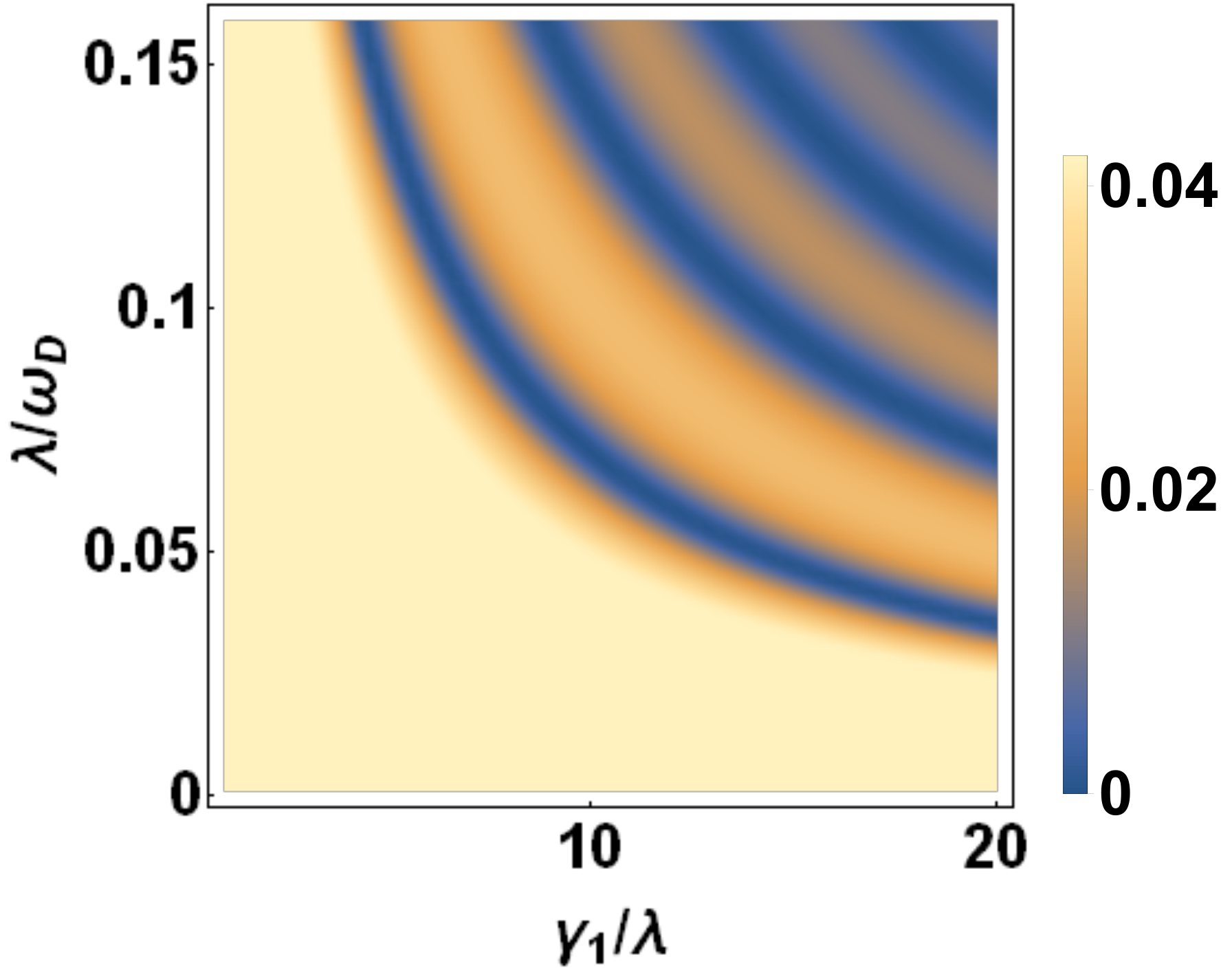}
    \includegraphics*[width=0.49\linewidth]{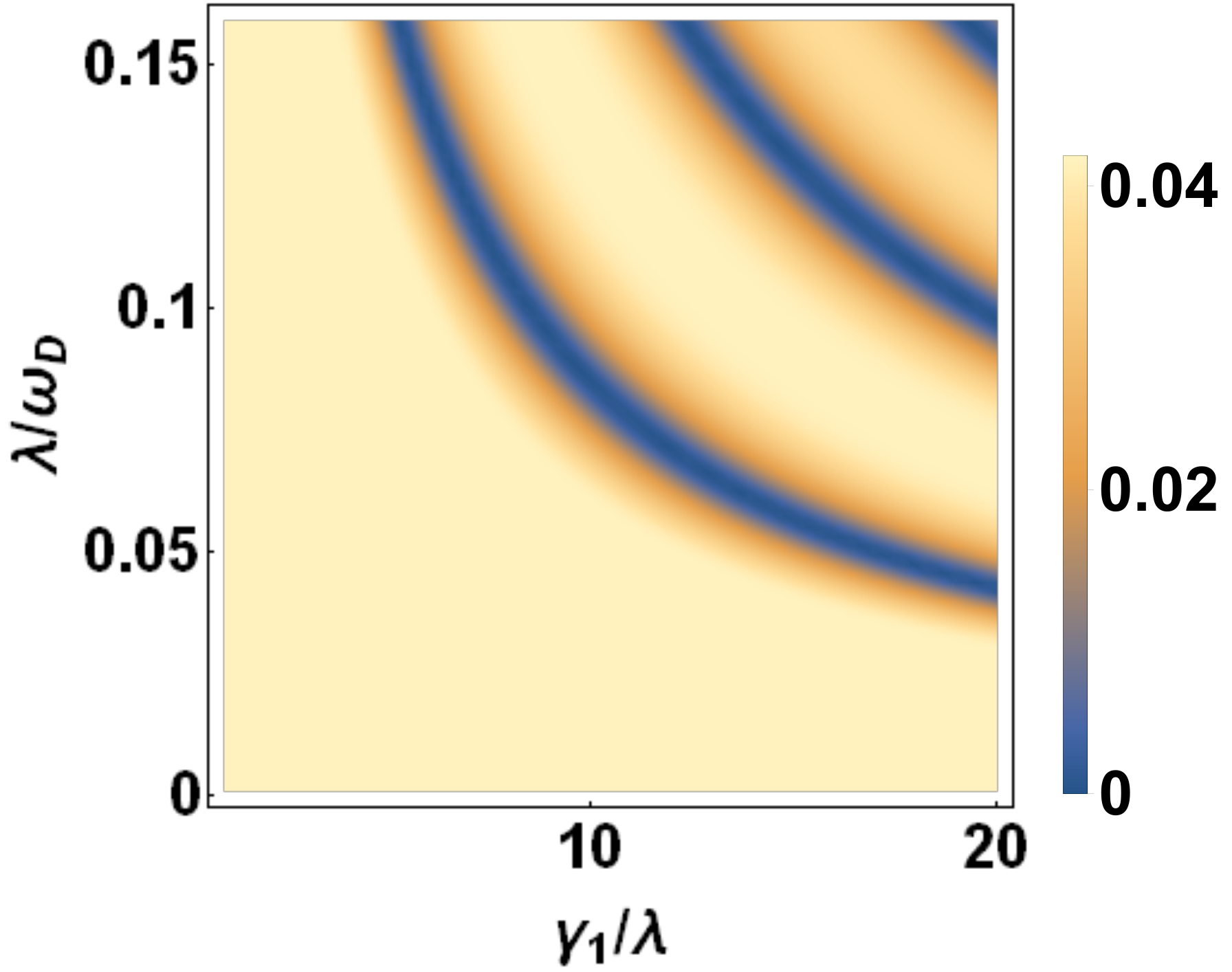}
\caption{Top left Panel: Plot of the smallest Floquet bandgap for
$H_F^{(1);s}$ as a function of $\lambda/(\hbar \omega_D)$ and
$\gamma_1/\lambda$. Top right panel: A similar plot corresponding to
$H_F^{(1);c}$. Bottom Left (right) panel: Similar plot corresponding
to the exact Floquet Hamiltonian for the discrete square pulse
(continuous cosine drive) protocols. For all plots $\gamma=-0.35$
and $m=-0.08$. See text for details.}
    \label{fig:flph}
\end{figure}

The plot of the smallest bandgap (the smallest difference in
quasienergy between the lowest positive and highest negative
quasienergy bands) of the Floquet Hamiltonian is shown in Fig.\
\ref{fig:flph}. The top panel shows the band gap as obtained using
the expressions of $H_F$ from FPT as given in Eqs.\
\ref{eqn:floham1s}, \ref{eqn:floham1c} and \ref{floham2c}. The top
left (right) panel shows the Floquet perturbation theory results for
discrete square pulse (continuous cosine drive) protocols. These
plots are compared to the exact results plotted in the respective
bottom panels. The plots shows remarkable match in the high ($\hbar
\omega_D \gg \gamma_1$) and intermediate frequency regime ($\hbar
\omega_D \sim \gamma_1$) which clearly reflects the accuracy of FPT
in this regime.

The plots in both the top and bottom panel indicate the presence of
regimes with extremely small gaps in the Floquet spectrum as a
function of $\gamma_1$ or $T$. These regimes appear around drive
frequencies $T_c$ and amplitude $\gamma_{1c}$ which satisfy
$\sqrt{2} \gamma_{1c} T_c/\hbar= 2 n \pi$ for the discrete protocol
and along $\sqrt{2} \gamma_{1c} T_c/\hbar= \pi\alpha_n$ for the
continuous protocol as shown in Fig.\ \ref{fig:flph}. This feature
is easy to understand from the expressions of the first order
Floquet Hamiltonians given in Eqs.\ \ref{eqn:floham1s} and
\ref{eqn:floham1c}; for these drive frequencies and amplitude both
of these Hamiltonians reduce to
\begin{eqnarray}
H_F^{\ast} &=&
\frac{1}{2}\left[(a_1+a_2)(\Gamma_2+\Gamma_4)+ia_5(\Gamma_2\Gamma_3+\Gamma_3\Gamma_4)\right],\nonumber\\
\label{hfstar}
\end{eqnarray}
which, as we shall see in Sec.\ \ref{2pipoint}, supports gapless
bulk modes. Further, as shall be analyzed in details in Sec.\
\ref{2pipoint}, a relative sign change of the mass of the gapped
edge modes occur at both the $x$ and one of the $y$ edges of the
system as the system traverses through these nearly gapless points.
The spectrum of the exact Floquet Hamiltonian, in contrast, retains
a small gap across these points; we shall discuss this feature in
details in Sec.\ \ref{2pipoint}.
\begin{figure}
    \vspace{-1\baselineskip}
    \includegraphics*[width=0.48\linewidth]{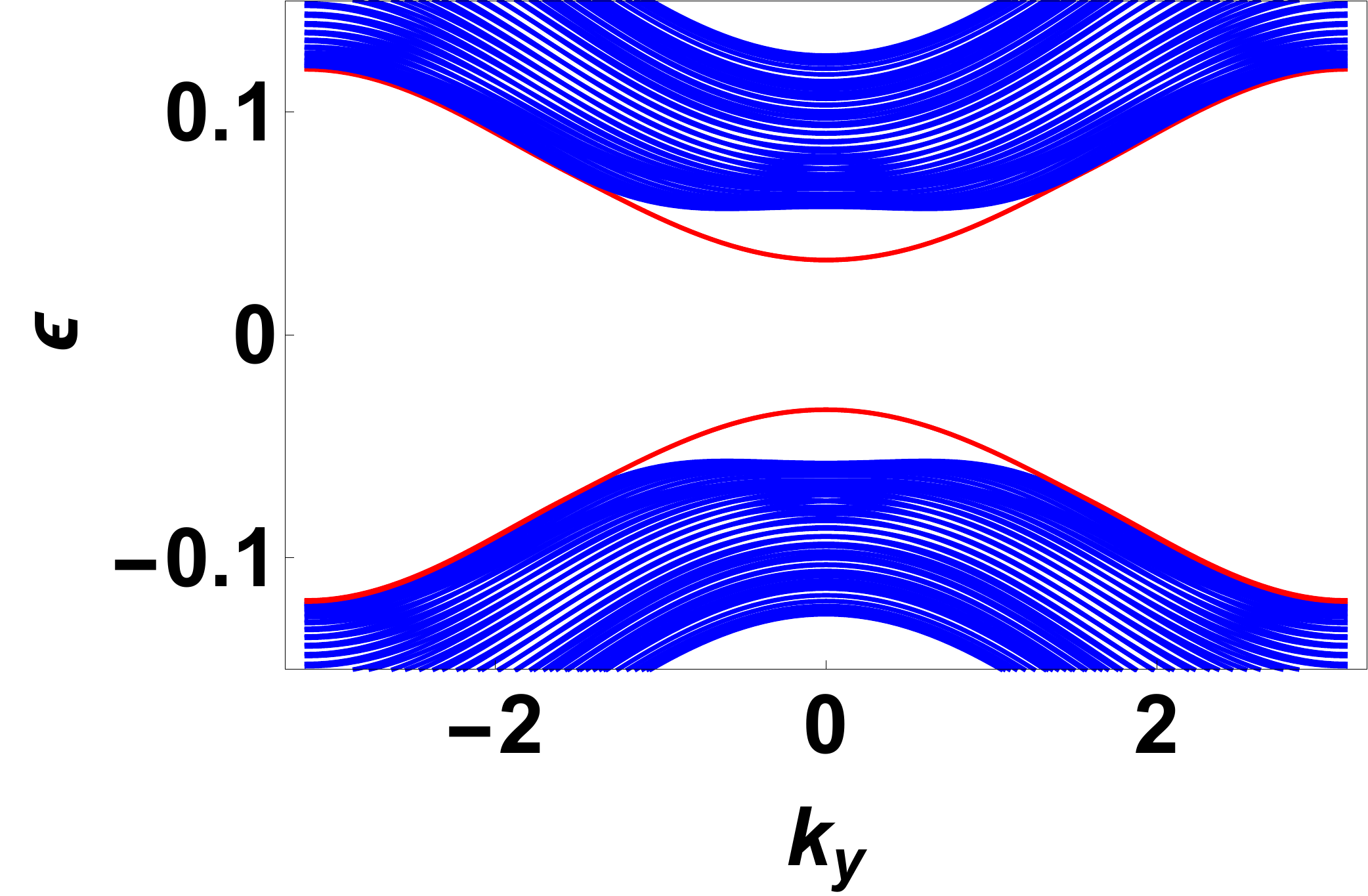}
    \includegraphics*[width=0.48\linewidth]{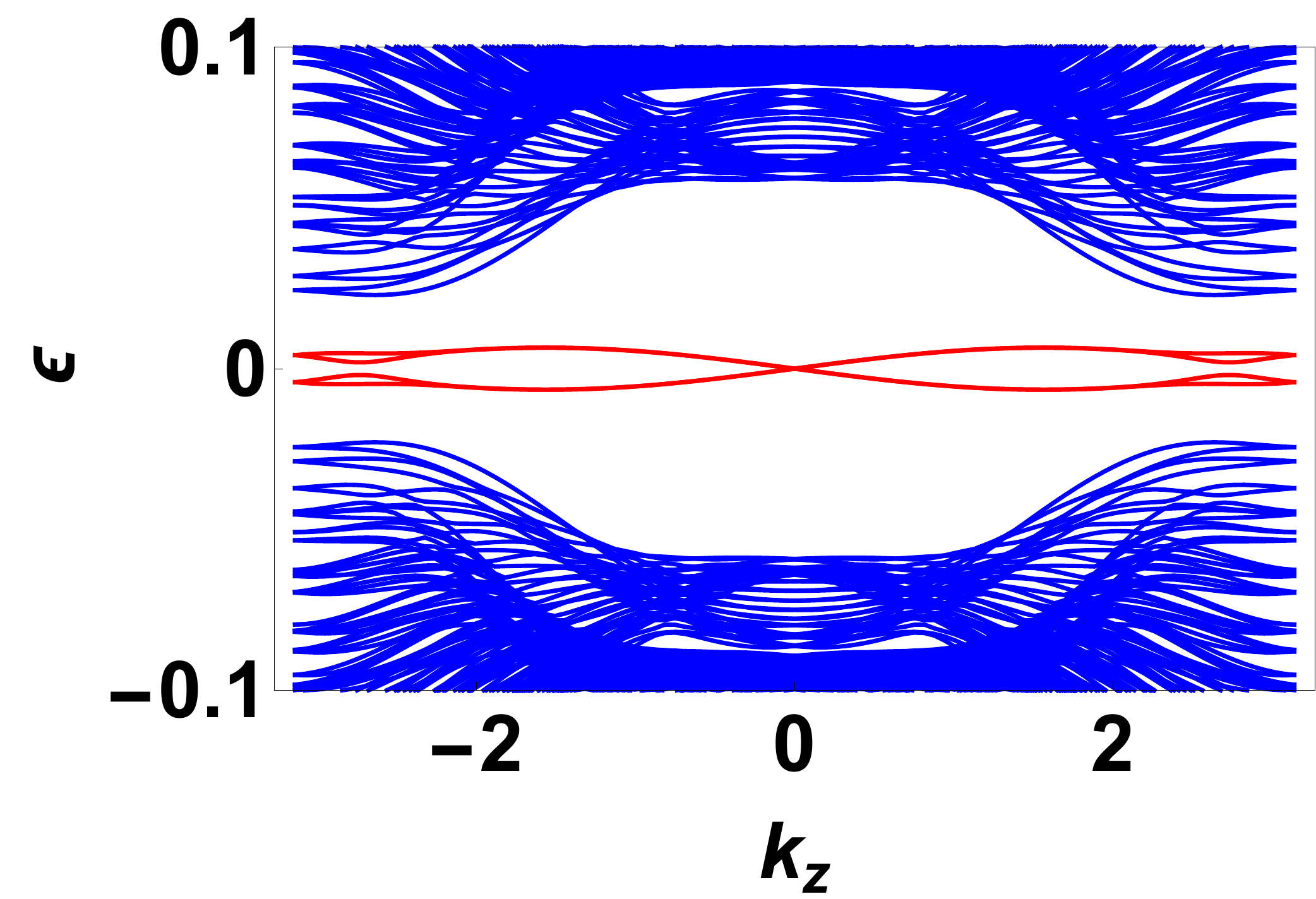}
    \includegraphics*[width=0.48\linewidth]{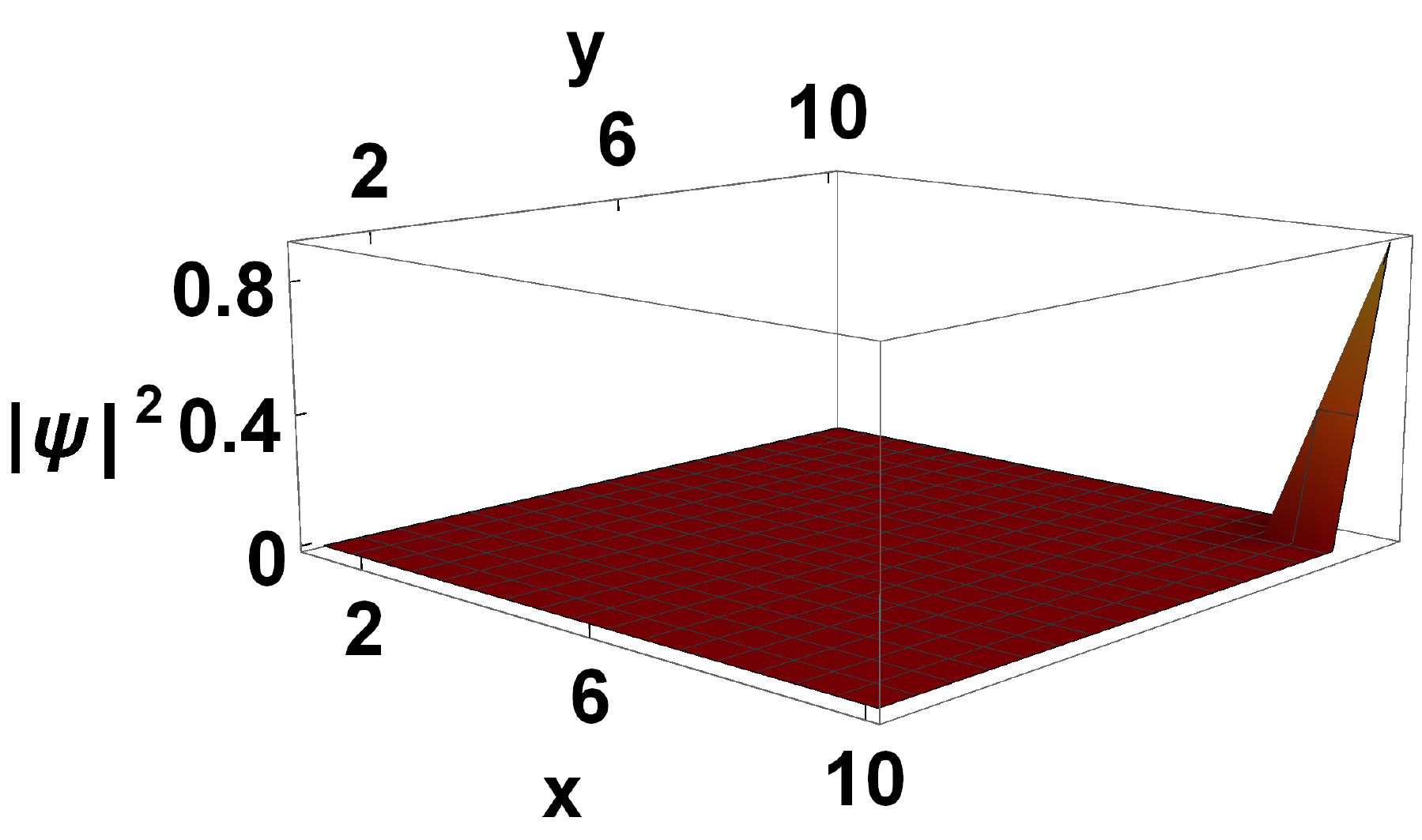}
    \includegraphics*[width=0.48\linewidth]{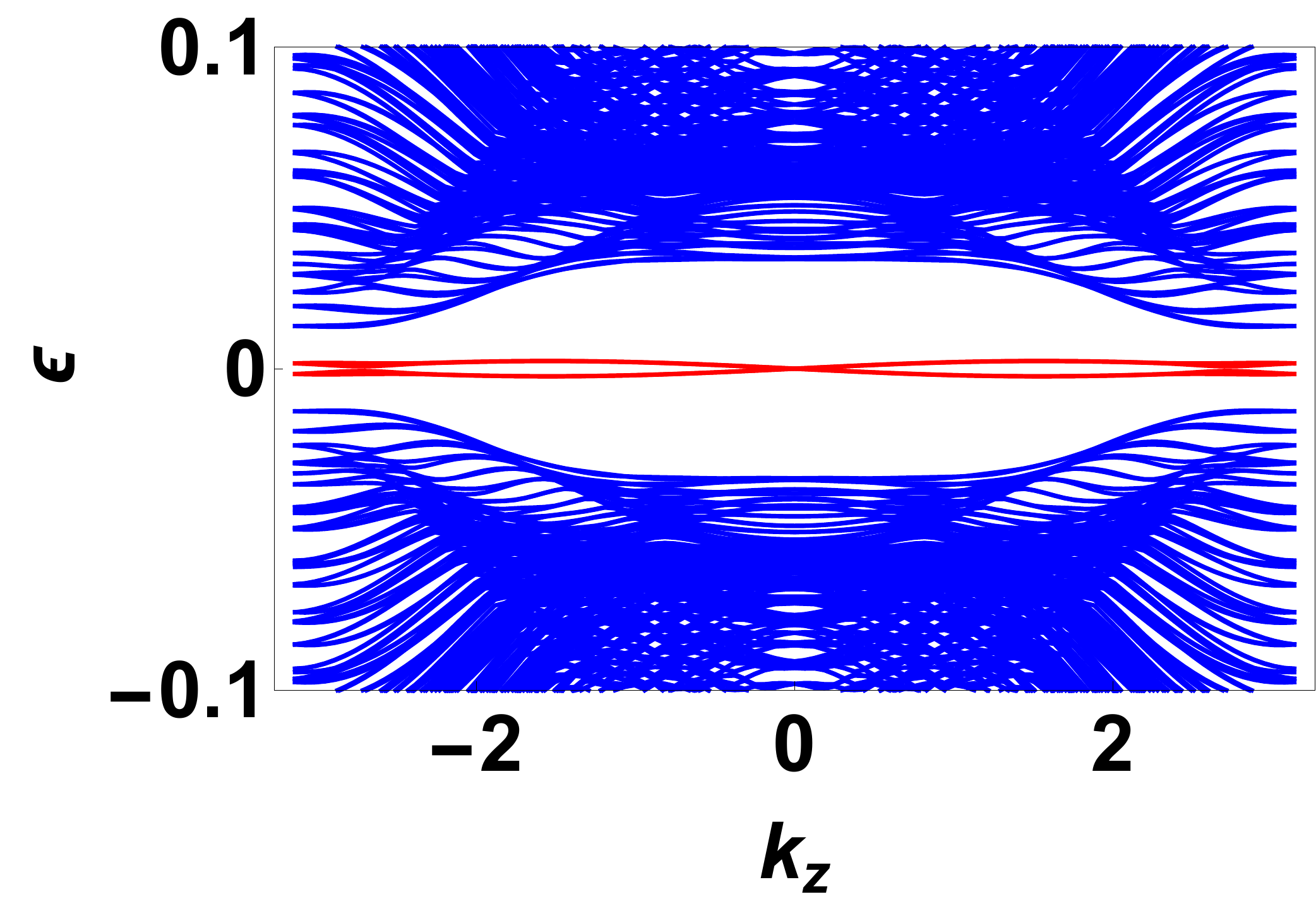}
\caption{Top left panel: Plot of energy spectrum of the first order
Floquet Hamiltonian $H_F^{(1);s}$ as a function of $k_y$ for $k_z=2$
with  PBC along $z$ and $y$ and OBC along $x$ for $\sqrt{2} \gamma_1
T/\hbar= \pi$. Top right panel: Surface bands of $H_F^{(1);s}$
plotted as a function of $k_z$ using OBC along $x$ and $y$. The red
bands show Floquet hinge localized modes. Bottom left panel: Plot of
$|\psi(x,y;k_z=1)|^2$ as a function of $x$ and $y$ for the Floquet
hinge localized mode. Bottom right panel: Surface bands of
$H_F^{(1);c}$ plotted as a function of $k_z$ using OBC along $x$ and
$y$ for $\hbar \omega_D/\lambda=3.3$. For all plots $\gamma=-0.35$,
$m=-0.08$.}
    \label{fig:drivencase}
\end{figure}

For the square pulse protocol, the magnitude of the gap becomes
smaller as the drive frequency is lowered from $T= 2
\pi\hbar/(\sqrt{2}\gamma_1)$ as can be seen from the left panels of
Fig.\ \ref{fig:flph}. This feature can be easily understood by
noting that the terms in the Floquet Hamiltonian $H_{F}^{(1);s}$
which are to be added to $H_F^{\ast}$ to obtain a spectral gap have
amplitudes $\sim 1/(\gamma_1 T)$ and thus decrease rapidly with
decreasing frequency. In contrast, for the continuous protocol, as
shown in the right panel, the gap at higher frequencies is much more
robust. This can be understood from the expression of $H_F^{(1);c}$
which indicates that the amplitude of the terms added to
$H_F^{\ast}$ (Eq.\ \ref{hfstar}) for a finite spectral gap is
proportional to $J_0(\sqrt{2}\gamma_1 T/(\pi\hbar))$; their
amplitude therefore decreases more slowly with decreasing frequency.

Thus we find that the Floquet Hamiltonians derived using FPT support
near gapless regimes as a function of both $\gamma_1$ and $T$. We
also note that the predictions based on first order FPT results are
validated from numerical computations of the bulk spectrum gap of
the  exact Floquet Hamiltonians as shown in the bottom plots of
Fig.\ \ref{fig:flph}. In the next subsections, we shall analyze the
properties of these gapped and near-gapless phases for specific
drive frequencies. We close this subsection by noting that the
second order Magnus expansion can not explain these transitions as
shown in the Appendix. The second order Floquet Hamiltonian obtained
from the Magnus expansion yields Floquet phases with large gaps for
both continuous and discrete protocols at all frequencies.

\subsection{Gapped phases of the first order Floquet Hamiltonian}
\label{pipoint}

In this section, we shall discuss the property of the gapped phases.
For this purpose we shall use the square pulse protocol with a time
period $T'$ which satisfies $\sqrt{2} \gamma_1 T'/\hbar = \pi$. The
reason for this choice is that the Floquet Hamiltonian $H_F^{(1);s}$
is particularly simple at this point making several aspects of the
phase analytically tractable.

For $T=T'$, the perturbative Floquet Hamiltonian can be written as
\begin{eqnarray}
H_F^{(1);s}(T') &=& H_F^{\ast} -\frac{\sqrt{2} i\hbar}{\pi} \left((a_1-a_2)\Gamma_2\Gamma_4 \right. \nonumber\\
&& \left.+(\Gamma_2+\Gamma_4)(a_3\Gamma_3+a_4\Gamma_1)+ia_5\Gamma_3
\right). \label{piham}
\end{eqnarray}
The bulk spectrum of $H_F^{(1);s}(T')$ is found to be gapped as can
be seen from  the top left panel of Fig.\ \ref{fig:drivencase} where
the energy spectrum is plotted as a function of $k_y$ for $k_z=2$
with open boundary condition (OBC) along $x$-direction and periodic
boundary condition (PBC) along $z$ and $y$. This shows gapped
surface states modes similar to the undriven case but does not show
topologically protected zero energy surface modes.
\begin{figure}
    \vspace{-1\baselineskip}
    \includegraphics*[width=0.48\linewidth]{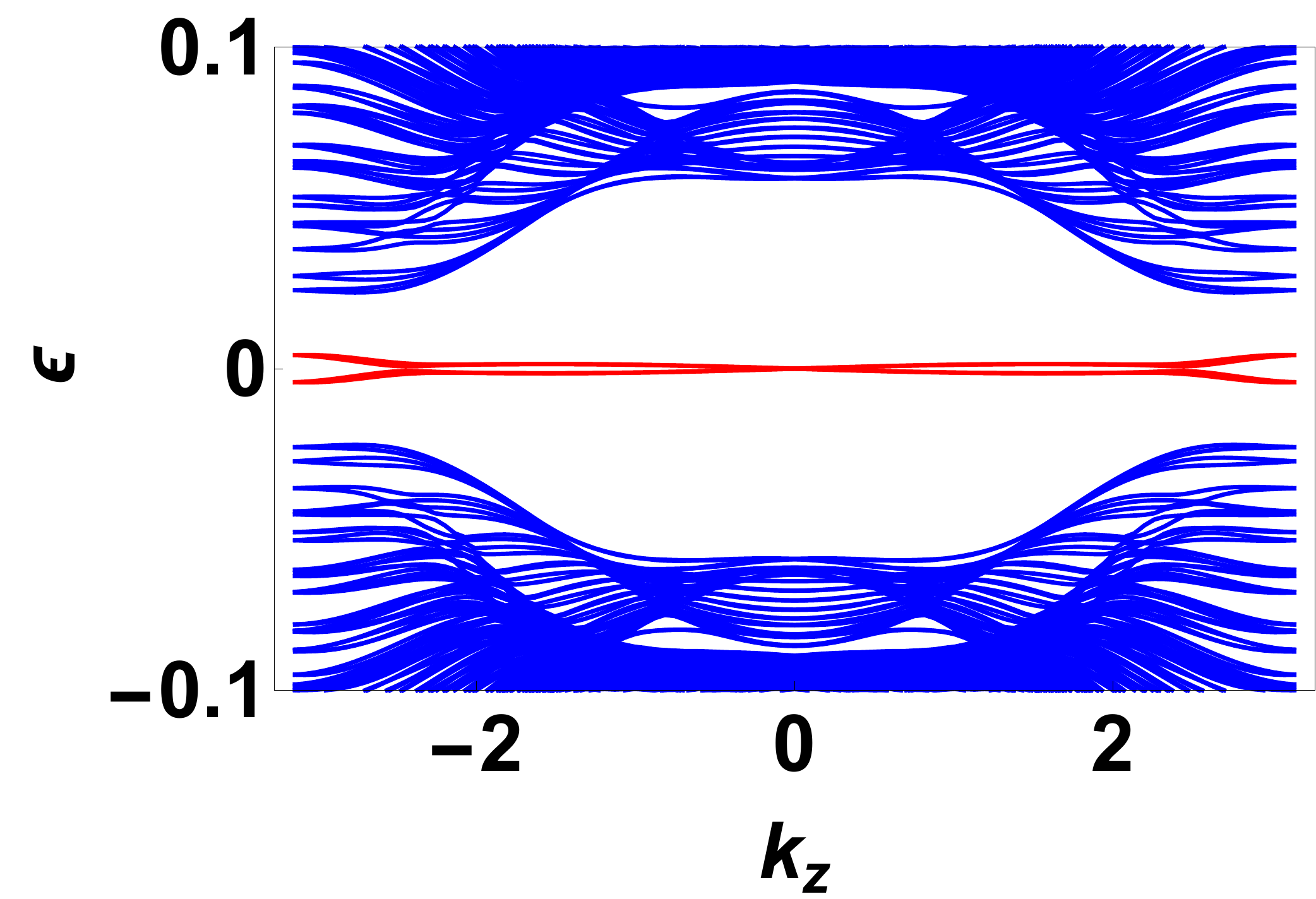}
    \includegraphics*[width=0.48\linewidth]{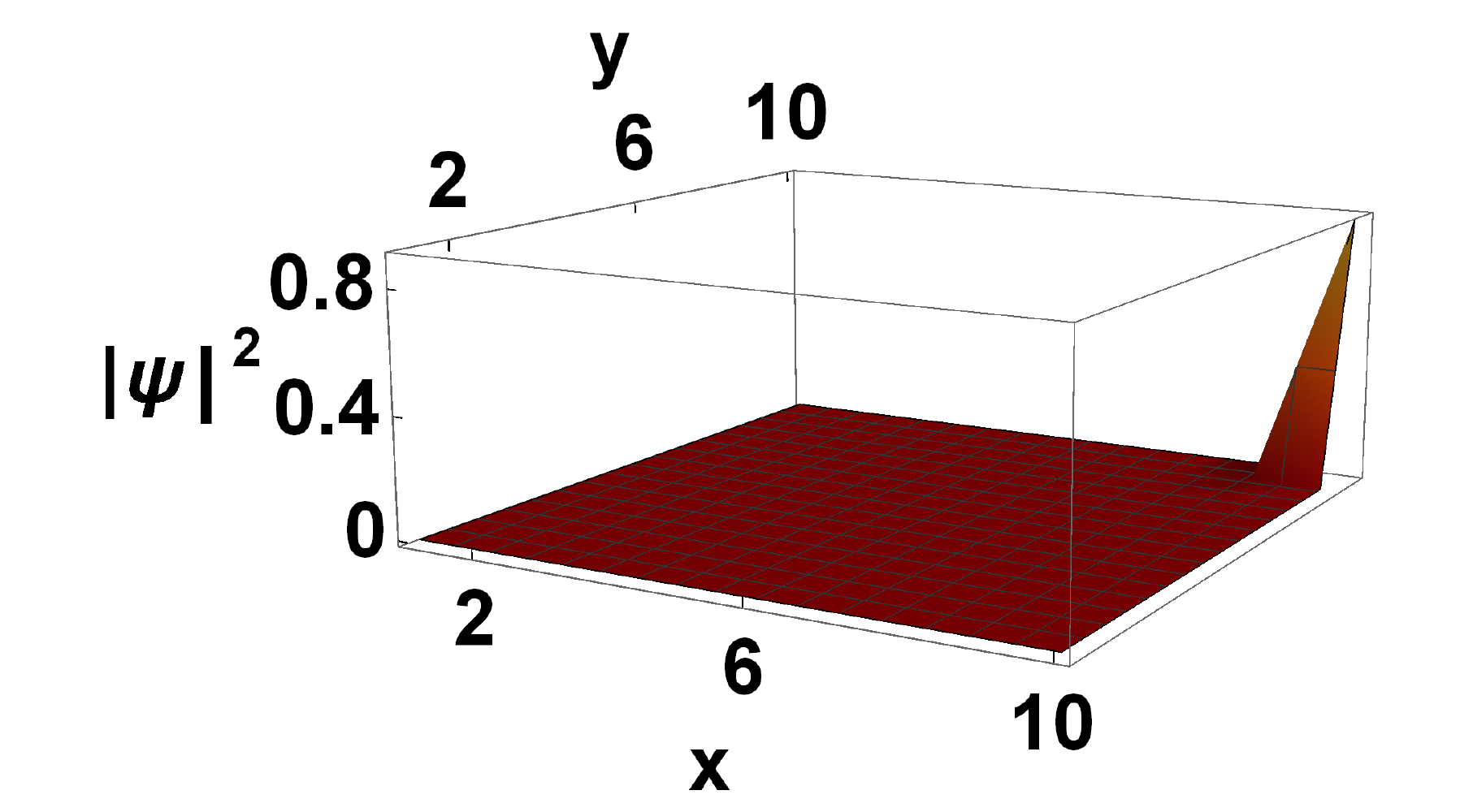}
\caption{Left Panel: Surface bands of exact Floquet Hamiltonian for
a square pulse protocol plotted as a function of $k_z$ using OBC
along $x$ and $y$ with $\sqrt{2} \gamma_1 T/\hbar=\pi$. Right Panel:
Plot $|\psi(x,y; k_z=1)|^2$ as a function of $x$ and $y$ for the
hinge mode. All parameters are same as those in Fig.\
\ref{fig:drivencase}}
    \label{drexact}
\end{figure}
To check for the possibility of the topologically protected hinge
modes, next we consider OBC along $x$ and $y$ directions in
conjunction with PBC along $z$ and plot the band spectrum in the top
right panel of Fig.\ \ref{fig:drivencase} as a function of $k_z$.
Interestingly, we find finite energy dispersive modes at the center
of the bulk bands for $k_z\ne 0$. This is in contrast to the
standard quadrupolar insulators which host gapless hinge modes. To
identify the nature of this energy mode, we plot probability density
of the corresponding eigenstate, $|\psi(x,y;k_z)|^2$, in the $x-y$
plane for $k_z =1$ as shown in the bottom left panel Fig.\
\ref{fig:drivencase}. Clearly, the finite energy mode appears to be
the hinge mode of driven Hamiltonian. Thus the Floquet Hamiltonian
seems to support dispersing hinge modes at finite $k_z$. The energy
of the hinge modes becomes zero at $k_z=0$ when $m$ is finite or at
all values of $k_z$ which allow hinge modes when $m_0=0$. The bottom
right panel of Fig.\ \ref{fig:drivencase} shows the presence of
hinge modes in the gapped phase of the Floquet Hamiltonian obtained
using a continuous protocol for representative value of drive
parameters. We find analogous dispersive hinge modes although the
dispersion turns out to be flatter compared to its discrete protocol
counterpart.

The hinge modes obtained from the analysis of $H_F^{(1); s}$ can be
compared to the ones obtained from exact numerics for the same
parameter value. The dispersion of these modes are shown in the left
panel of Fig.\ \ref{drexact} while the probability density
$|\psi(x,y; k_z=1)|^2$ for the hinge states are plotted in the right
panel of the figure. We find the presence of hinge modes in the
spectrum of exact Floquet Hamiltonian for the square pulse protocol
which is qualitatively consistent with the results obtained from
$H_F^{(1);s}$. We note that the exact hinge modes display a much
flatter dispersion compared to their analytic counterparts;
moreover, they start to deviate from $E=0$ for $|k_z| \ge 2.6$
predicting absence of zero energy states beyond $-2.6 \le k_z\le
2.6$ for the chosen set of parameters. The latter feature is also
seen for $m_0=0$ where all modes with $|k_z| \le 2.6$ correspond to
$E=0$.

To understand such dispersing hinge modes, we first address the case
with $m=0$. This is followed by a perturbative treatment for a
small, non-zero value of $m$. The first-order Floquet Hamiltonian,
given in Eq.~(\ref{eqn:floham1s}) at this driving frequency is
given, for $m=0$, by (Eq.\ \ref{piham})
\begin{eqnarray}
H_F^{(1);s}(T';\vec
k)&=&\frac{1}{2}(2\gamma_z+\lambda\cos{k_x}+\lambda\cos{k_y})(\Gamma_2+\Gamma_4)\nonumber\\
&& -i\frac{\sqrt{2}\hbar}{\pi}\lambda \Big[(\Gamma_2+\Gamma_4)(\sin{k_x}\Gamma_3+\sin{k_y}\Gamma_1)\nonumber\\
&& +(\cos{k_x}-\cos{k_y})\Gamma_2\Gamma_4 \Big], \label{eqn:HFm0}
\end{eqnarray}
where we have used Eq.\ \ref{eqn:ham0} for expressions of $a_i$ for
$i=1..4$ and $\gamma_z$. The analysis of hinge modes can then be
cast as a solution of the edge problem for $H_F^{(1);s}(T'; k_x=\pm
k_y,k_z)$. For the sake of definiteness we choose $k_x=k_y=k$ here;
along this line
\begin{equation}
H_F^{(1);s}(T';k,k_z)=\frac{1}{2}(2\gamma_z+2\lambda\cos{k})\Lambda_1-i\frac{\sqrt{2}\hbar}{\pi}\lambda\sin{k}\Lambda_1\Lambda_2,
     \label{eqn:HFdiag}
\end{equation}
where $\Lambda_1=\Gamma_2+\Gamma_4$ and
$\Lambda_2=\Gamma_1+\Gamma_3$. Eq.\ \ref{eqn:HFdiag} corresponds to
a 1D hopping Hamiltonian with its two ends at the two diagonally
opposite corners of the original lattice (treating $k_z$ as a
parameter). Thus the end modes of this Hamiltonian will correspond
to any hinge localized mode of the original problem.

To find the end mode, we need a solution of
\begin{eqnarray}
H_F^{(1);s}(T';k=-i \partial_{\xi},k_z) \phi(\xi)  &=& E\phi(\xi),
\label{sche1}
\end{eqnarray}
with $\phi(\xi=0)=0$ where the edge is at $\xi=0$. Here we have
chosen a semi-infinite line occupying $\xi>0$ and have used the
identification $k=-i\partial_\xi$ with
$\xi=\frac{1}{\sqrt{2}}(x+y)$. To this end, we consider the
wavefunction $\phi_\pm$ given by
\begin{eqnarray}
\phi_\pm(\xi) &=& e^{-\alpha\xi}e^{\pm i\beta\xi}\chi.   \label{wavtrial}
\end{eqnarray}
Substituting Eq.\ \ref{wavtrial} in Eq.\ \ref{sche1}, we find
\begin{align}
(\gamma_z+\lambda\cos{(i\alpha\pm\beta)})\Lambda_1\chi-i\frac{\sqrt{2}\hbar}{\pi}\lambda&\sin{(i\alpha\pm\beta)}\Lambda_1\Lambda_2\chi\nonumber\\&~~~~~~~~=E\chi.
\end{align}

Next, we try a solution of the form $\Lambda_2\chi=p\chi$ (where
$p=\pm \sqrt{2}$ constitute the two doubly degenerate eigenvalues of
$\Lambda_2$) and $E=0$. This leads to
\begin{equation}
\left((\gamma_z+\lambda\cos{(i\alpha\pm\beta)})-i\frac{\sqrt{2}\hbar}{\pi}\lambda\sin{(i\alpha\pm\beta)}p
\right)\Lambda_1\chi=0.  \label{e0sol}
\end{equation}
For Eq.\ \ref{e0sol} to hold, we need to equate its real and
imaginary parts to zero. This gives us the conditions
\begin{eqnarray}
\gamma_z+\lambda\cosh{\alpha}\cos{\beta}=-\frac{\sqrt{2}\hbar}{\pi}\lambda p\sinh{\alpha}\cos{\beta},\nonumber \\
\sinh{\alpha}\sin{\beta}=-\frac{\sqrt{2}\hbar}{\pi}p\cosh{\alpha}\sin{\beta},
\label{condeq}
\end{eqnarray}
which needs to be satisfied by $\alpha>0$ and $\beta$. If
$\beta=\pi$, this therefore leads to the condition
\begin{eqnarray}
\gamma_z\pm\lambda\cosh{\alpha}=\mp\frac{\sqrt{2}\hbar}{\pi}\lambda
p\sinh{\alpha}.
\end{eqnarray}
This condition, however, is not satisfied for any $(k_z,\alpha)$
pair corresponding to our chosen set of parameters. In contrast, for
$\beta\neq0,\pi/2,\pi$, we need
\begin{eqnarray}
\tanh{\alpha} &=& -\frac{\sqrt{2}\hbar}{\pi}p, \quad
\cos{\beta}=-\frac{\gamma_z}{\lambda}\cosh{\alpha}.
\end{eqnarray}
Also, we note that no solution exists for $\beta=\pi/2$ provided
$\gamma_z\neq0$.

For a localized solution at this edge, $\alpha$ should be positive;
so $p$ should be chosen to be the negative eigenvalue of
$\Lambda_2$, namely $-\sqrt{2}$ and $|\gamma_z/\lambda|<$ sech
$\alpha$. This provides the allowed range of $k_z$ for zero-energy
hinge modes for $m=0$. For our chosen set of parameter values, this
second condition implies that our analysis does not hold for
$|k_z|>2.6$. We note here that a similar analysis carried out for
the diagonally opposite hinge (for which $\alpha<0$) would yield a
similar solution but with $p$ chosen to be positive eigenvalues of
$\Lambda_2$.

The analysis above indicates that there are two linearly independent
solutions for $\chi_{1,2}$ given by
\begin{eqnarray}
\psi_{1,2}(\xi) &=& \mathcal{N}e^{-\alpha\xi}\sin{(\beta\xi)}\chi_{1,2},  \label{edgesol1}\\
\chi_{1(2)} &=&  \left( -\frac{i}{\sqrt{2}}, +(-) \frac{i}{\sqrt{2}},0(1),1(0)\right)^T. \nonumber
\end{eqnarray}
which satisfies $\psi(\xi=0)=0$. Here $\chi_{1,2}$ denotes the
eignvectors of $\Lambda_2$ corresponding to $p=-\sqrt{2}$. We note
that naively one may conclude the existence of two hinge modes per
hinge from such a solution which contradicts exact numerics which
yields one such state. It is to be stressed that our solution does
not necessarily mean the existence of two such modes since the edge
mode needs to respect $C_{4}^{z}$, $M_y$, $M_x$ and ${\mathcal I}$
symmetries. This may indeed lead to choice of a specific linear
combination of the two solutions leading to the correct number of
modes. However, a detailed analysis of this requires a solution
constituting all four edge modes; we do not attempt it in this work.
\begin{figure}
 \vspace{-1\baselineskip}
\includegraphics*[width=0.98\linewidth]{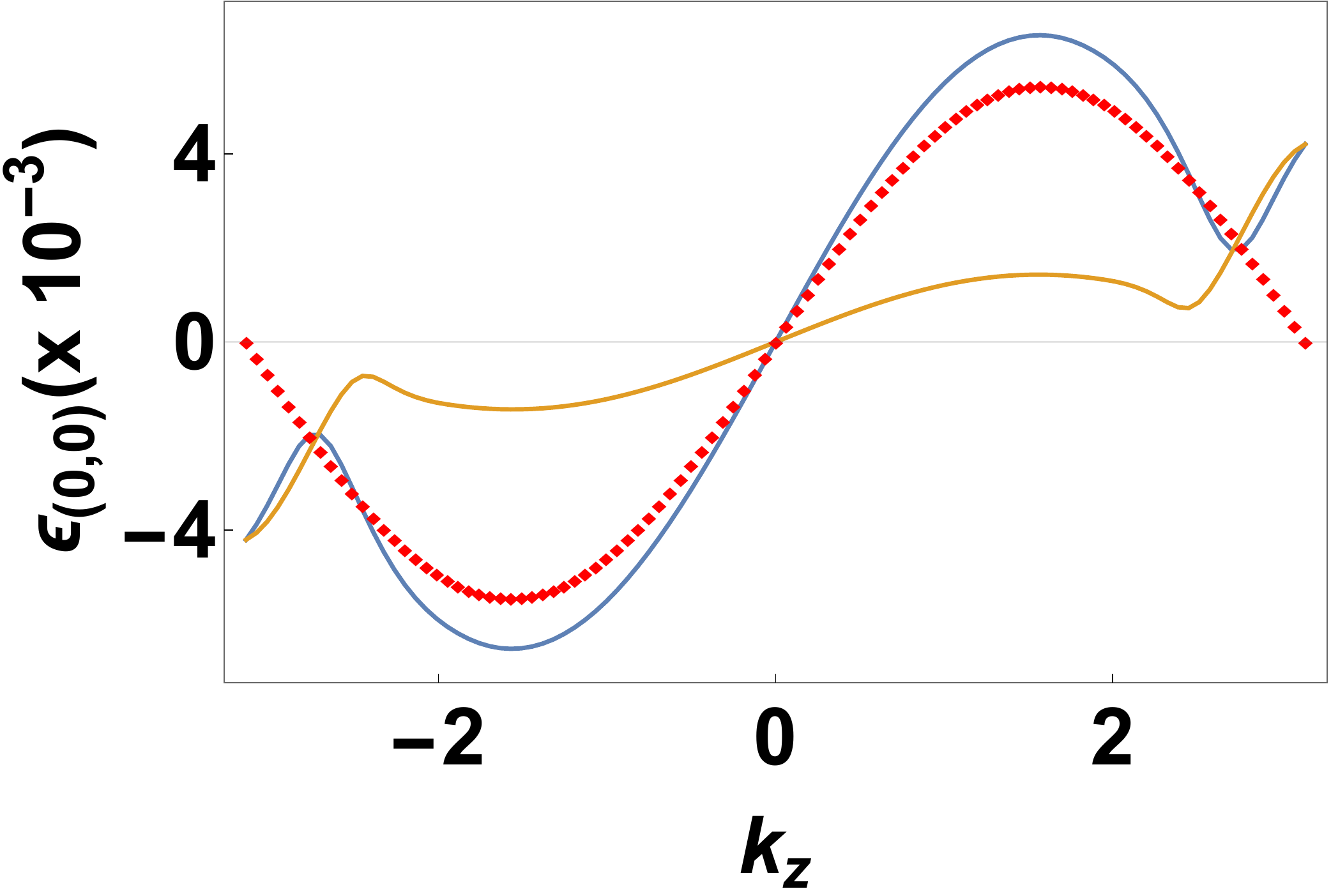}
\caption{Plot of the hinge mode dispersion as a function of $k_z$
using analytical solution ($\mu_1$ from Eq.\ \ref{edgesol2} with
$m=-0.08$) [red squares], numerical solution of $H_{F}^{(1);s}$
using OBC along $x,y$ and PBC along $z$ (blue line) and $H_F$
obtained from exact numerics (yellow line). The analytical result
shows near exact match with the numerical result obtained from
$H_{F}^{(1);s}$ for small $k_z$. All parameters are same as those in
bottom left panel of Fig.\ \ref{drexact}.}
    \label{corfig}
\end{figure}

Instead, to understand the dispersion of the hinge modes in the
presence of a finite $m_0$, we now switch to the case of non-zero
$m_0$. We find that allowing the presence of a non-zero $a_5$ in
Eq.\ \ref{eqn:floham1s} introduces an additional term in the
$H_F^{(1);s}(T';k,k_z)$ (Eq.\ \ref{eqn:HFdiag}) given by
\begin{eqnarray}
H_m &=&
m_0\sin{k_z}\left(i(\Gamma_2\Gamma_3+\Gamma_3\Gamma_4)-\frac{\sqrt{2}}{\pi}\Gamma_3
\right).
\end{eqnarray}
Projecting this term in the space spanned by $\chi_{1,2}$ and
diagonalizing, one obtains two eigenvalues and eigenvectors as
\begin{eqnarray}
\mu_{a} &=& (-1)^{a} m_0\sin{k_z} (\pi+(-1)^a 1)/\pi \nonumber\\
\psi_{m,a} &=&  \frac{1}{\sqrt{2}} \left( \psi_1 +(-1)^{a+1} \psi_2
\right) \label{edgesol2}
\end{eqnarray}
where $a=1,2$. The mode which remain close to $E=0$ at small $m_0$
corresponds to $a=1$. As shown in Fig.\ \ref{corfig}, this mode
shows remarkable match with the numerical spectrum obtained from
$H_F^{(1);s}$ at $T=T'$ for small $k_z$; however, it fails to
reproduce the up turn of the spectrum at large $k_z$. It also shows
qualitatively similar behavior to the corresponding exact hinge mode
spectrum for small $k_z$ which produces a dispersing spectrum with
much flatter dispersion. This prompts us to choose the linear
combination $\psi \sim \psi_1 + \psi_2$ as the hinge mode solution
for $m_0=0$; the other mode corresponding to $a=2$ seems to be an
artifact of treating a single hinge which we ignore. We also note
that our analysis not only explains the dispersion of the hinge
modes for finite $m_0$ but also predicts that for $m_0=0$, there are
no zero energy hinge states for $k_z$ satisfying $ {\rm sech} \alpha
>|(\gamma_z/\lambda)|$ (which translates to $|k_z|>2.6$ for our
chosen parameters). This matches with the result of exact numerics
where the hinge modes for $m_0=0$ starts to deviate from zero energy
beyond $|k_z| \sim 2.6$ for the chosen set of parameters.

\subsection{Gapless points in the first order Floquet spectrum}
\label{2pipoint}

In this section, we shall analyze the Floquet Hamiltonian at the
gapless point using the first order perturbative Hamiltonian which
allows some analytic insight into the nature of these points. For
the discrete square pulse protocol such points occur at
$\sqrt{2}\,\gamma_1\, T/\hbar=2n \pi$ while for the continuous
protocol, they occur at $\sqrt{2}\gamma_1 T/\hbar= \pi \alpha_n$.
For both these points, the first order perturbation theory yields
$H=H_F^{\ast}$. In this section, we shall analyze the bulk and the
surface properties of $H_F^{\ast}$.

\subsubsection{Bulk modes}
\label{bm} We begin our analysis with the bulk modes of $H_F^{\ast}$
using PBC in all directions. This yields
\begin{align}
\epsilon_{\pm,\pm}^{}= \pm \frac{1}{\sqrt{2}}  ||a_1 +a_2| \pm |a_5||.
\label{eqn:nodes}
\end{align}
It turns out that the spectrum contains zero energy curves which
constitutes stacks of Weyl nodes for $k_z\ne 0, \pi$ and Dirac nodes
for $k_z=0, \pi$. The former constitutes a crossing of two bands
leading to Weyl nodes at the crossing points. At $k_z=0,\pi$, both
the positive and negative quasienergy bands are two-fold degenerate;
hence in this case, one has a crossing of all four bands leading to
Dirac nodes.

\begin{figure}
    \vspace{-1\baselineskip}
    \includegraphics*[width=0.45\linewidth]{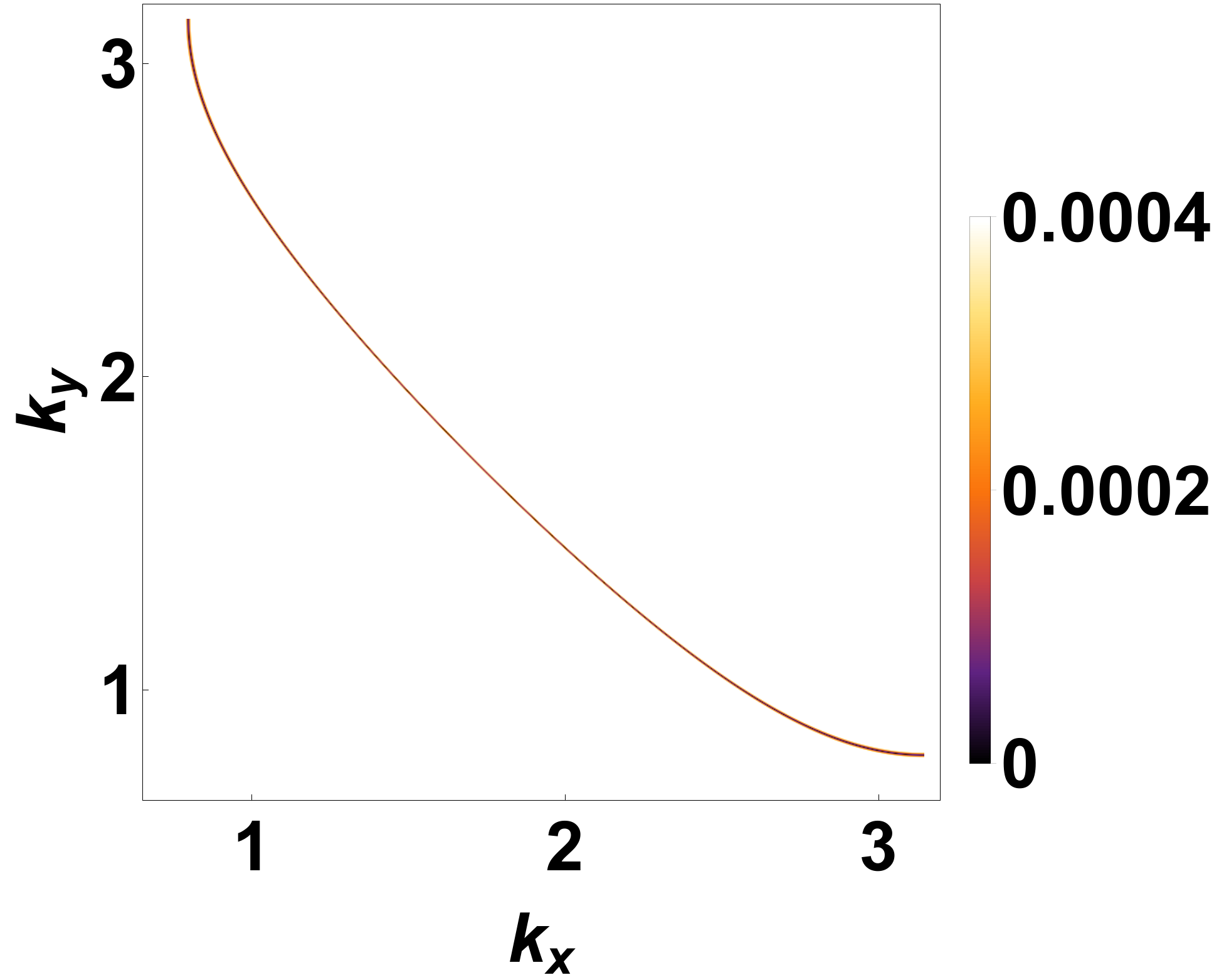}
    \includegraphics*[width=0.45\linewidth]{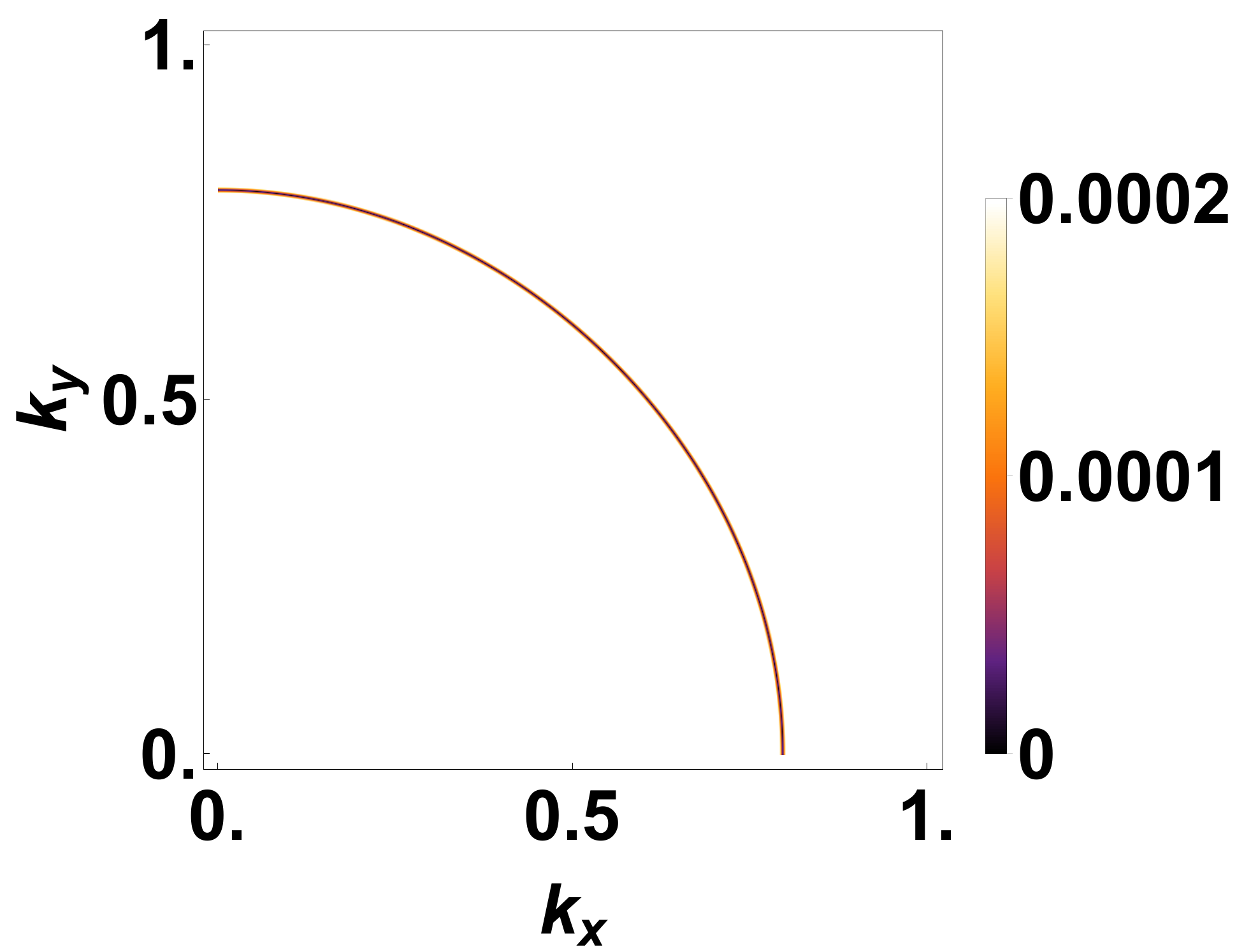}
    \includegraphics*[width=0.45\linewidth]{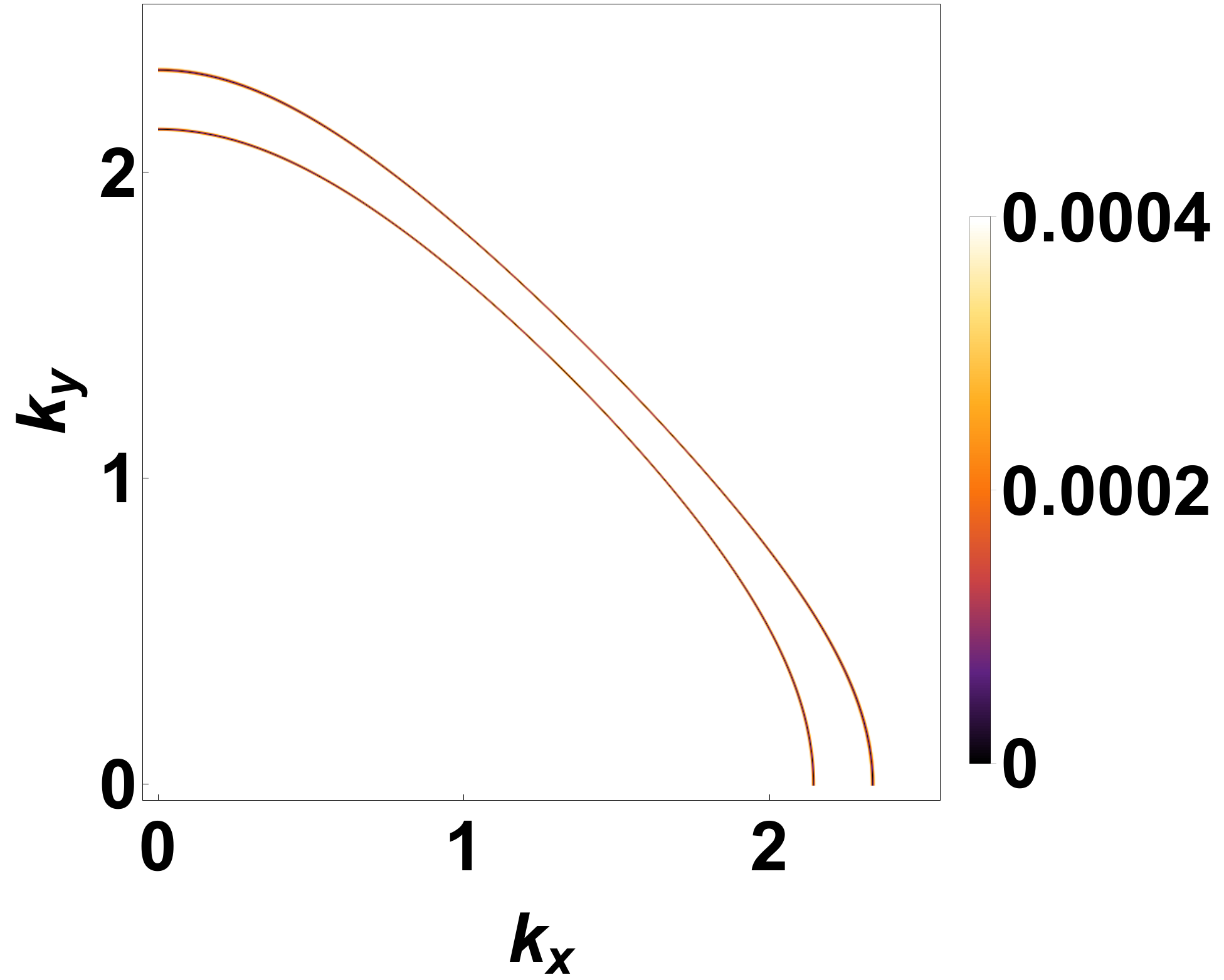}
    \includegraphics*[width=0.45\linewidth]{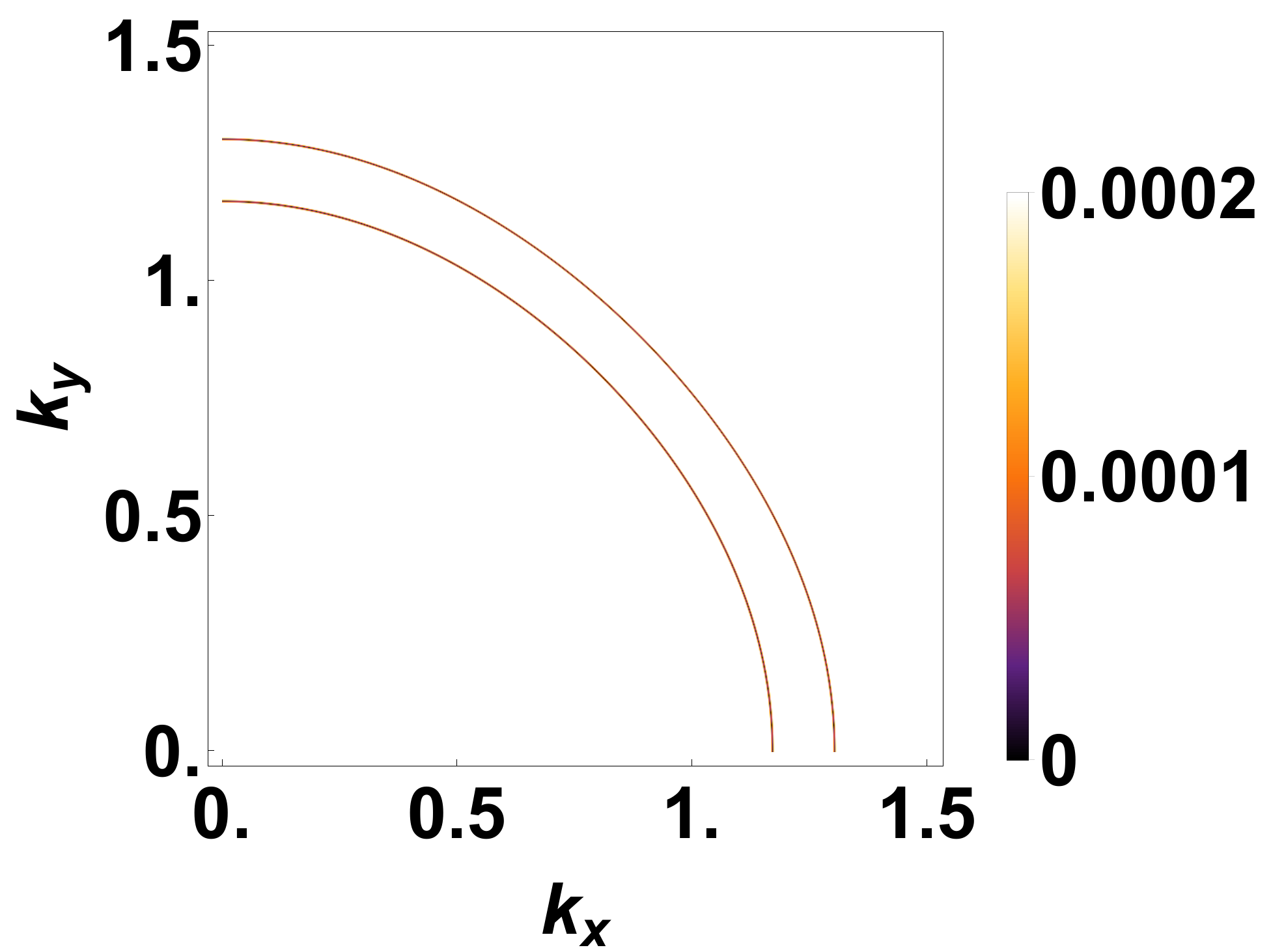}
\caption{Contour plot of $\epsilon_{+,-}$ in the $k_x-k_y$ plane.
Top left panel: For $k_z=0$, the two-fold degenerate conduction and
valence bands meet along the arcs, giving rise to line node Dirac
semimetals. Right panel: For $k_z=\pi$. Bottom left panel: For
$k_z=1.25$ where two bands cross leading to line node Weyl
semimetal. Right panel: For $k_z=2.25$. For all plots
$\gamma=-0.35$, and $m=-0.08$.}
    \label{fig:nodalring}
\end{figure}

The generic condition on the momenta $(k_x,k_y,k_z)$ for these band
crossings is given by
\begin{eqnarray}
&&- \sqrt{1+m^2} \le \mu(k_x,k_y) \le \sqrt{1+m^2}, \label{croscon} \\
&& \cos k_z^{\pm}  = \frac{\mu(k_x,k_y)}{1+m^2}\pm \frac{|m|}{1+m^2}
\sqrt{1+m^2 -\mu^2(k_x,k_y)}, \nonumber
\end{eqnarray}
where we have used Eqs.\ \ref{eqn:nodes} and \ref{eqn:ham0} and
$\mu(k_x,k_y) = (2\gamma + \cos k_x + \cos k_y)$. The band spectrum
corresponding to Eq.\ \ref{eqn:nodes}, shown in Fig.\
\ref{fig:nodalring}, demonstrates these nodes for different values
of $k_z$. Notice that for fixed $k_z$, the bulk gap closes in the
$k_x-k_y$ plane along a single or multiple curves, giving rise to
nodal lines/rings Weyl (for $k_z \ne 0,\pi$) or Dirac semimetals
(for $k_z=0,\pi$) \cite{young_PRL15}. The shape of the lines/rings
depends on the values of $k_z,\gamma$ and $m$ and is determined by
Eq.\ \ref{croscon}. We note that the Dirac line nodes on
mirror-invariant planes $k_z=0$ and $k_z=\pi$ are protected by
$M_z$, $I$ and $T_0$ symmetries which leads to the two-fold
degeneracy of the bulk bands. For $k_z \ne 0, \pi$, these symmetries
are broken; this leads to lifting of the degeneracies and a generic
crossing between two non-degenerate bands leading to Weyl nodes.

\begin{figure}
    \vspace{-1\baselineskip}
    \includegraphics*[width=0.45\linewidth]{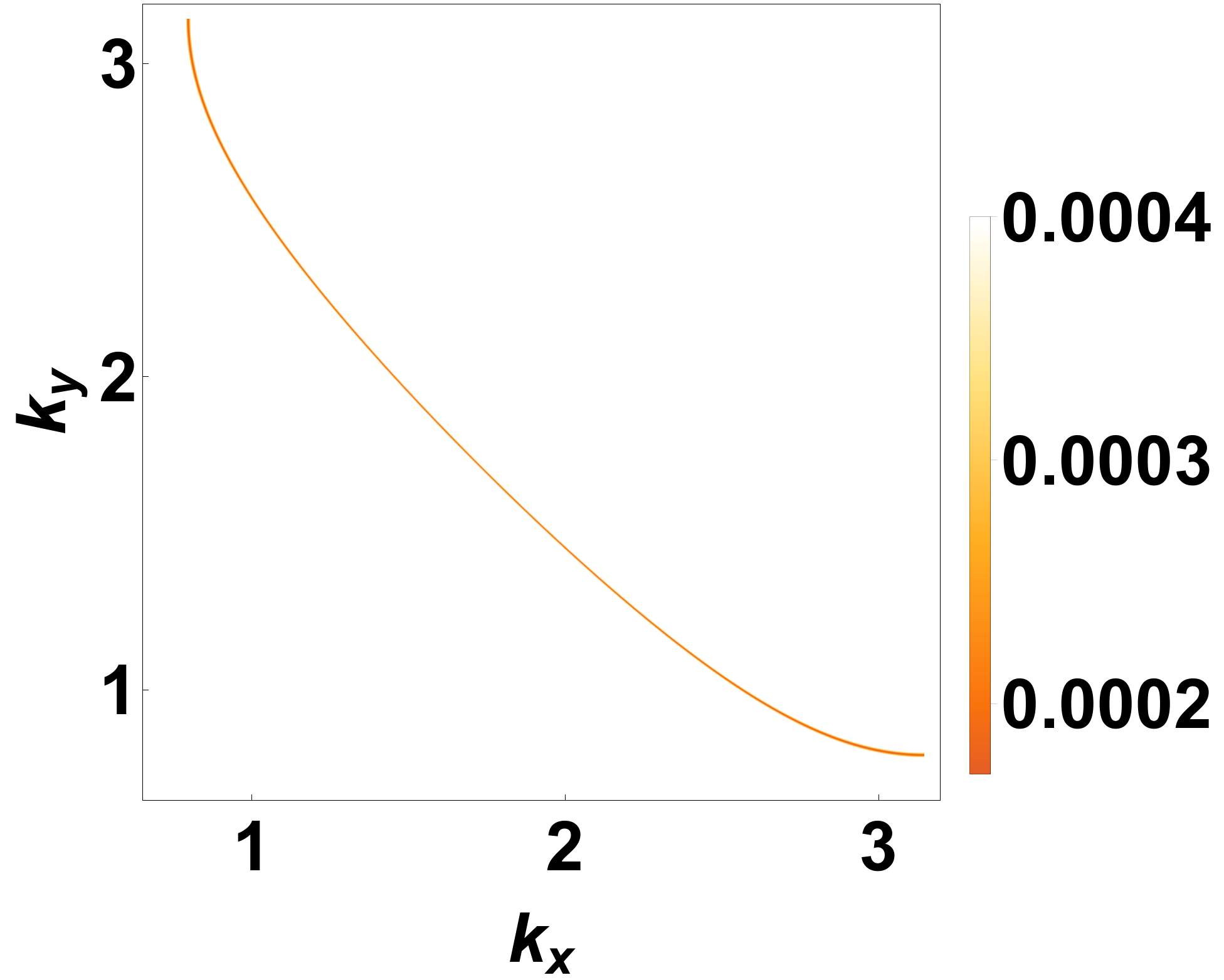}
    \includegraphics*[width=0.45\linewidth]{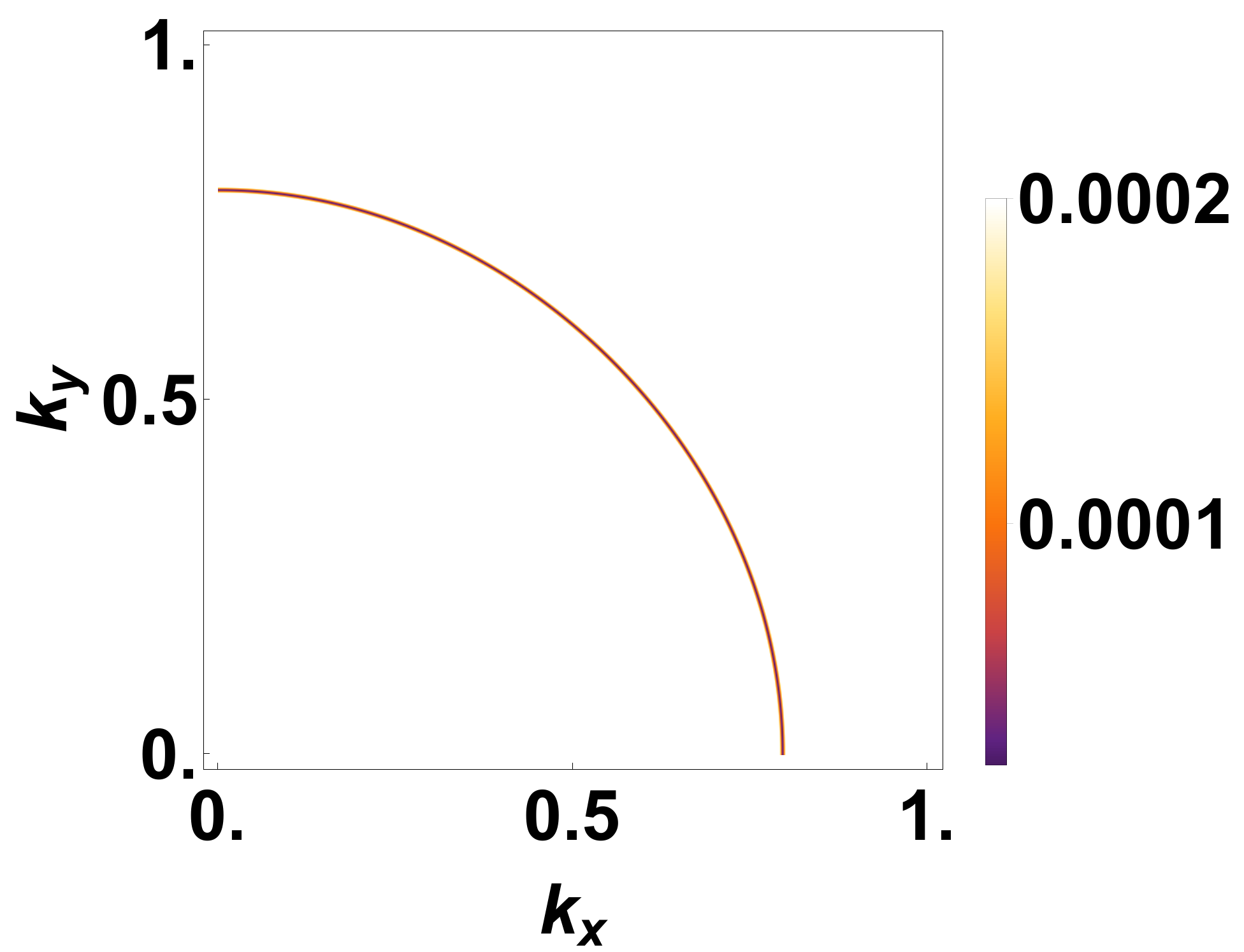}
    \includegraphics*[width=0.45\linewidth]{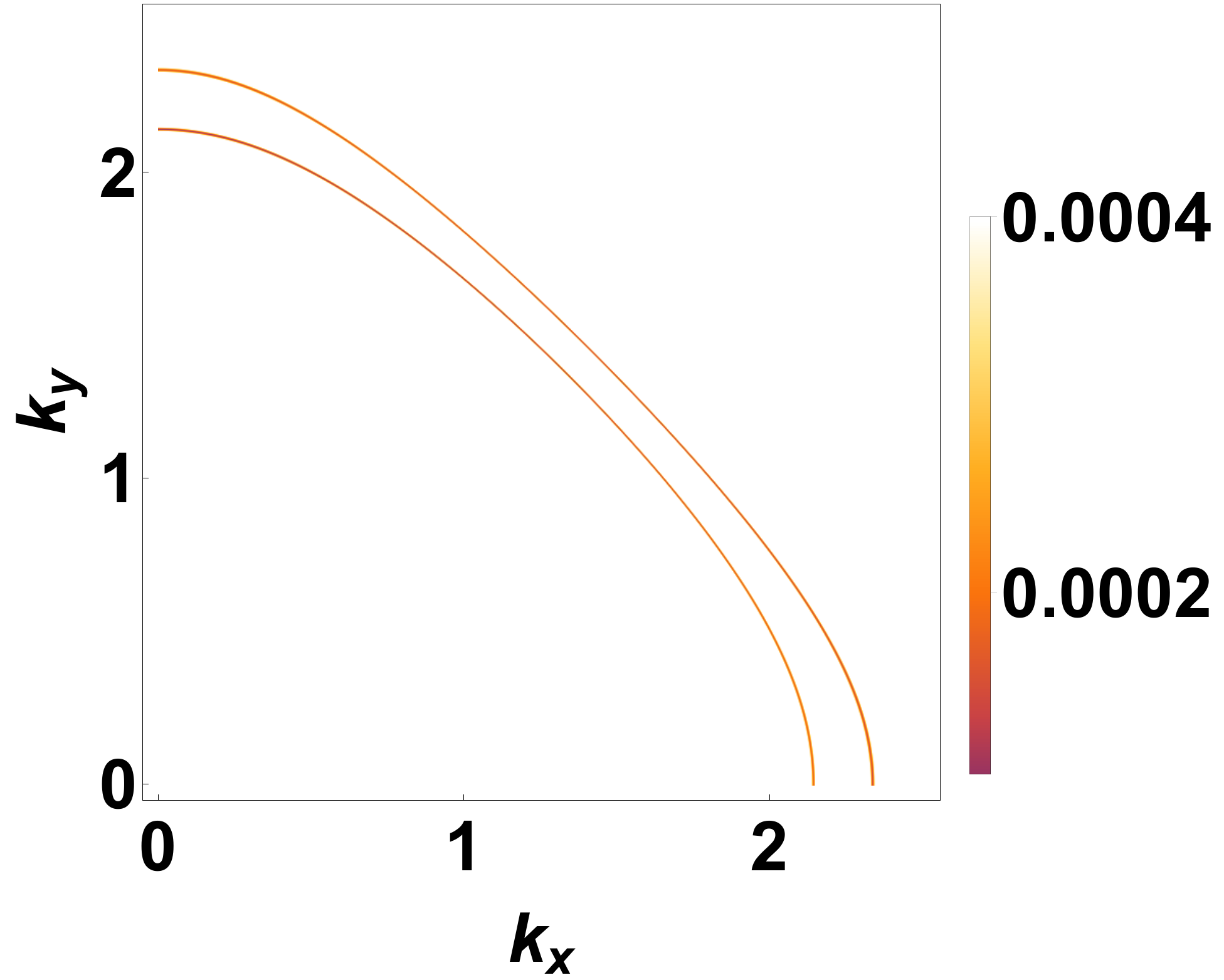}
    \includegraphics*[width=0.45\linewidth]{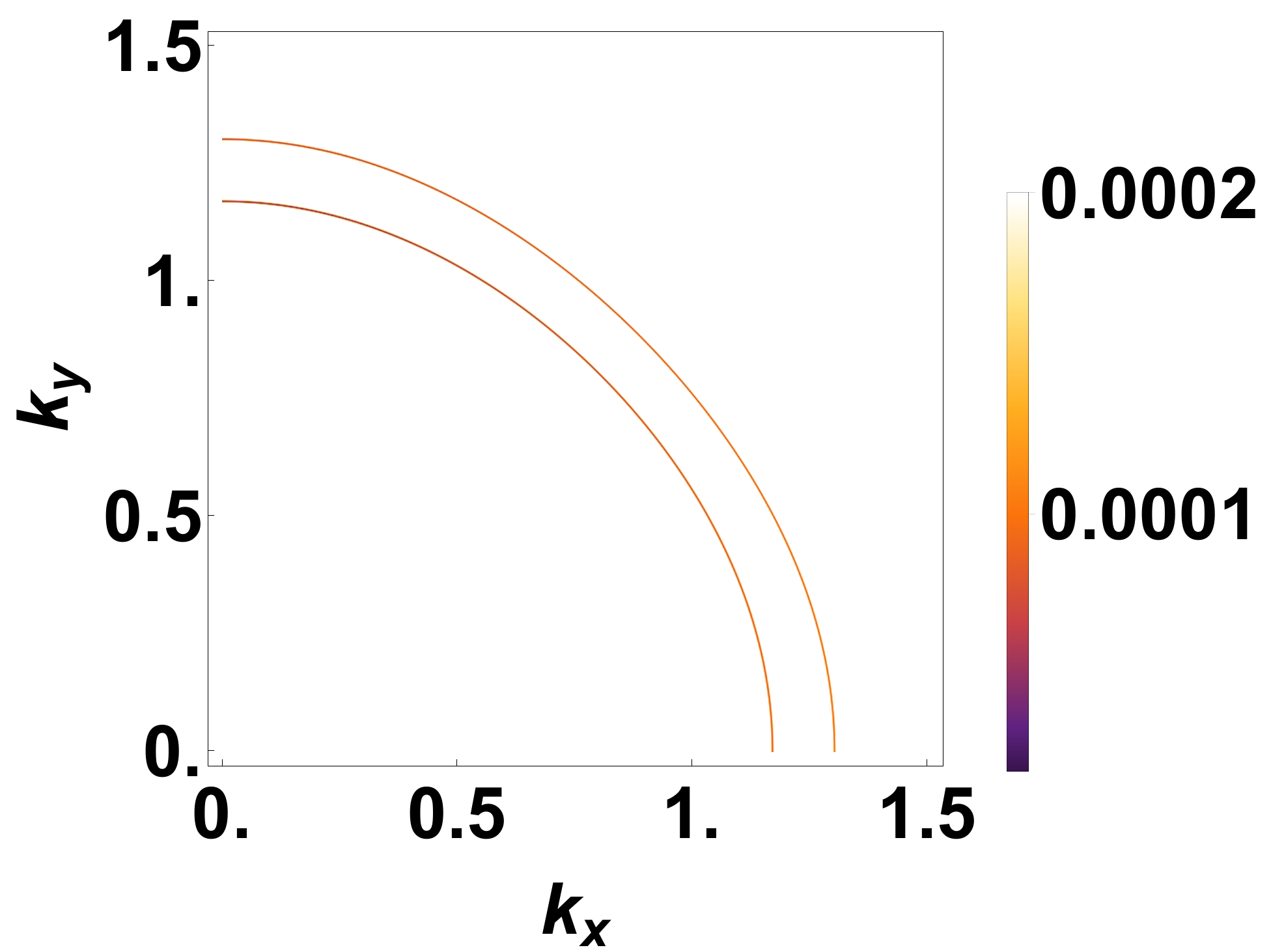}
\caption{Contour plot of $\epsilon_{+,-}$ for the exact Floquet
Hamiltonian in the $k_x-k_y$ plane for the square pulse protocol
with $\sqrt{2} \gamma_1 T/\hbar= 2 \pi$. All parameters are same as
the corresponding panels in Fig.\ \ref{fig:nodalring}. Note that the
gaps do not close exactly leading to an avoided level crossing.}
    \label{fig:ex}
\end{figure}

As explained earlier at the closing of Sec. \ref{sec:floquet_PT},
the gaps for the bulk Floquet mode of the exact Hamiltonian do not
close. This can be seen by looking at the Floquet spectrum in Fig.\
\ref{fig:ex}. These plots are almost identical except for the
presence of a tiny gap in the spectrum in the regions where the
first order Floquet theory yields gapless Weyl or Dirac nodes. Thus
the band crossings of the first order theory becomes avoided level
crossings for the exact Floquet Hamiltonian. The effect of this
reduction of the bulk Floquet gap on the hinge modes shall be
discussed in Sec.\ \ref{dyncor}. For both Fig.\ \ref{fig:nodalring}
and \ref{fig:ex}, we have set the plot range so as to only highlight
the contours along which the bulk band gap is the smallest.

\subsubsection{Surface and hinge modes}
\label{sm} In this section, we shall analyze the surface and hinge
modes around the drive frequencies at which the bulk spectrum gap
for the first order Hamiltonian closes. Thus we shall be concerned
with drive frequencies $\sqrt{2}\gamma_1 T/\hbar= 2 n\pi +\delta$
for the discrete protocol and $\sqrt{2} \gamma_1 T/\hbar= \pi
\alpha_n +\delta$ for the continuous protocol, where $|\delta| \ll 2
\pi$. For both these protocols, the effective Floquet Hamiltonian in
this region, for $m=0$, reads
\begin{eqnarray}
H_{\rm eff} &=& H_F^{\ast} + \delta c_0 [(a_1-a_2)
(\Gamma_4-\Gamma_2) +2 (a_3\Gamma_3 + a_4\Gamma_1)],\nonumber\\
\label{effham}
\end{eqnarray}
where $c_0$ is a protocol dependent constant which takes values
$c_{0s}=1/(4\pi)$ for the discrete protocol and
$c_{0c}=J_{-1}(\alpha_n) /\pi$ for the continuous protocol. The bulk
modes corresponding to PBC are easily calculated to be
\begin{align}
\epsilon_{\pm}=\pm\frac{1}{\sqrt{2}}\bigg[
(a_1+a_2)^2+4\delta^2c_0^2((a_1-a_2)^2+2(a_3^2+a_4^2))\bigg]^{1/2},
\end{align}
which show a band touching at $\delta=0$ for $\vec k$ values which yields gapless modes.

In what follows we shall first analyze the surface modes of $H_{\rm
eff}$ as a function of $\delta$. To this end, we use the continuum
limit of this lattice Hamiltonian by expanding around the point
$(k_x,k_y)=(0,0)$ and keeping terms till second order in the
momenta. When dealing with the $x=0$ surface specifically, $k_y$
will remain a good quantum number, so that, we can neglect the
negligible $k_y^2$ terms and write the Hamiltonian as a sum of two
parts, viz
\begin{equation}
H(-i\partial/\partial x,k_y,k_z)=H^I + H^{II}.
\end{equation}
Here $H^I$ carries the whole $x$ dependence and is given by
\begin{align}
H^I =
&\frac{1}{2}(\gamma_z+\lambda)(\Gamma_2+\Gamma_4)+\frac{\lambda}{4}\frac{\partial^2}{\partial
x^2}(\Gamma_2+\Gamma_4)\nonumber\\&+\frac{\delta\lambda}{8\pi}\frac{\partial^2}{\partial
x^2}(\Gamma_4-\Gamma_2)-i\frac{\delta\lambda}{2\pi}\frac{\partial}{\partial
x}\Gamma_3,
\end{align}
while $H^{II}$, which carries the entire $k_y$ dependence, can be
written as
\begin{equation}
H^{II}=\frac{1}{2}(\gamma_z+\lambda)(\Gamma_2+\Gamma_4)+\frac{\delta\lambda}{2\pi}k_y
\Gamma_1.
\end{equation}
Here and in rest of the section, we choose to work with $c_{os}$ for
clarity. The form of Eq.\ \ref{effham} guarantees that this does not
lead to a loss of generality.

For a surface localized state, we choose as our ansatz
$\psi_{k_y,k_z}(x)\sim \exp{(\alpha x)}e^{i(k_yy+k_zz)}\sin{(\beta
x)}\Phi$, satisfying the boundary condition $\psi(x=0)=0$. We note
that this corresponds to a choice of a semi-infinite system along
$x$ occupying $x>0$ for $\alpha<0$. Plugging in this ansatz in $H^I
\psi_{k_y,k_z}(x)$, one gets
\begin{widetext}
\begin{eqnarray}
H^I \psi_{k_y,k_z}(x) &=& \sin{(\beta x)}\bigg[\bigg(\frac{1}{2} (\gamma_z+\lambda)+\frac{\lambda}{4} (\alpha^2-\beta^2)\bigg)(\Gamma_2+\Gamma_4)+\frac{\delta\lambda}{8\pi}(\alpha^2-\beta^2)(\Gamma_4-\Gamma_2)-i\frac{\delta\lambda}{2\pi}\alpha\Gamma_3\bigg]\Phi\nonumber\\
&&+ \cos{(\beta
x)}\bigg[\frac{\lambda}{2}\alpha\beta(\Gamma_2+\Gamma_4)+\frac{\delta\lambda}{4\pi}\alpha\beta(\Gamma_4-\Gamma_2)-i\frac{\delta\lambda}{2\pi}\beta\Gamma_3\bigg]\Phi.
\label{surfeq1}
\end{eqnarray}
\end{widetext}
For $\psi_{k_y,k_z}(x)$ to be an eigenfunction of $H^I$, we need two
conditions on $\Phi$. First, the coefficient of the $\cos \beta x$
term in Eq.\ \ref{surfeq1} must vanish. Second, $\Phi$ should be an
eigenvector of the matrix appearing in the coefficient of $\sin
\beta x$.  The first condition implies that $\Phi$ should satisfy
\begin{eqnarray}
\left[\alpha(\Gamma_2+\Gamma_4)+\alpha b/2
(\Gamma_4-\Gamma_2)-ib\Gamma_3\right] \Phi=0,
\end{eqnarray}
where $b=\delta/\pi$. The value of $\alpha$ satisfying this and the
localization condition which necessitates a negative value of
$\alpha$ is $\alpha=-\sqrt{\frac{2b^2}{4+b^2}}$. This yields two
degenerate eigenvectors $\Phi_{1,2}$. Projecting the $\Gamma$
matrices in this null space spanned by $\Phi_{1,2}$, one finds that
the coefficient of the $\sin \beta x$ term in this space is
proportional to $\sigma_x$. Thus, to satisfy the second condition,
the basis vectors of this nullspace should be chosen as
$\chi_\pm=\frac{1}{\sqrt{2}}\begin{pmatrix} 1\\ \pm
\text{sgn}(b)\end{pmatrix}$. where ${\rm sgn}(b)$ denotes the sign
of $b$. We note that the choice of these eigenvectors need to be
carefully done to maintain the same definition of $\sigma_x$ for
$b>0$ and $b<0$.  Projecting $H^{II}$ in this space gives the
surface Hamiltonian at this surface to be
\begin{equation}
H^{\delta>0}_{x=0}=-\frac{\lambda |b|}{2}k_y
\sigma_y-(\gamma_z+\lambda) \frac{b}{\sqrt{2(4+b^2)}}\sigma_z.
\end{equation}
Thus the mass term changes sign with $b$. An exactly similar
analysis can be carried out for other edges.

\begin{figure}
    \vspace{-1\baselineskip}
    \includegraphics*[width=0.48\linewidth]{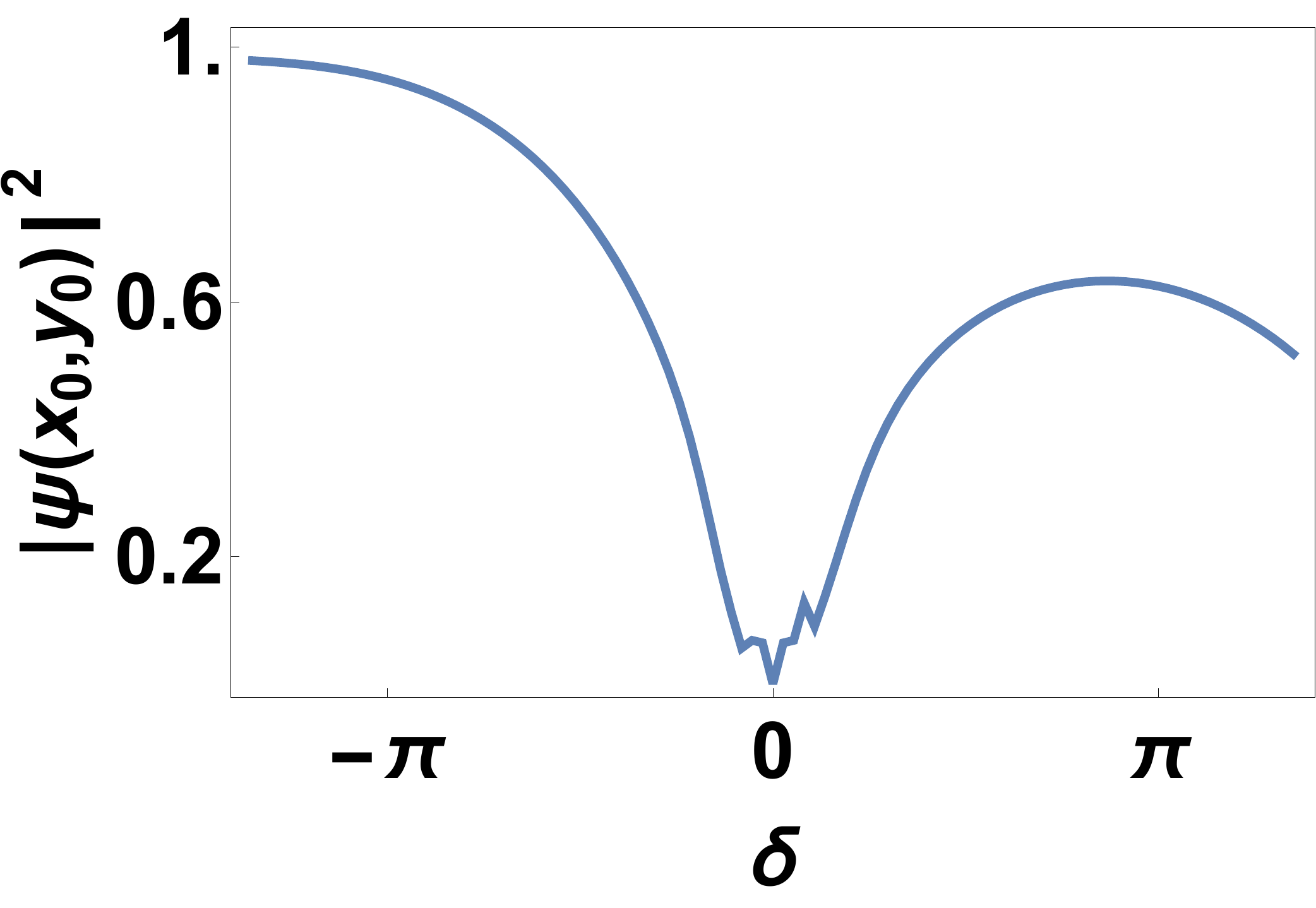}
    \includegraphics*[width=0.48\linewidth]{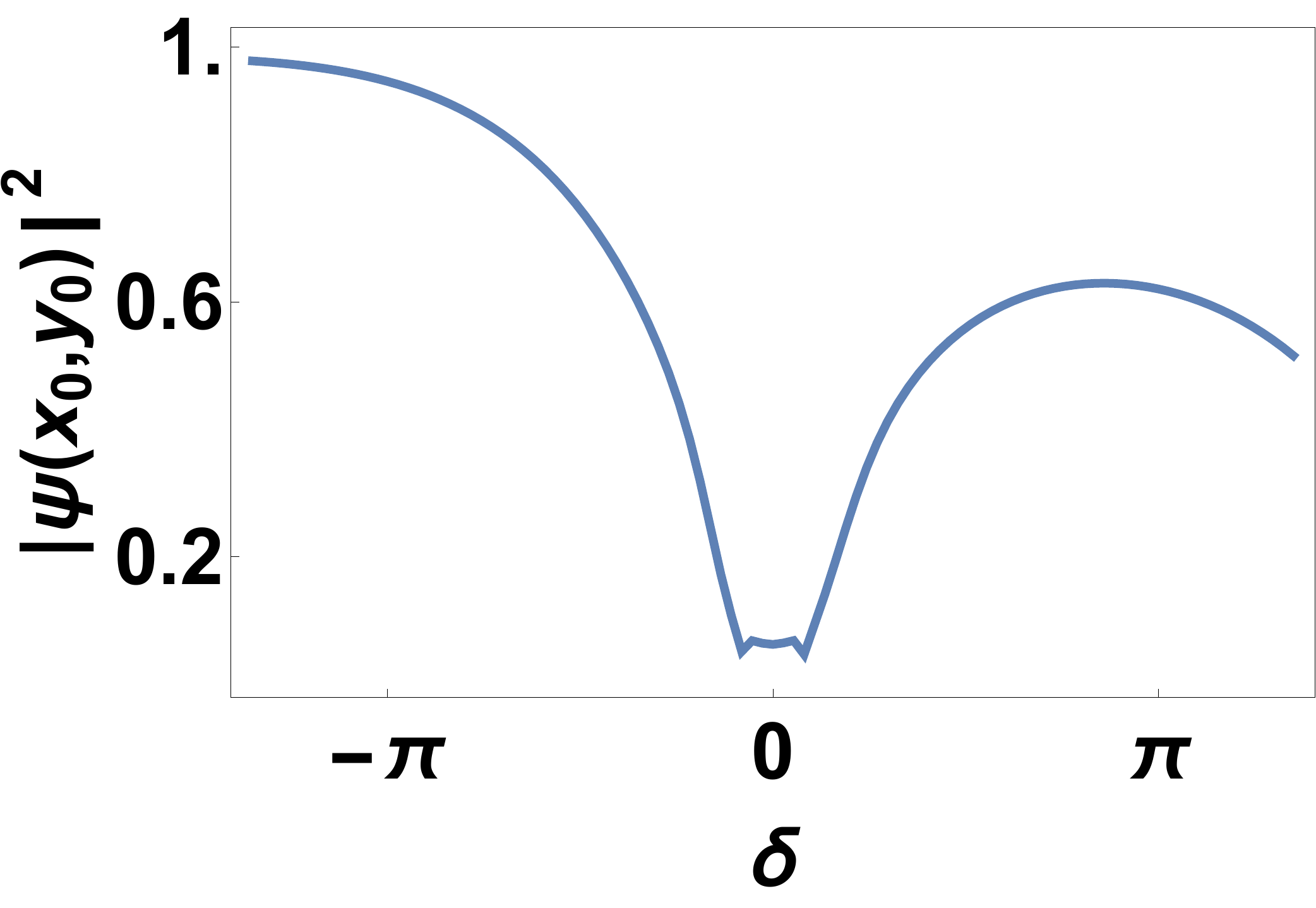}
\caption{Left Panel: Plot of the probability density of the hinge
mode, $|\psi(x_0,y_0;k_z=1)|^2$, as a function of $\delta$ as
obtained from first order perturbative Floquet Hamiltonian for a
square pulse protocol. Right panel: Same as the left panel but using
exact Floquet Hamiltonian. For all plots, $\gamma_1/\lambda=20.0$,
$\gamma=-0.35$, and $m=-0.08$. The chosen hinge corresponds to
$x_0=y_0=1$ and we note that the behavior remains the same if we
focus on any one of the four hinges.}
    \label{figcorn}
\end{figure}

Next, we study the hinge modes for small $\delta$. To this end we
plot the probability density $|\psi(x=x_0,y=y_0)|^2$ (where
$x_0=y_0=1$ is one of the hinges) as a function of $\delta$. We find
that the hinge modes leak into the bulk around $\delta=0$ where the
gap in the first order Floquet Hamiltonian closes as can be seen
from Fig.\ \ref{figcorn}. We also find that the probability density
$|\psi(x_0=1,y_0=1;k_z=1)|^2$ for the hinge modes dips to a value
close to zero for the exact Floquet spectrum where the gap remains
finite at all drive frequencies. Thus we find that the hinge modes
of $H_F$ leak significantly into the bulk at specific drive
frequencies for which the first order Floquet Hamiltonian vanish. We
shall explore the dynamical consequence of such hybridization in the
next section.

\section{Dynamics of the hinge modes}
\label{dyncor}

In this section, we study the dynamics of hinge modes for two
representative drive frequencies for the square pulse protocol. The
first corresponds to $\sqrt{2}\gamma_1 T/\hbar= \pi, 3 \pi$ where
the Floquet Hamiltonian has large gap while the second corresponds
to $\sqrt{2} \gamma_1 T/\hbar= 2\pi$ where they are almost gapless.
In what follows, we shall study the time evolution of the
wavefunction hinge modes localized at each hinge of the sample. The
probability amplitude of the driven wavefunction can be probed
experimentally as we discuss below; hence it serves as a diagnostic
tool of the Floquet phases outlined in the previous sections.

For a fixed $k_z$, $H_0$ (Eq.\ \ref{eqn:ham0}) supports four
degenerate states at zero energy when Eq.\ \ref{cond1} is satisfied.
With appropriate linear combination of these states, we obtain four
zero energy states localized at the four hinges of the sample. We
start with one of these hinge modes whose wavefunction is given by
$|\psi(0)\rangle$ as the initial state and study its evolution under
the square pulse protocol. We consider a system with $L=10$ units
cells along $x$ and $y$ and numerically compute the exact
stroboscopic time-evolution operator given by Eq.\ \ref{unitevol1}
using OBC along $x$ and $y$ and PBC along $z$. We obtain $|\psi
(nT)\rangle = U^n(T,0) |\psi(0)\rangle$. For each of the four hinge
unit cells, chosen to be at $(x,y)= (1,1)$, $(1,L)$, $(L,1)$ and
$(L,L)$, we designate the initial state $|\psi(0)\rangle$ which is
localized in the $\alpha^{\rm th}$ hinge of the system as
$|\psi^{\alpha}(0)\rangle$.

Next, we define column vectors $|\phi_i^\beta\rangle$ having weight
on the $i^{\rm th}$ site ($i=1,2,3,4$) of a unit cell $\beta$ in the
$x-y$ plane. We then probe the evolution of the quantity,
\begin{eqnarray}
\Phi^{\beta}(nT) &=& \sum_{i=1}^4
|\langle\phi_i^{\beta}|\psi(nT)\rangle|^2= \sum_{i=1}^4
\Phi_i^{\beta}(nT).  \label{phidefeq}
\end{eqnarray}
as a function of $n$. The value of this quantity gives an estimate
of the weight of the state in any given unit cell $\beta$. Our
definition ensures that $\Phi^{\beta}(0) \sim \delta_{\alpha \beta}$
is localized within the sites of the unit cell in the $\alpha^{\rm
th}$ hinge. We note that $\Phi_i^{\beta}(nT)$, being proportional to
the local electronic density of states, can be directly probed
experimentally through scanning tunneling microscopic (STM)
measurements.
\begin{figure}
    \vspace{1\baselineskip}
    \includegraphics*[width=0.49\linewidth]{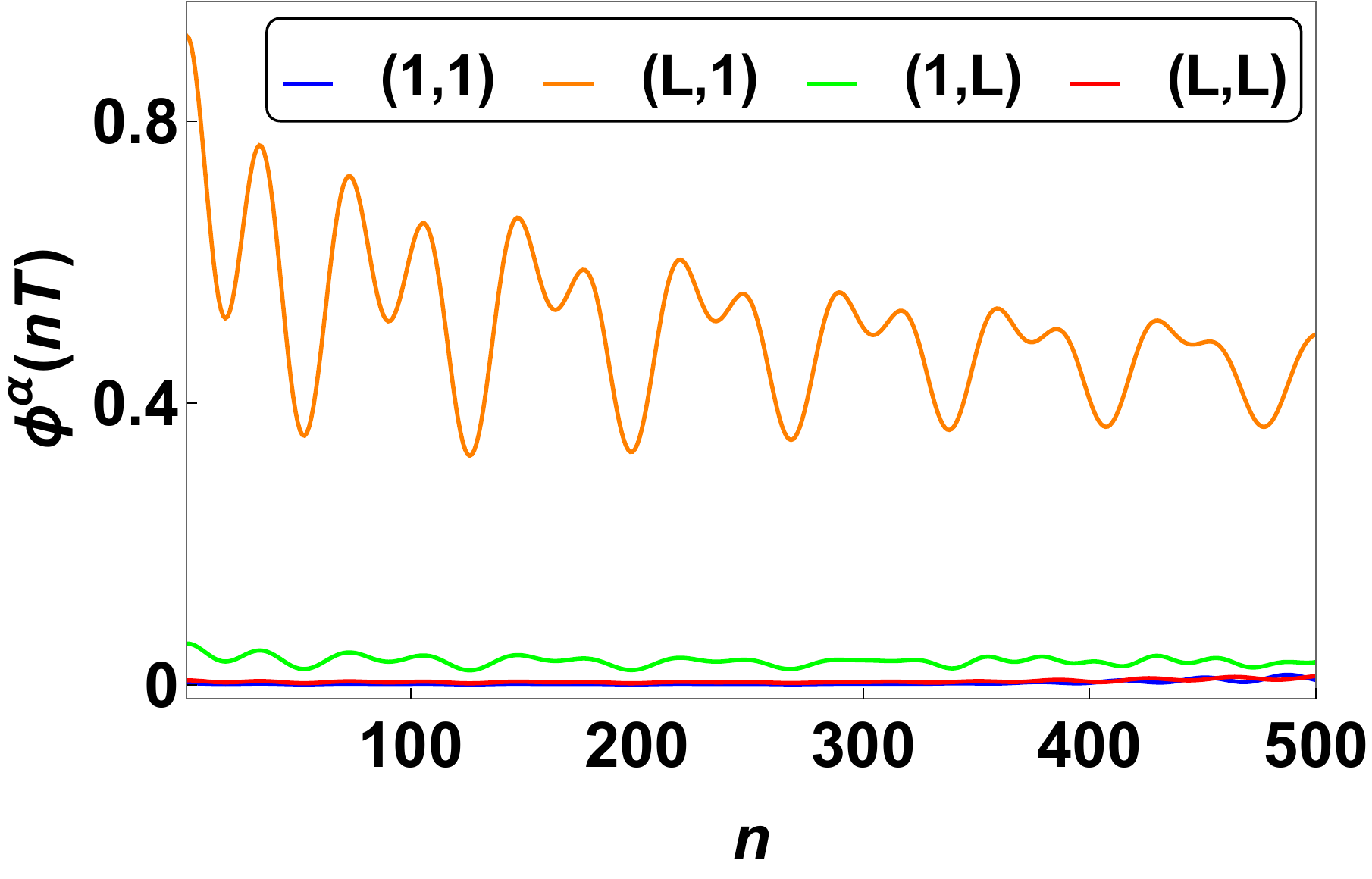}
    \includegraphics*[width=0.49\linewidth]{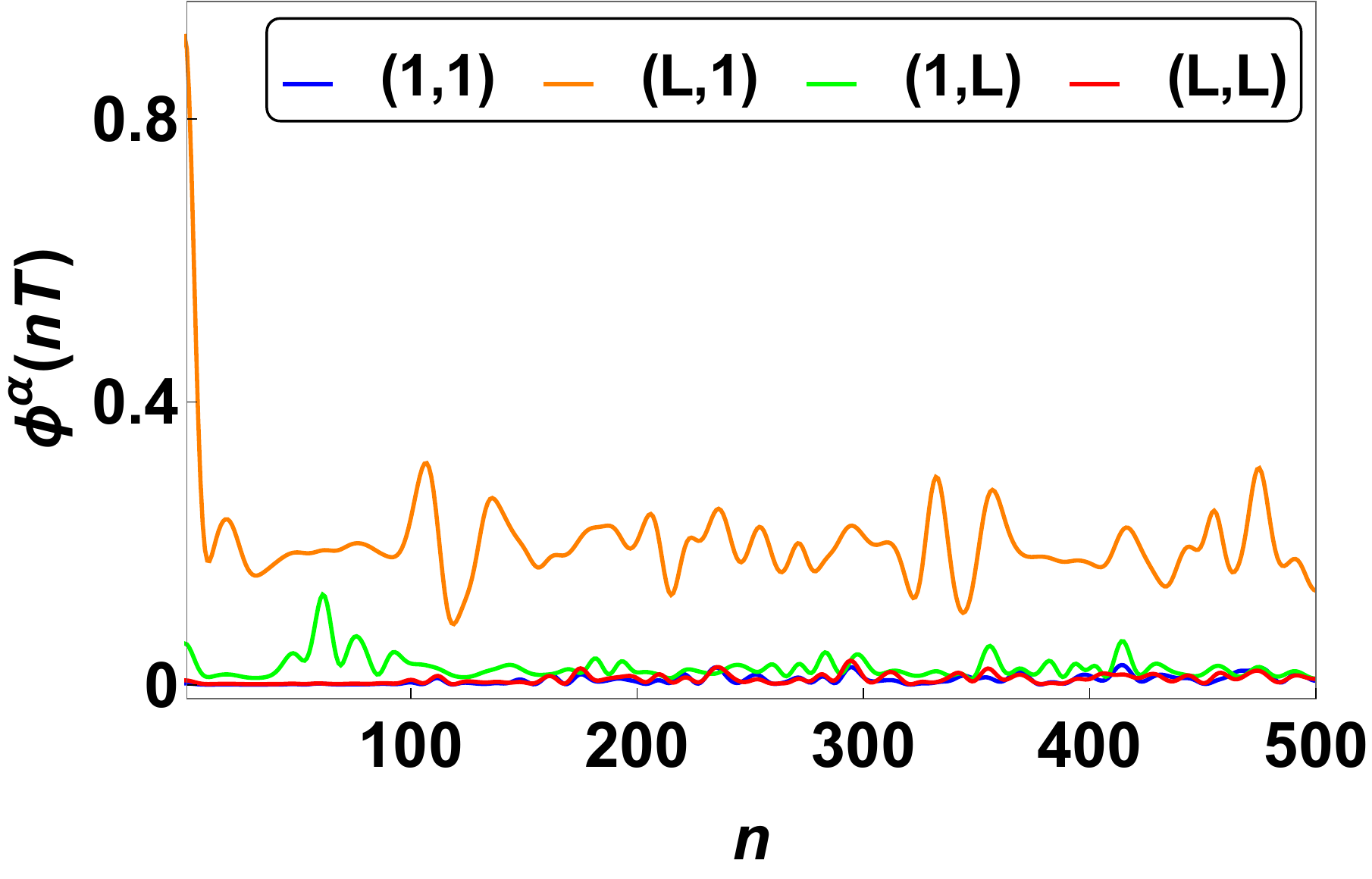}
    \includegraphics*[width=0.49\linewidth]{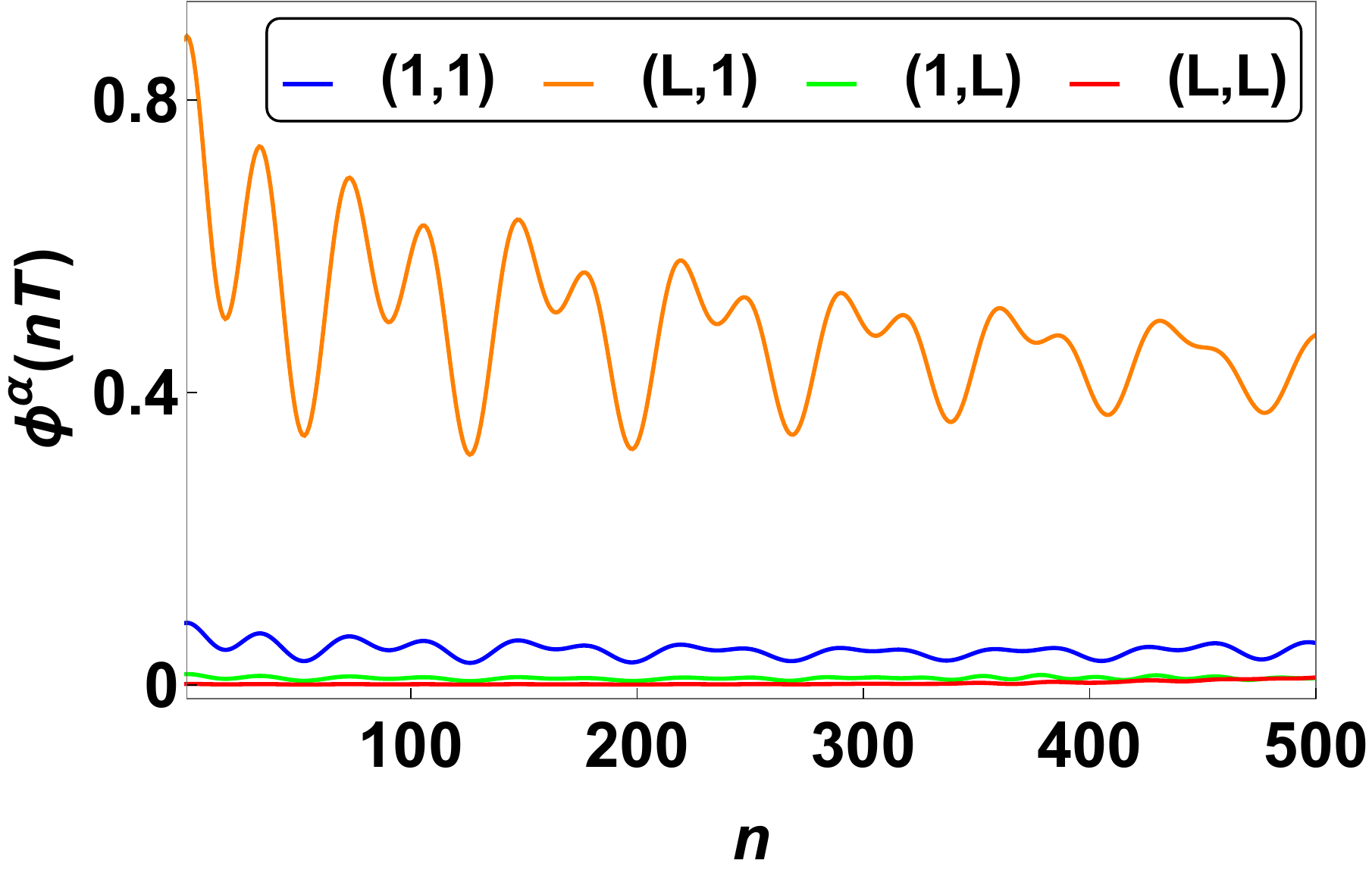}
    \includegraphics*[width=0.49\linewidth]{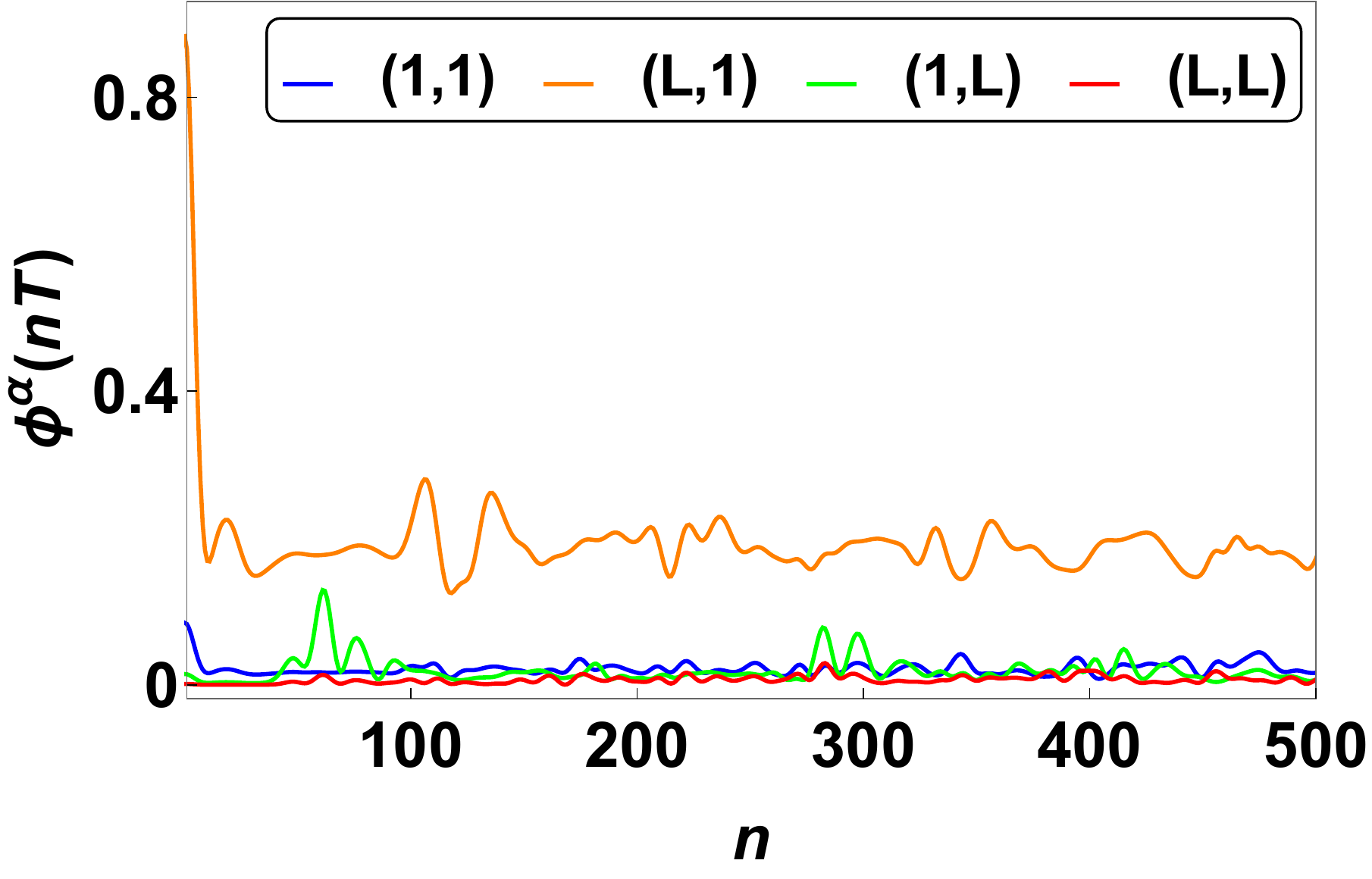}
\caption{Top Panels: Stroboscopic evolution of $\Phi^{\alpha}(nT)$
where $\alpha$ denotes the unit cell at each of the hinges $(1,1)$,
$(1,L)$, $(L,1)$ and $(L,L)$ for $m=0$ as a function of $n$ for
$\sqrt{2} \gamma_1 T/\hbar =\pi $ (left panel) and  $\sqrt{2}
\gamma_1 T/\hbar =3 \pi $ (right panel). Bottom panels: Same as
corresponding top panels but with $m=-0.08$. For all plots
$\gamma=-0.35$, $k_z=0.5$, and $\gamma_1/\lambda=20$.}
    \label{fig:hingedynamics}
\end{figure}

\begin{figure}
    \vspace{1\baselineskip}
    \includegraphics*[width=0.49\linewidth]{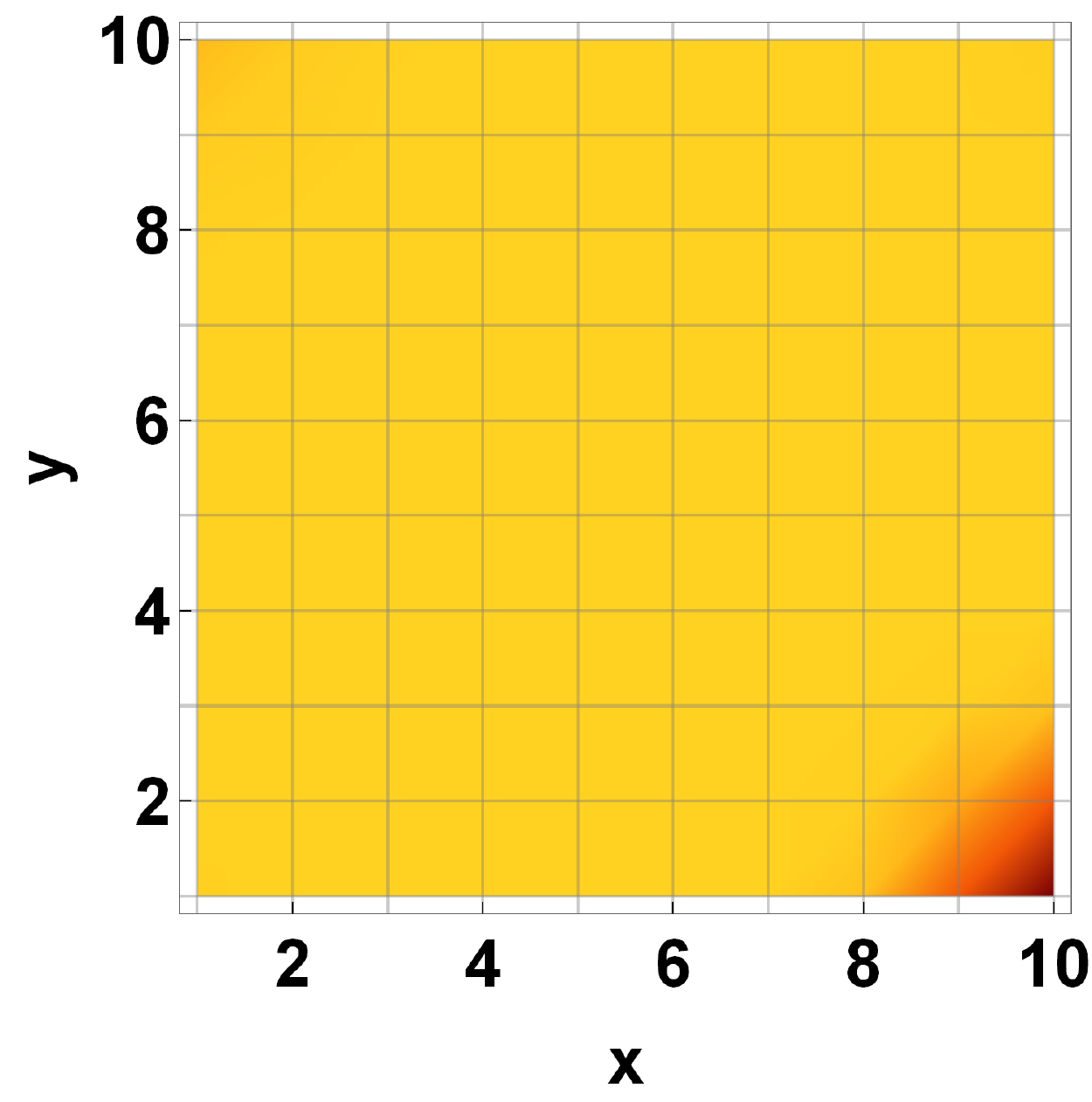}
    \includegraphics*[width=0.49\linewidth]{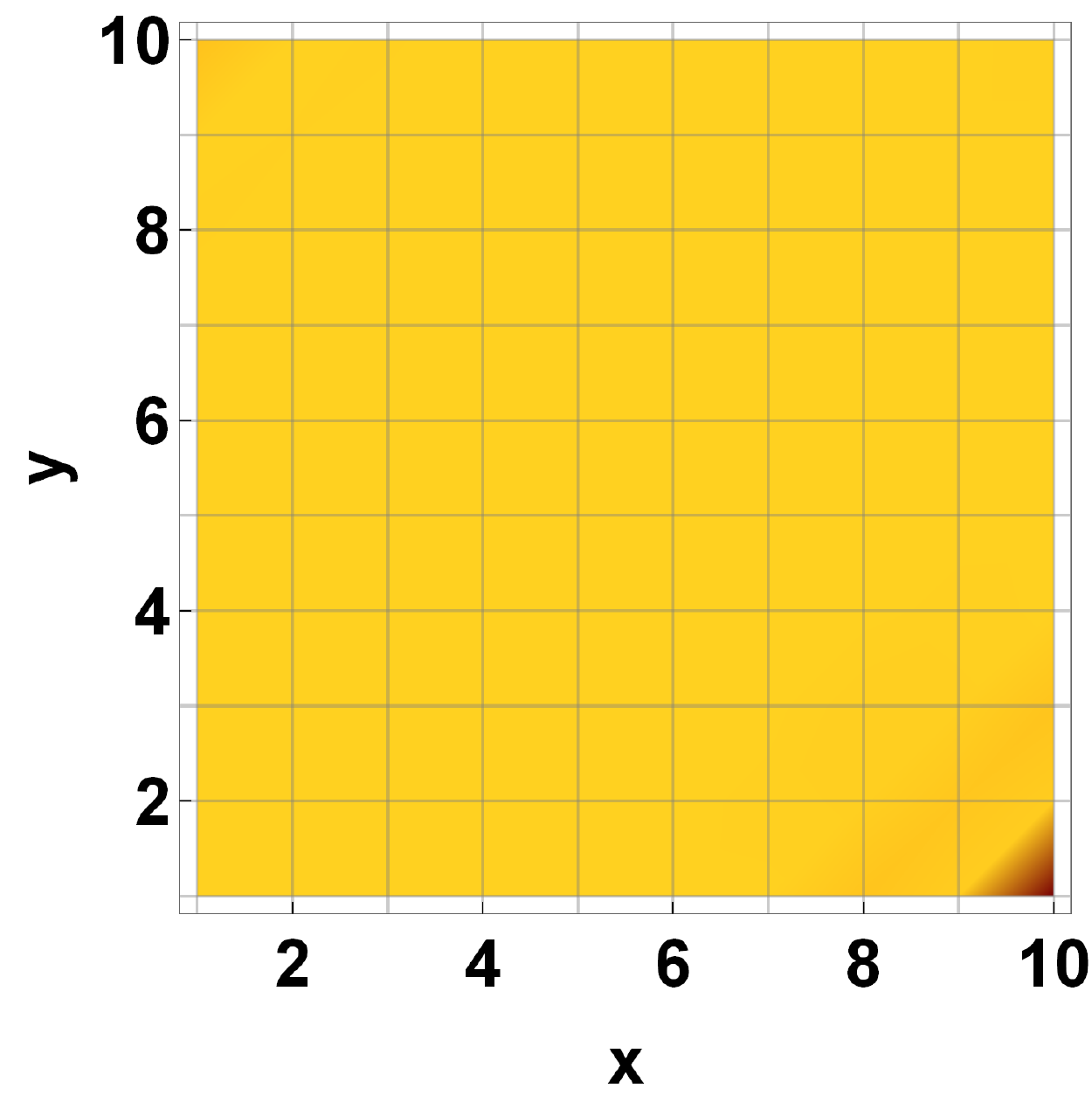}
    \includegraphics*[width=0.49\linewidth]{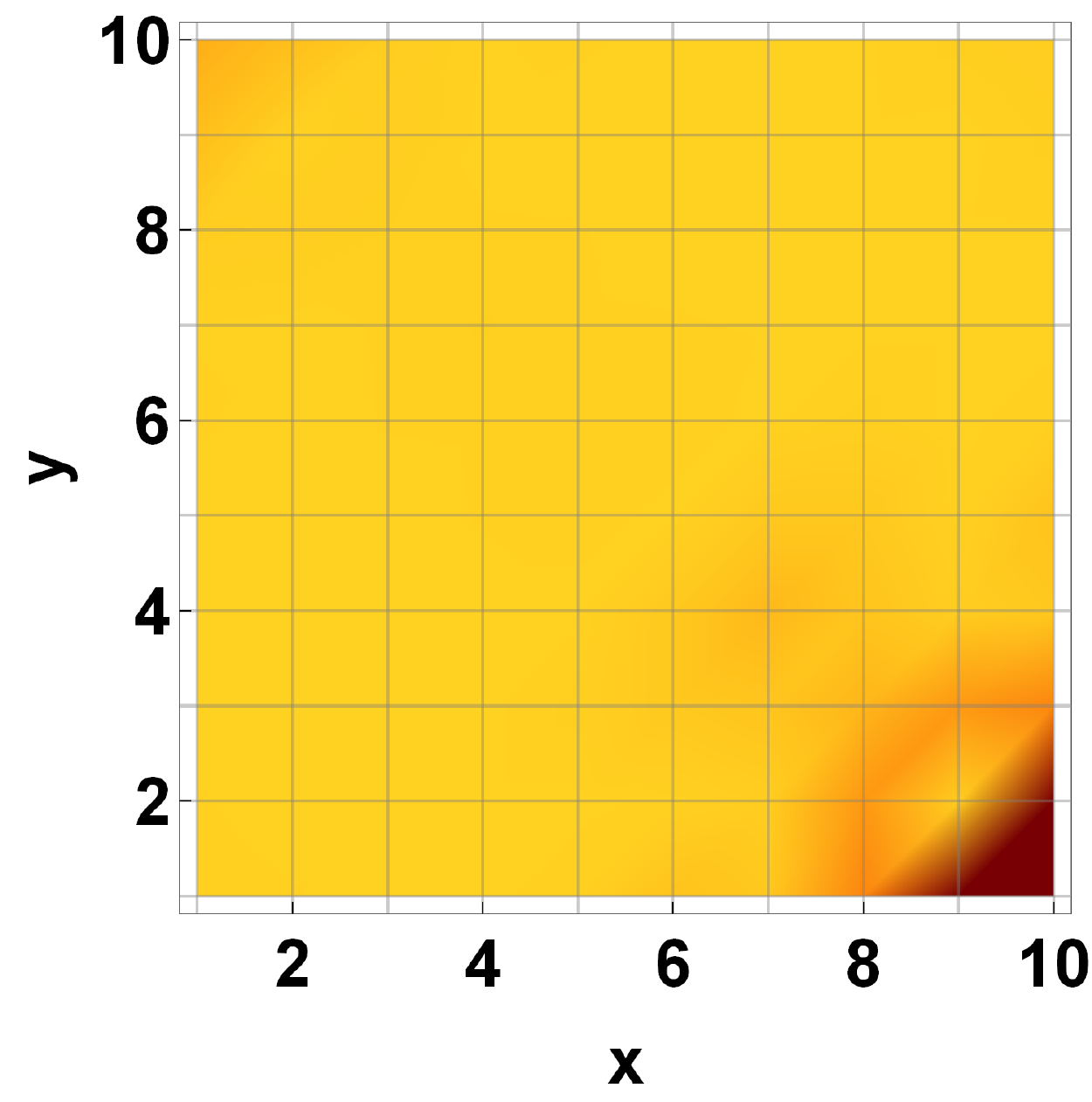}
    \includegraphics*[width=0.49\linewidth]{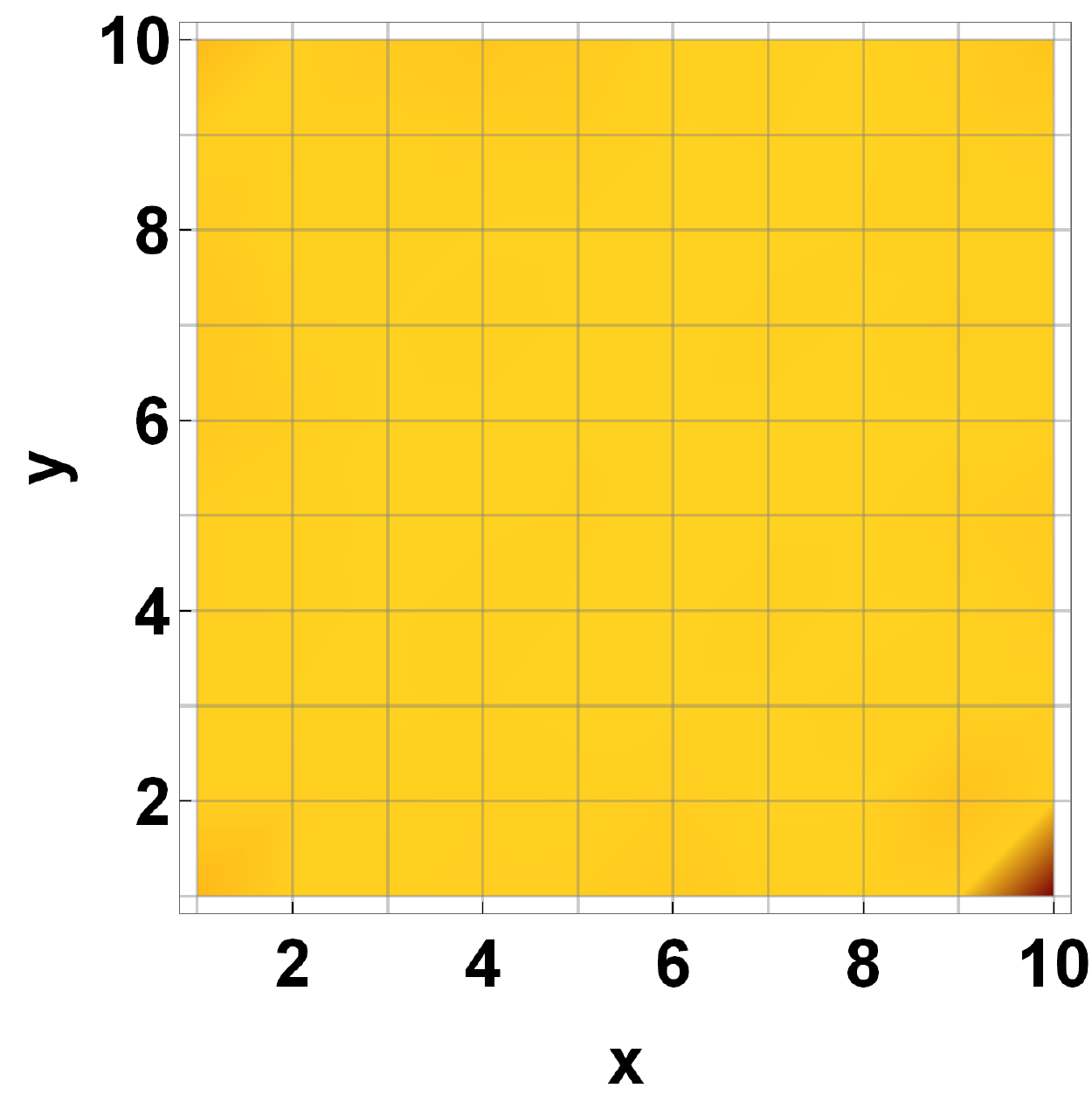}
    \caption{Plot of the spatial profile of $\Phi^{\beta}(nT)$
with $\sqrt{2}\gamma_1 T/\hbar=\pi$ for a system with $L=10$ unit
cells each along $x$ and $y$ in the $x-y$ plane. The top panels
correspond to $n=20$ (left) and $32$ (right) while the bottom panels
correspond to $n=126$ (left) and $3267$ (right). All other
parameters are same as top panels of Fig.\
\ref{fig:hingedynamics}.}
    \label{fig:sppi}
\end{figure}

Fig.\ \ref{fig:hingedynamics} illustrates the evolution of spectral
weight of the hinge state initially localized at site $(L,1)$ for
$\sqrt{2}\gamma_1 T=\pi, 3\pi$ for both $m=0$ and a non-zero
$m=-0.08$. For both values of $m$ we find qualitatively similar
behavior; $\Phi^{\beta}(nT)$ assume appreciable non-zero value for
$\beta \sim \alpha$ for all $n$. This indicates that the state at
any stroboscopic time $t=nT$  remains mostly localized around the
hinge at which it had an initial large overlap. For $\sqrt{2}
\gamma_1 T/\hbar=3\pi$, the state delocalizes to a greater extent
which is due to the presence of a smaller bulk Floquet gap. This can
be further understood from the spatial contour of the hinge state
shown in Fig.\ \ref{fig:sppi} after representative number ($n$) of
drive periods. We find that the weight of the hinge state always
remains localized to the hinge where it was initially localized;
this behavior is consistent with having a gapped Floquet spectrum at
the bulk.

\begin{figure}
    \vspace{1\baselineskip}
     \includegraphics*[width=0.49\linewidth]{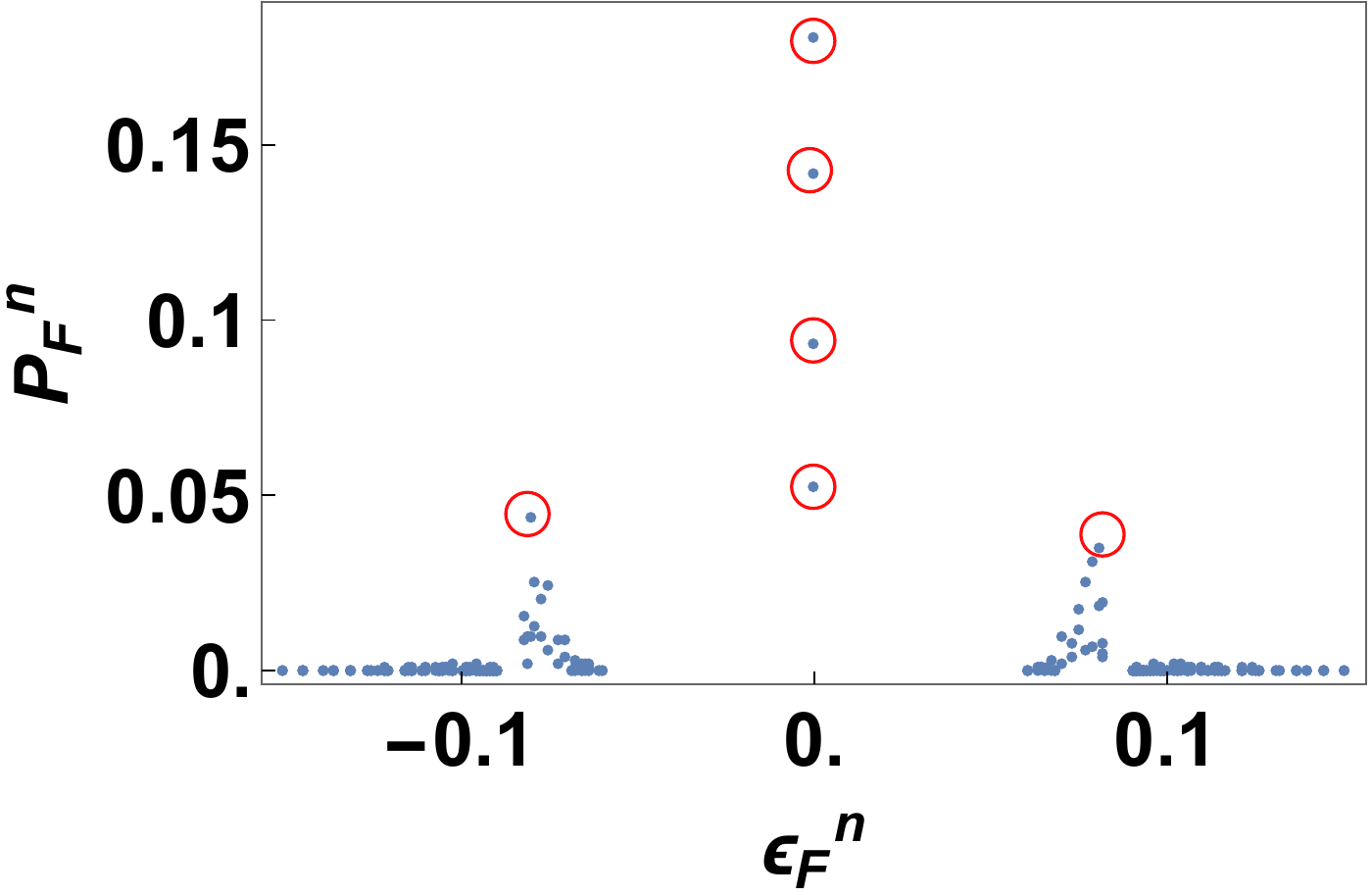}
    \includegraphics*[width=0.49\linewidth]{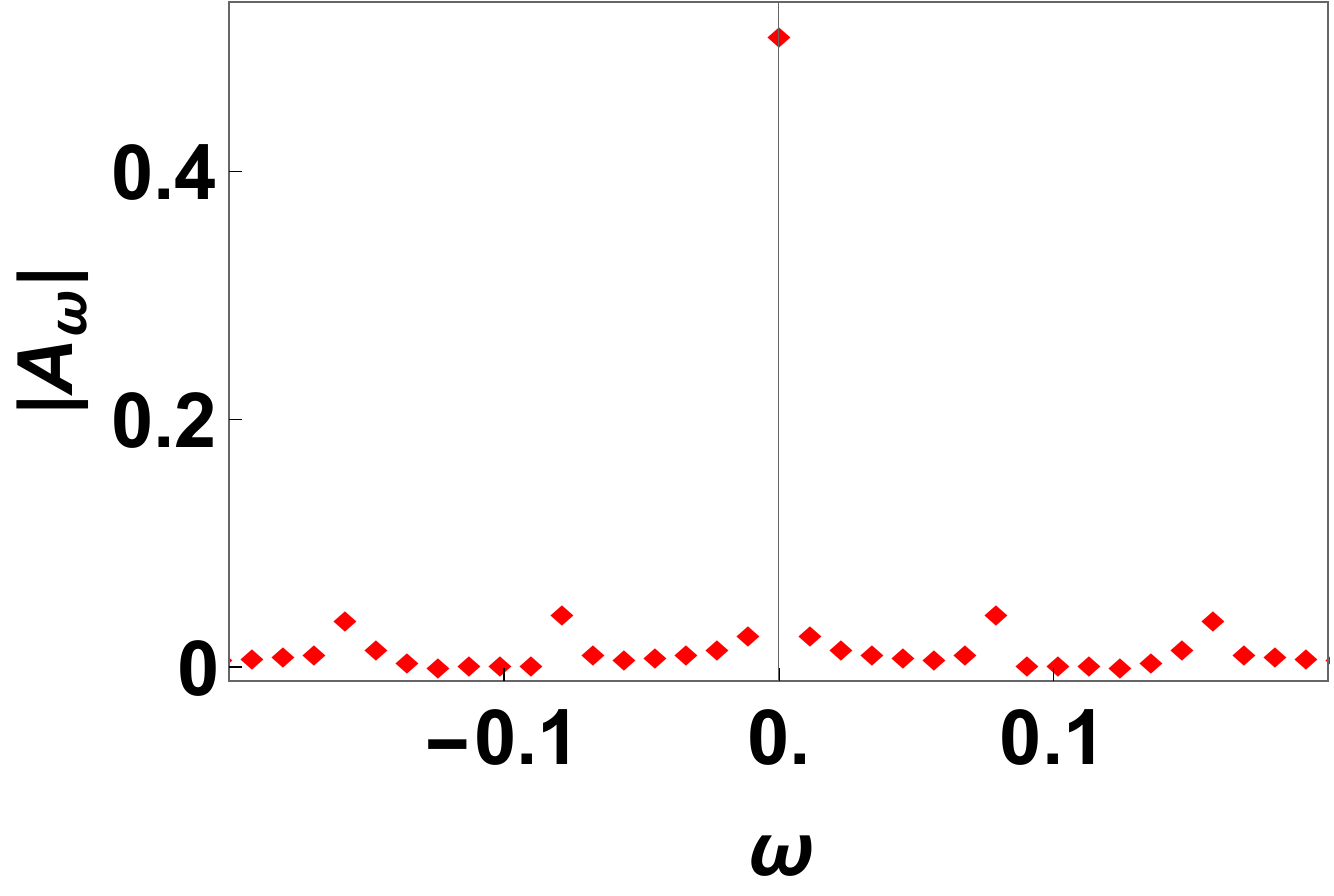}\\
    \includegraphics*[width=0.49\linewidth]{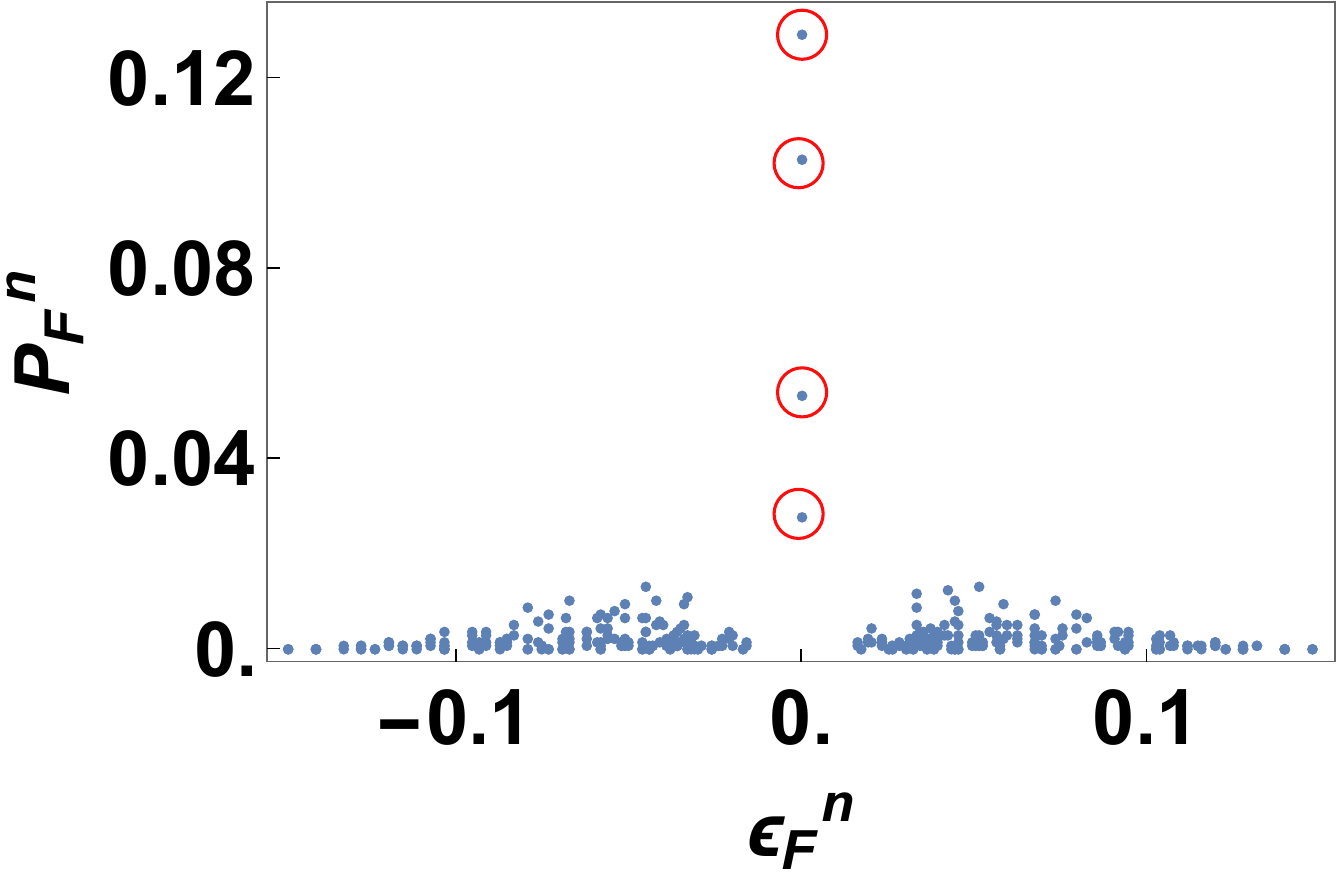}
    \includegraphics*[width=0.49\linewidth]{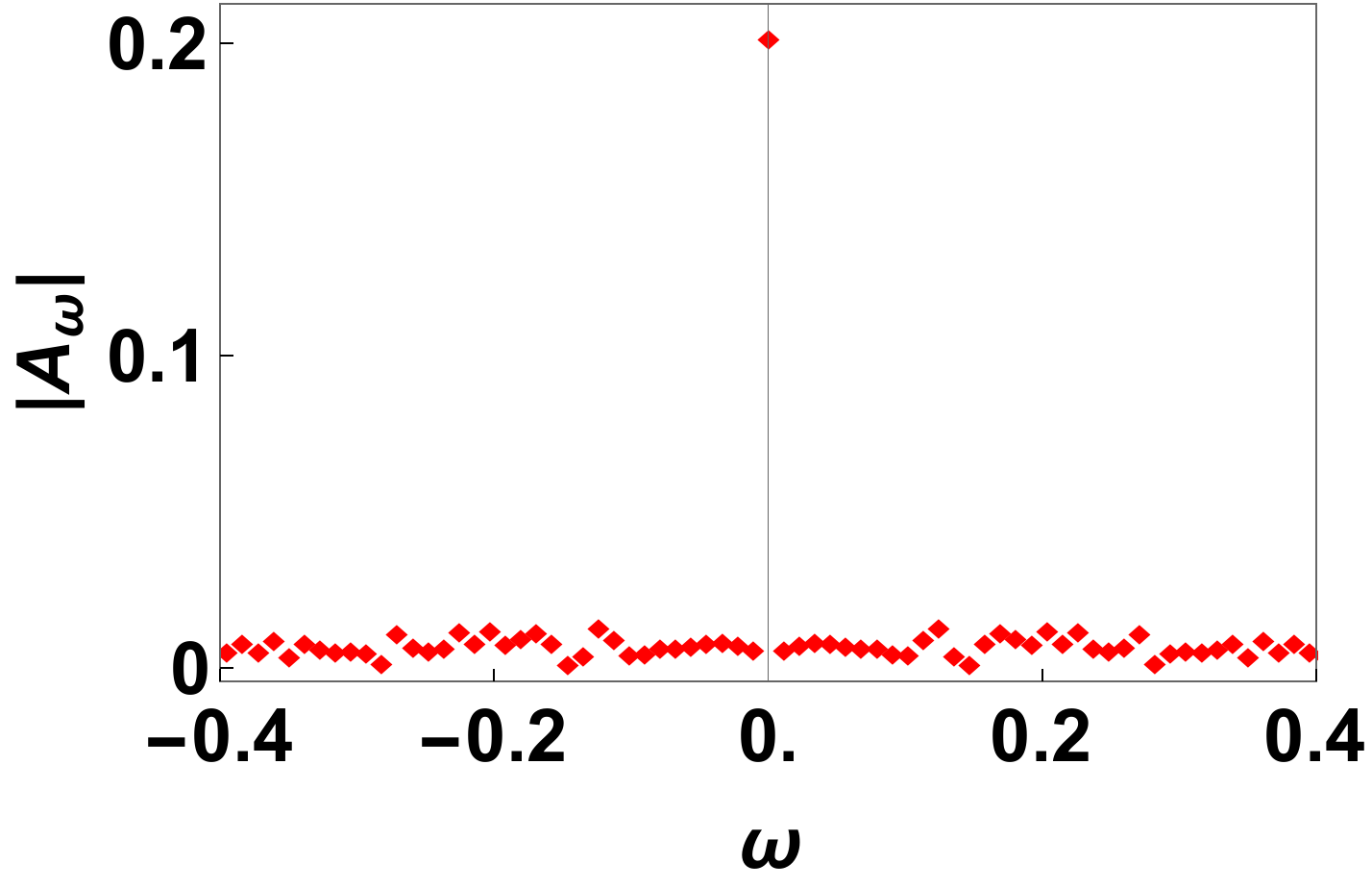}
\caption{Top left panel: Plot of overlap probability  $ P_F^n =
|\langle\xi_F^n|\psi^{\alpha}(0)\rangle|^2$ of the initial state
$|\psi^{\alpha}(0)\rangle$ localized at the hinge $\alpha$ at
$(L,1)$ with the Floquet eigenstates $|\xi_F^n\rangle$ for $\sqrt{2}
\gamma_1 T/\hbar =\pi $ and $m=0$. Top right panel: Fourier modes of
the dynamics for $\sqrt{2} \gamma_1 T/\hbar =\pi $ and $m=0$. Bottom
panels: Same as corresponding top panels but for $\sqrt{2}\gamma_1
T/\hbar = 3 \pi$. All other parameters are same as in Fig.\
\ref{fig:hingedynamics}.}
    \label{fig:floqspectralanalysis}
\end{figure}

The dynamical behavior of the hinge modes can be further understood
by examining the overlap of our initial state with the Floquet
eigenstates, $P_F^n = |\langle\xi_F^n|\psi^{\alpha}(0)\rangle|^2$ as
shown in the top left panel of Fig.\ \ref{fig:floqspectralanalysis}.
The Floquet eigenstates having appreciable overlap with the initial
states are encircled in red; these include the four zero energy
states (ZES). These coefficients do not change with time and thus
provide a base average value about which the fluctuation occurs.
This value is larger for higher drive frequency where the gap is
larger. The analysis of the Fourier modes shown in the right panels
of Fig.\ \ref{fig:floqspectralanalysis} yields constituent
frequencies of these fluctuations. These turn out to be consistent
with the difference in quasienergy values on which the initial state
has substantial projections. We note that for $\sqrt{2}\gamma_1
T/\hbar =3\pi$, there is a rapid dissipation of the state in the
bulk. The difference in this case stems from the fact that the
Floquet ZES are separated from the bulk by a reduced energy gap,
resulting in a faster decay. The corresponding Fourier weights of
the mode, $A(\omega)$, shown in the bottom right panel of Fig.\
\ref{fig:floqspectralanalysis}, indicates the presence of multiple
Fourier modes with small overlap due to which the dynamics appears
incoherent. We have checked that this behavior remains qualitatively
similar even when a small finite $m$ is switched on.

\begin{figure}
\vspace{1\baselineskip}
\includegraphics*[width=0.49\linewidth]{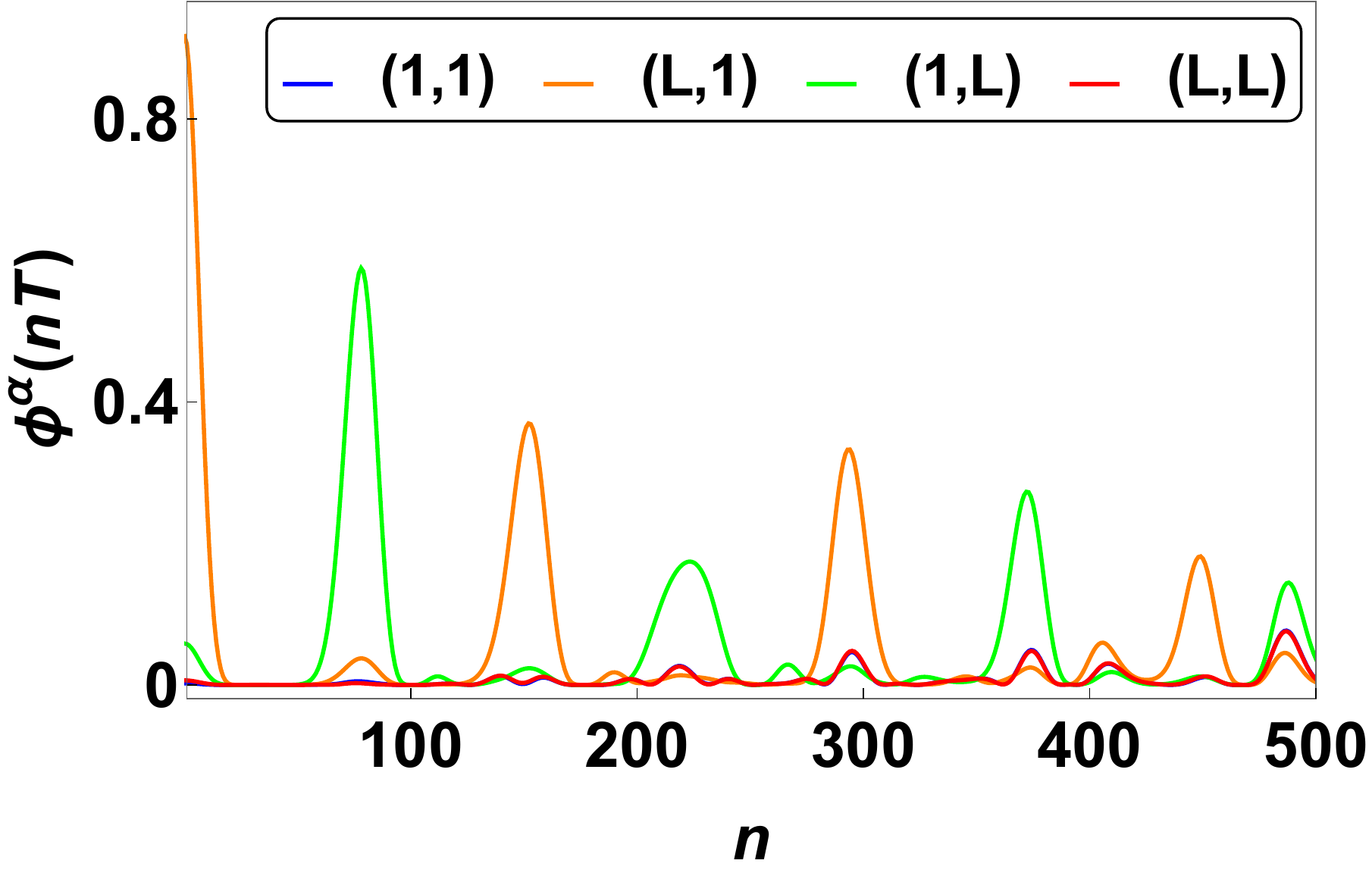}
\includegraphics*[width=0.49\linewidth]{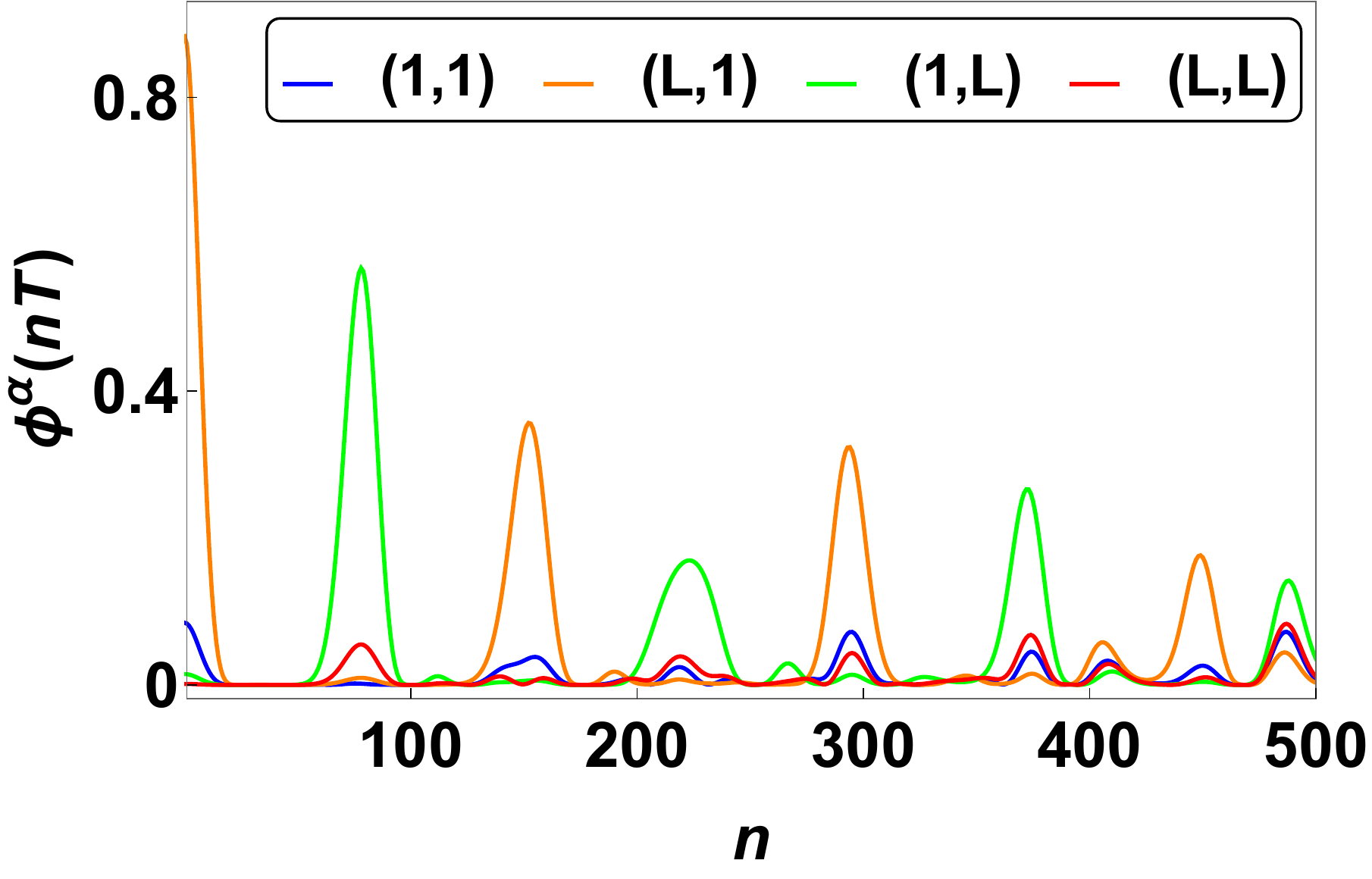}
\includegraphics*[width=0.49\linewidth]{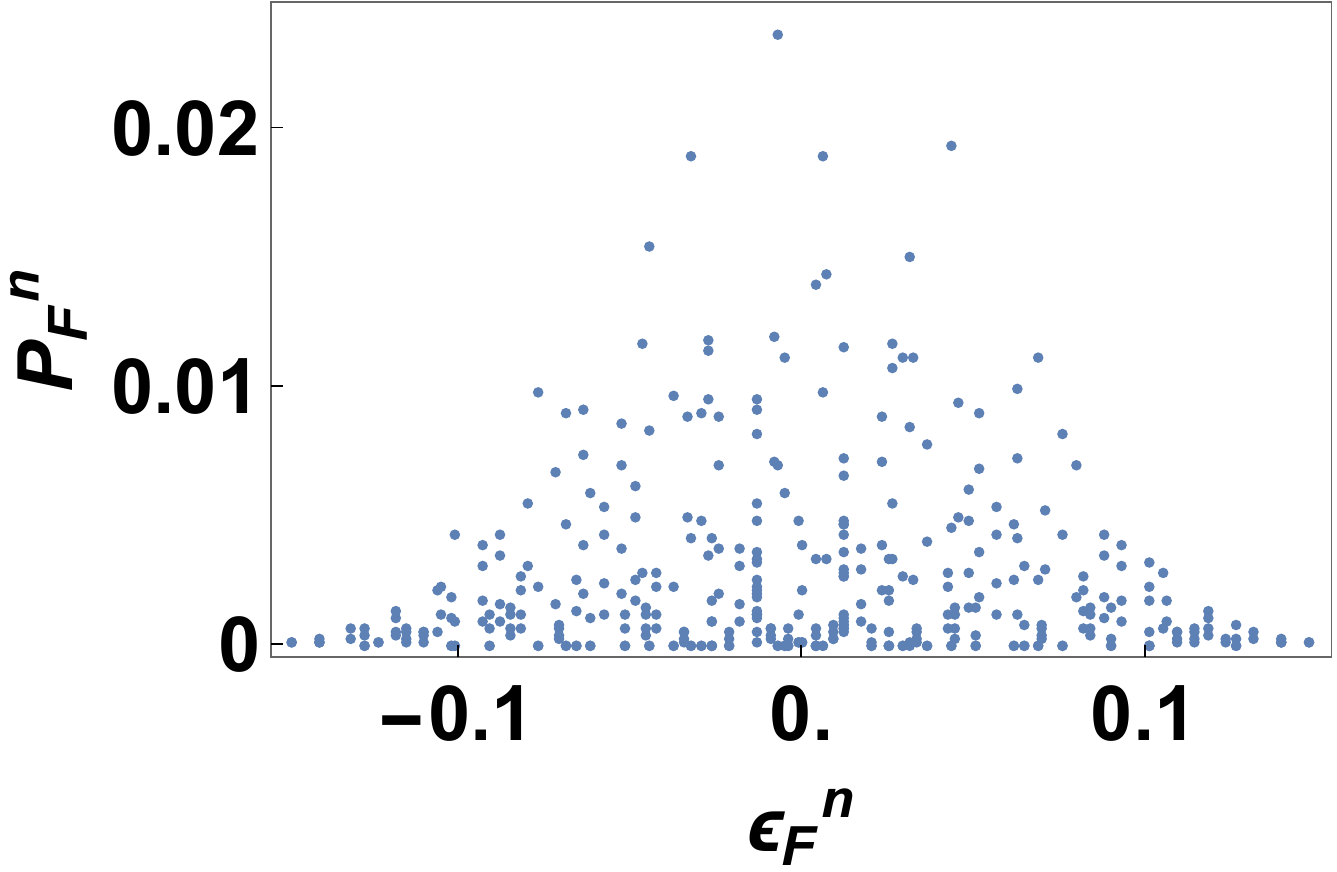}
\includegraphics*[width=0.49\linewidth]{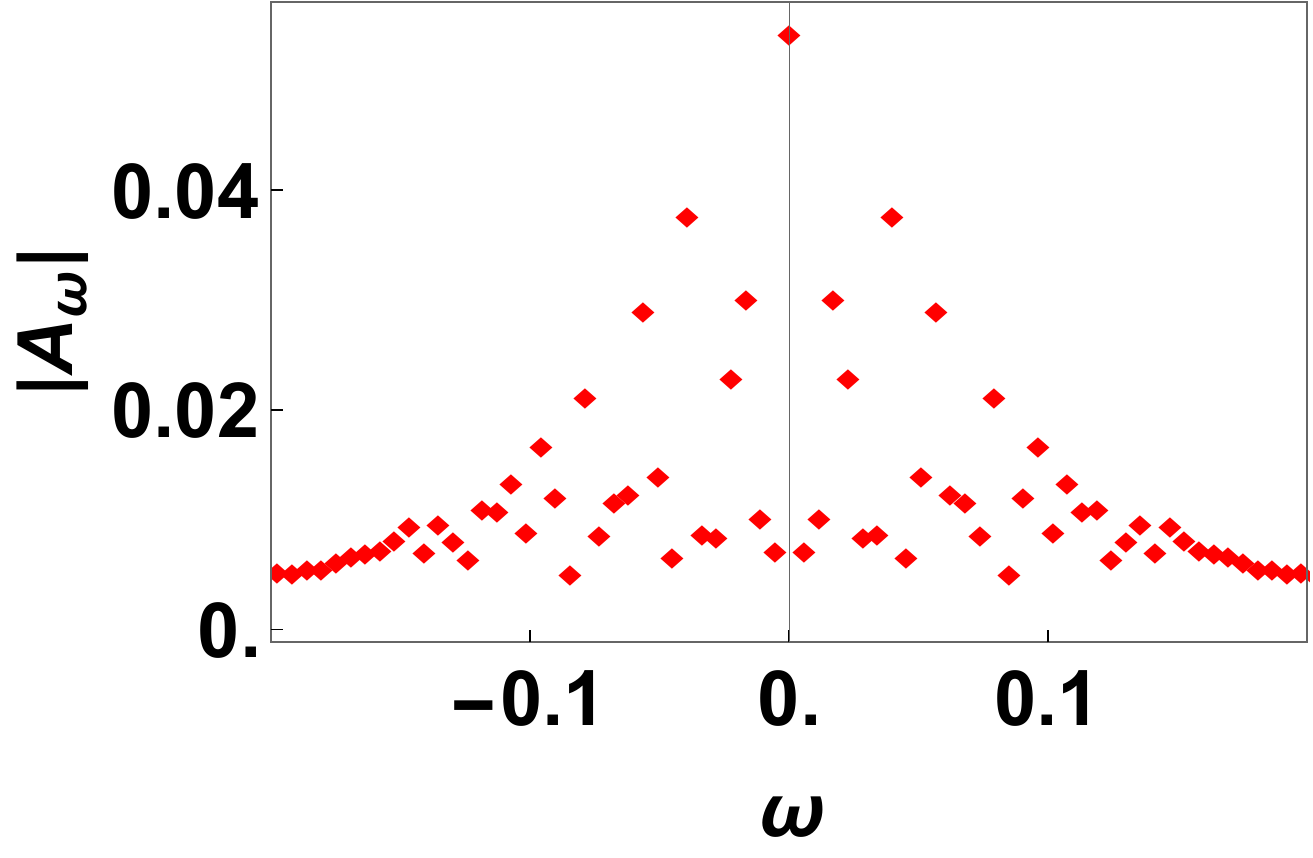}
\caption{Top panels:  Stroboscopic evolution of $\Phi^{\alpha}(nT)$
where $\alpha$ denotes the unit cell at each of the hinges $(1,1)$,
$(1,L)$, $(L,1)$ and $(L,L)$ as a function of $n$ for $\sqrt{2}
\gamma_1 T/\hbar =2\pi $ with $m=0$ (left panel) and  $m=-0.08$
(right panel). Bottom panels: Plot of the overlap $P_F^n$ (left
panel) and the Fourier modes $A(\omega)$ (right panel). Both plots
correspond to $\sqrt{2} \gamma_1 T/\hbar =2\pi $ and $m=0$. All
other parameters are same as in Fig.\ \ref{fig:hingedynamics}.}
\label{fig:2pihingedynamics}
\end{figure}

For $\sqrt{2}\gamma_1 T/\hbar=2\pi$, where the bulk Floquet spectrum
is almost gapless, the dynamics of the hinge mode is qualitatively
different, as shown in Fig.\ \ref{fig:2pihingedynamics}. We find
that the hinge mode shows almost coherent transport between the
diagonally opposite hinges as shown in the top panels of Fig.\
\ref{fig:2pihingedynamics}. The weight of the state starts being
localized at the unit cell of hinge $(L,1)$ and reaches the hinge
$(1,L)$ after $n \sim 80$ cycles; after $n\sim 160$ cycles of the
drive, the weight of the state again becomes localized at the hinge
where its initial weight was large. We note that this does not
necessarily mean that the wavefunction of the driven hinge state
after $n \sim 160$ cycles exhibits large overlap with the initial
wavefunction; there exists significant difference in the
distribution of weights of these wavefunctions within the hinge unit
cell. As shown in the bottom panels of Fig.\
\ref{fig:2pihingedynamics}, the spectral weight $P_F^n$  has a large
overlap with several bulk Floquet modes and has finite weight in
several Fourier modes. This also is in sharp contrast to that found
for $\sqrt{2}\gamma_1 T/\hbar=\pi$ where the bulk Floquet gap is
large; for the latter case, the wavefunction of the hinge mode has
significant overlap with only a few Floquet modes.

\begin{figure}
    \vspace{-1\baselineskip}
    \includegraphics*[width=0.48\linewidth]{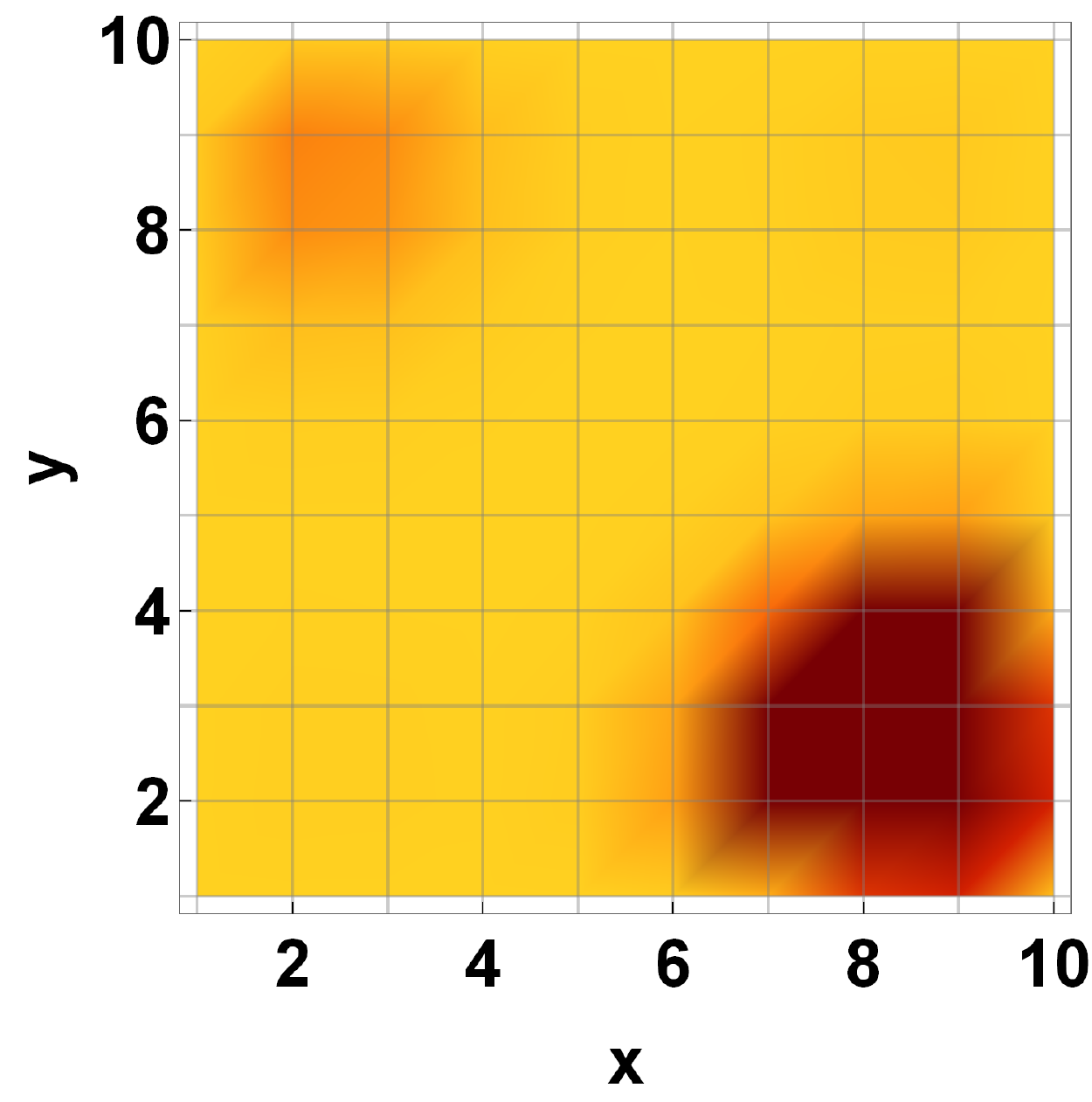}
    \includegraphics*[width=0.48\linewidth]{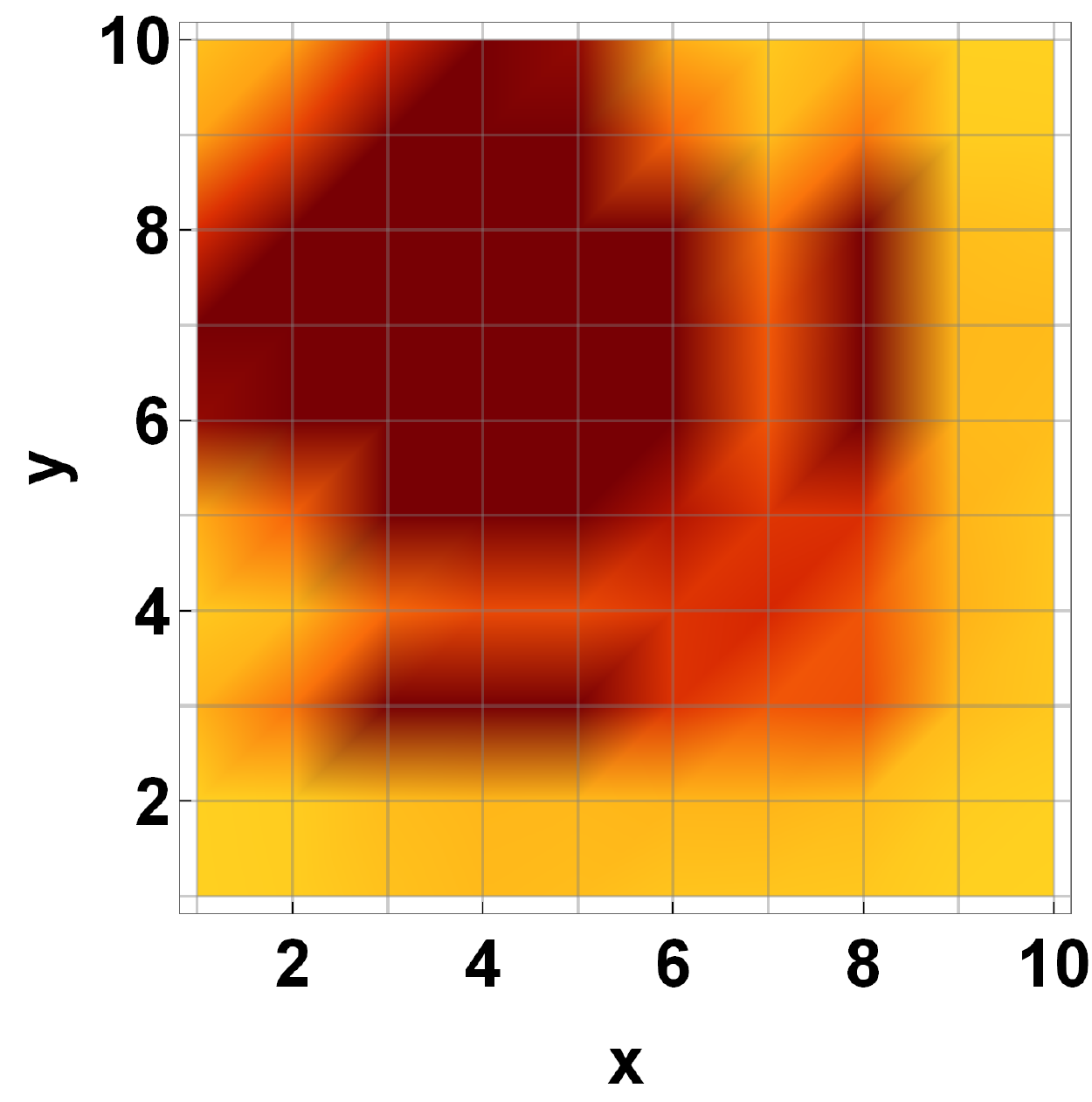}
    \includegraphics*[width=0.48\linewidth]{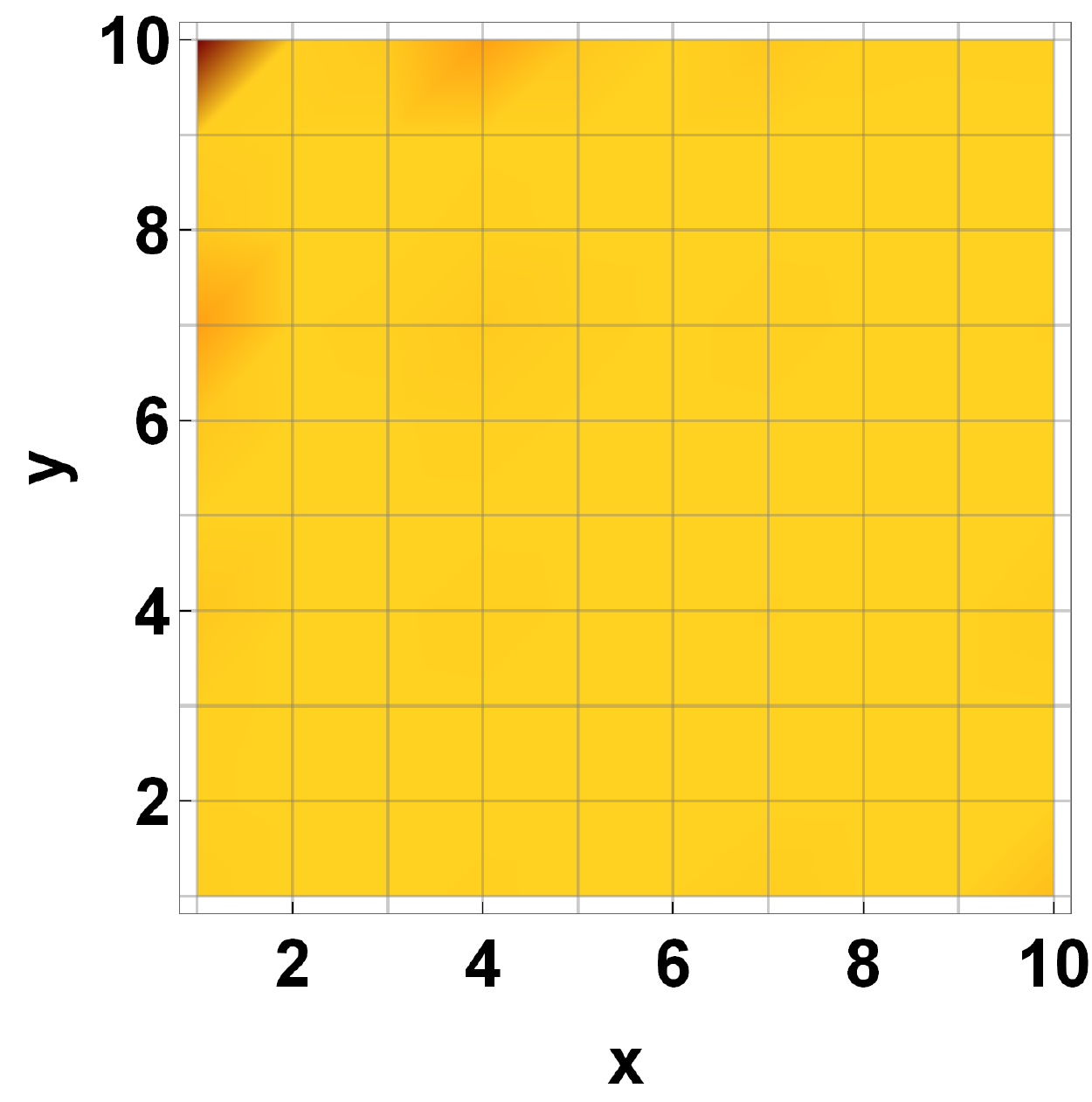}
    \includegraphics*[width=0.48\linewidth]{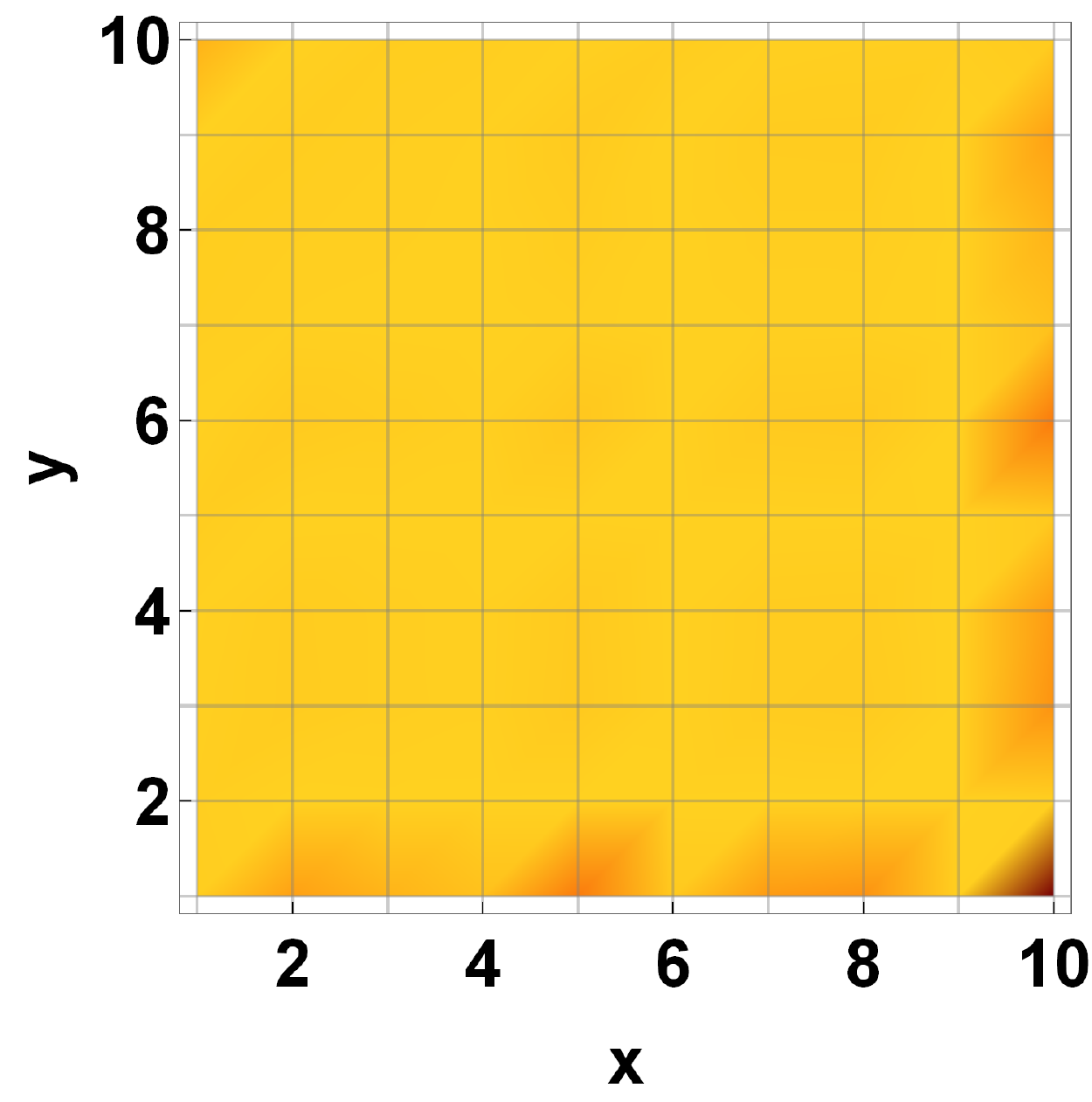}
\caption{Plot of the spatial profile of $\Phi_i^{\beta}(nT)$ with
$\sqrt{2}\gamma_1 T/\hbar=2\pi$ for a system with $L=10$ unit cells
each along $x$ and $y$ in the $x-y$ plane: (a) $n=20$, (b) n=$52$,
(c) $n=80$, and (d) n=$153$. All other parameters are same as in
Fig.\ \ref{fig:hingedynamics}.}
    \label{fig:snapshot}
\end{figure}

The difference in dynamics of the hinge mode for $\sqrt{2} \gamma_1
T/\hbar=2 \pi$ and $\sqrt{2} \gamma_1 T/\hbar= \pi$ can be further
understood from the spatial profile of $\Phi^{\beta}(nT)$ for
representative values of $n$ as shown in Fig.\ \ref{fig:snapshot}.
We find that at intermediate times $0<n<80$, the amplitude of the
driven state is spread out in the bulk as can be seen from the top
right panel of Fig.\ \ref{fig:snapshot}. In contrast, for $n \simeq
80 p$, where $p$ is integer, they are localized in one of the two
diagonally opposite hinges with spread along the respective
surfaces. All of these features indicate qualitatively different
hinge mode dynamics for $\sqrt{2} \gamma_1 T/\hbar= 2 \pi$.

\begin{figure}
    \vspace{-1\baselineskip}
    \includegraphics*[width=0.48\linewidth]{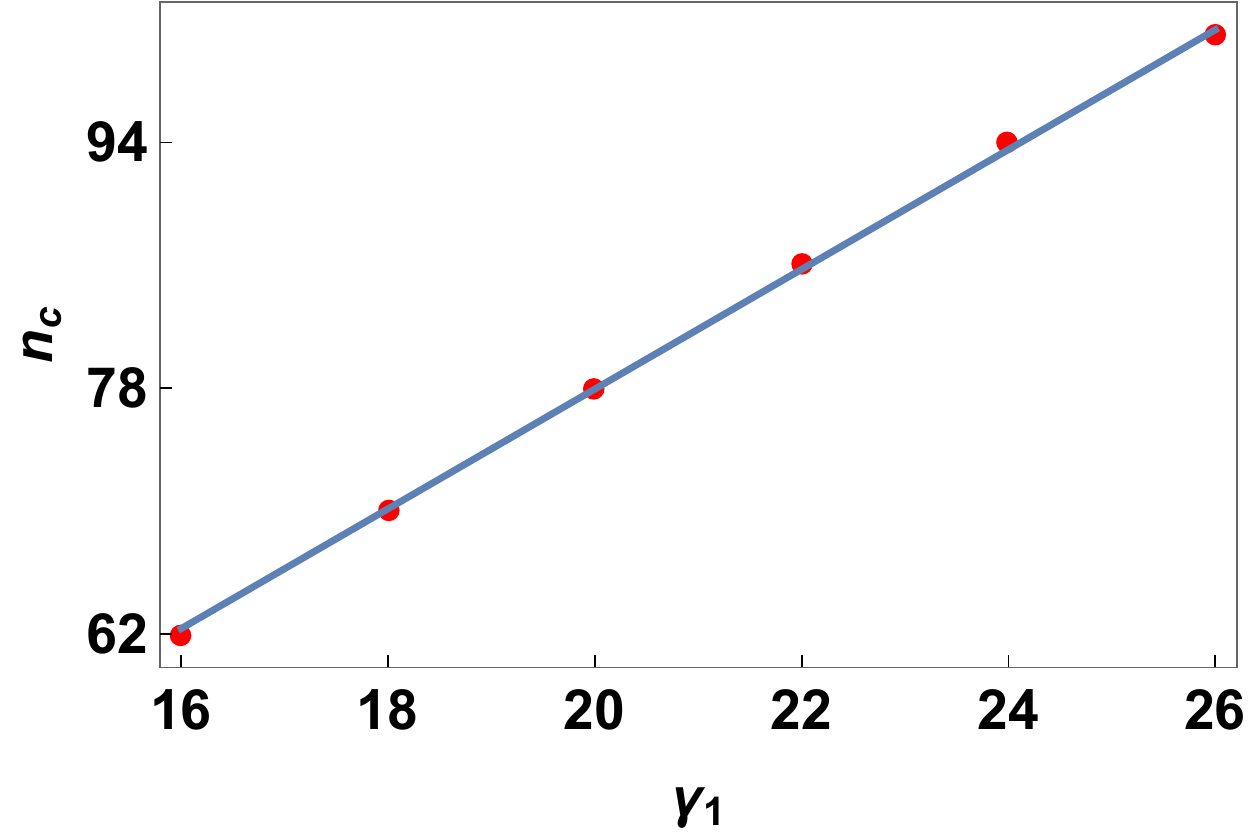}
    \includegraphics*[width=0.48\linewidth]{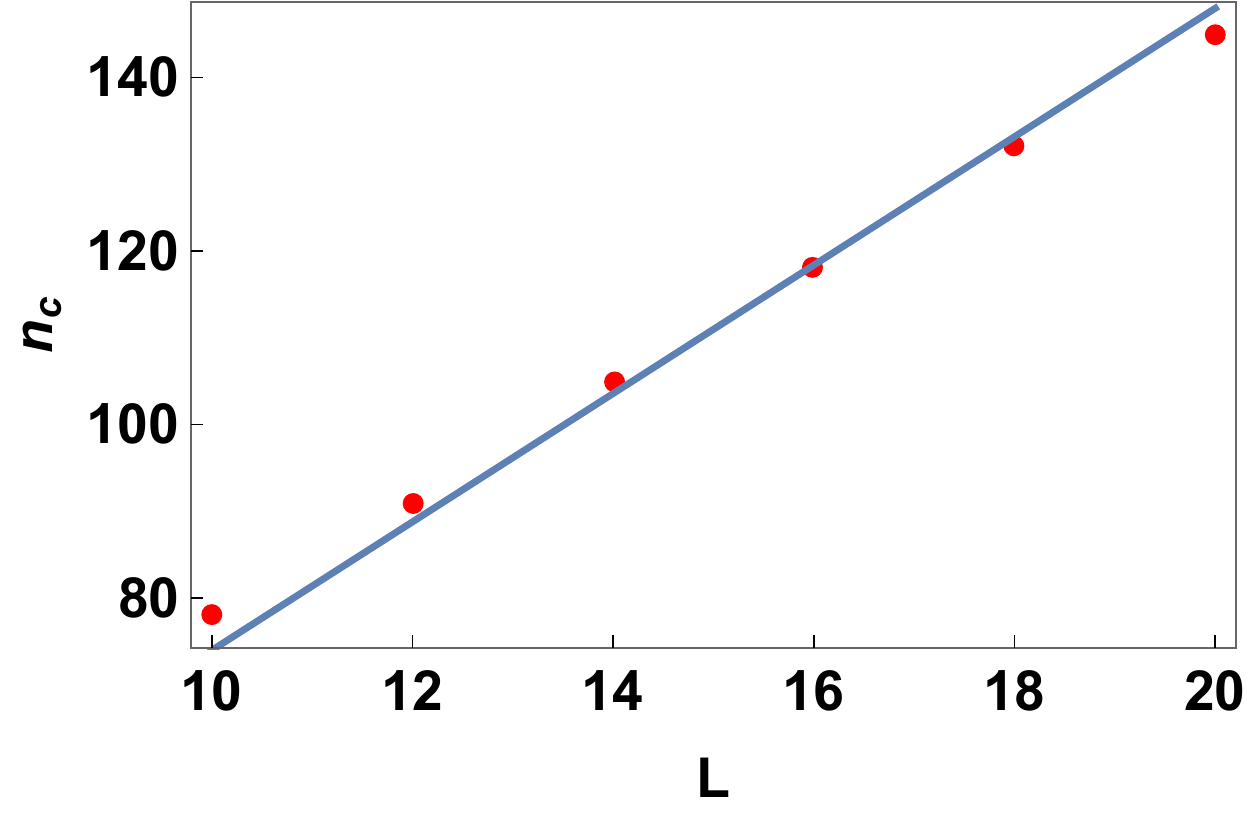}
\caption{Plot of $n_c$ as a function of $\gamma_1$, with $\lambda=1$
for $L=10$ (left panel) and as a function of $L$ for
$\gamma_1/\lambda=20$ (right panel). Both the plots indicate linear
dependence of $n_c$ in accordance with the prediction of the saddle
point analysis. All other parameters are same as in Fig.\
\ref{fig:hingedynamics}. See text for more details.}
    \label{ncplot}
\end{figure}

To obtain a qualitative analytic understanding of the behavior of
these hinge modes, we note that, in contrast to their counterparts
for $\sqrt{2}\gamma_1 T/\hbar=\pi, 3\pi$, several bulk Floquet modes
have high overlap with the initial state representing the hinge
mode. Thus the state representing the driven hinge mode can be
written as
\begin{eqnarray}
|\psi(nT)\rangle &=& \sum_m c_{\alpha m} e^{-i E_m^F nT} |m\rangle,
\label{overlap}
\end{eqnarray}
where $c_{\alpha m}= \langle m|\psi_{\alpha}(0)\rangle$, $|m\rangle$
denote the bulk Floquet states, and $E_F^m$ are their quasienergies.
For having analytic insight into the problem, we now assume the bulk
Floquet modes are almost similar to the ones with periodic boundary
condition; in this, one can replace $|m\rangle$ with $|k\rangle$ and
$E_m^F$ with $E_a^F(\vec k)$. There are four such eigenvalues for
each $\vec k$ as given by Eq.\ \ref{eqn:nodes}; we label these with
the index $a$ assuming values $1,2,3,4$. The sum over $m$ can then
be replaced by an integral over $k$ and a sum over the index $a$ and
we obtain
\begin{eqnarray}
\psi(x,y; nT) &=& \sum_{a=1,4}\int \frac{d^2k}{(2\pi)^2} c^{a}_{\alpha} (\vec k) e^{i(\vec k \cdot \vec r- E_a^F({\vec k})nT)}  \nonumber\\
&=& \sum_{a=1,4} \int \frac{d^2k}{(2\pi)^2} c^a_{\alpha}(\vec k) e^{i n \Xi^a (x,y,:\vec k)}, \nonumber\\
\Xi^a (x,y;{\vec k}) &=& (k_x x +k_y y)/n -E_a^F(\vec k)T,
\label{keq1}
\end{eqnarray}
where ${\vec r}= (x,y)$ and $a$ denotes index for eigenvalues (Eq.\
\ref{eqn:nodes}) and the corresponding eigenvectors. For large $n$,
thus the contribution to $\psi(x,y;n)$ comes from coordinates which
satisfy $\partial_{k_x} \Xi^a(x,y;\vec
k)=\partial_{k_y}\Xi^a(x,y;\vec k)=0$. Using Eqs.\ \ref{eqn:nodes},
we find that these are given by
\begin{eqnarray}
\frac{x}{n} &=& \pm \frac{\lambda T}{\sqrt{2}} \sin k_x^0 , \quad
\frac{y}{n} = \pm \frac{\lambda T}{\sqrt{2}} \sin k_y^0.
\label{saddle1}
\end{eqnarray}
We now use this to find the shortest number of drive cycles at which
the state reaches the diagonally opposite hinge, we seek a solution
of Eq.\ \ref{saddle1} for $x=y=L$ and smallest possible $n_c>0$.
This yields $k_x^0=k_y^0=\pm \pi/2$ so that $n_c = L
\gamma_1/(\lambda\pi)$. For $L=10$ and $\gamma_1=20\lambda$, this
yields $n_c=64$ which is close to the numerical value of $n_c \sim
80$ (Fig.\ \ref{fig:snapshot}). This analytical result can be
validated by plotting $n_c$, obtained from exact numerics, as a
function $L$ and $\gamma_1$, as shown in Fig.\ \ref{ncplot}. We find
that in accordance with the analytic prediction, $n_c$ varies
linearly with both $L$ and $\gamma_1$. For $L=10$, the slope of the
plot of $n_c$ as a function of $\gamma_1$ is found to be $3.9$ while
the theoretical prediction turns out to be $L/(\lambda \pi) \simeq
3.2$. Similarly, for $\gamma_1/\lambda=20$, the slope of the best
fit for $n_c(L)$ is found to be $7.4$ while the theoretical
prediction is $\gamma_1/(\lambda \pi)=6.4$. This difference is
partly due to a finite $n_c$ which induces additional corrections to
the saddle point value. Thus the time period between revivals of the
hinge mode shown in the top left panel of Fig.\
\ref{fig:2pihingedynamics} can be qualitatively understood using
this approximate saddle point analysis; however, one needs to go
beyond this simple analysis to obtain more accurate value of $n_c$.

\section{Discussion}
\label{diss}

In this work, we have studied the Floquet dynamics of hinge modes of
second-order topological material modeled by free fermions hopping
on a cubic lattice. The equilibrium phase diagram of the model shows
topological transitions between gapped quadrupolar phase supporting
zero energy hinge modes and gapless Weyl semimetals phase.

Upon driving the model by varying one of its parameter periodically
with time, we find the existence of special drive frequencies at
which the bulk gap for the Floquet quasispectrum almost vanishes.
Such frequencies exists for both discrete square pulse and
continuous cosine drive protocols. Our work shows that this effect
can be analytically understood by computing the perturbative Floquet
Hamiltonian of the system using FPT. In the large drive amplitude
region, where FPT is expected to be accurate, a drive at these
frequencies leads to vanishing bulk gap for the first order
perturbative Floquet Hamiltonian. This happens at $\sqrt{2} \gamma_1
T/\hbar= 2n\pi$ for the discrete square protocol and
$\sqrt{2}\gamma_1 T/\hbar= \pi \alpha_n$ for the continuous cosine
protocol. Thus the bulk gap becomes small since it can only
originate from higher order terms in $H_F$; in the high drive
amplitude regime, such terms are expected to be small. Thus the
Floquet spectrum shows lines where the bulk quasienergy gap is
small. We note that this reduction of the gap is not captured by the
Floquet Hamiltonian obtained using second order Magnus expansion.

Our numerical analysis of the Floquet Hamiltonian can be extended to
other protocols. An obvious extension may occur when the fermions
are subjected to a periodically time-dependent vector potential
arising from the presence of incident radiation. However, in this
case, all terms of the Fermion Hamiltonian (Eq.\ \ref{eqn:ham0})
shall become time dependent. This makes the problem difficult to
address using analytic Floquet perturbation theory. Another
possibility is to use a protocol involving periodic kicks. In this
case, one can extend our formalism in a straightforward manner and
obtain results which are qualitatively similar to the square pulse
protocol provided the drive parameters are chosen appropriately.

Away from these special points, where the bulk gap is large, our
analysis finds Floquet hinge modes in the Floquet spectrum. We
provide an analytic expression for these Floquet hinge modes for the
discrete protocol with $\sqrt{2} \gamma_1 T/\hbar= \pi$; we find
that the analytical results agree qualitatively to exact numerics.
In contrast to their equilibrium counterpart, these hinge modes have
$k_z$ dependent dispersion as confirmed from both the first order
analytic Floquet Hamiltonian and exact numerics. The dispersion of
the hinge modes turns out to be flatter for continuous drive
protocols; also the analysis based on first order Floquet
Hamiltonian predicts stronger dispersion compared to that obtained
using exact numerics. In contrast, near the special drive
frequencies where the gap is small, the hinge modes leak into the
bulk; they become almost indistinguishable from the bulk when the
drive frequency matches these special frequencies.

The presence of the small Floquet quasienergy gap manifests in the
dynamics of the hinge modes. To study such dynamics we start with an
initial zero energy eigenstate of the equilibrium Hamiltonian $H_0$
which is localized at one of the hinge. We then study its dynamics
by driving the system with two representative frequencies. One of
these corresponds to the case where the bulk Floquet Hamiltonian is
gapped. For the square pulse protocol, we choose $\sqrt{2} \gamma_1
T/\hbar=\pi$. In this case we find that the hinge mode remains
localized in the vicinity of its original position. In contrast, for
systems driven with a frequency which satisfies $\sqrt{2} \gamma_1
T/\hbar= 2 \pi$, the hinge mode propagates in the bulk and displays
wavefront like propagation between diagonally opposite hinge. This
becomes apparent by computing the spatially resolved probability of
the driven wavefunction given by $\Phi^{\alpha}(nT)$. We find that
$\Phi^{\alpha}(nT)$ shows distinct revivals; their time dependence
represents motion of hinge modes between diagonally opposite hinges
of the sample. Our analysis shows that the period of such a motion
can be analytically understood within a saddle point analysis of the
driven wavefunction.

The experimental verification of our theory can be achieved via STM
measurements which track the local density of states for electrons
within an unit cell. For an initial zero energy state localized at
one of the hinge, the time variation of the local density of state
will clearly depend on $\Phi^{\alpha}$. Our prediction is that
starting from a hinge state localized at $(L,1)$, the dynamics with
$\sqrt{2} \gamma_1 T/\hbar= \pi$ will not show significant variation
of $\Phi^{\alpha'}(nT)$ for $\alpha'$ corresponding to the
diagonally opposite hinge $(1,L)$. In contrast $\Phi^{\alpha'}(nT)$
will show periodic variations for $\sqrt{2} \gamma_1 T/\hbar=2 \pi$
with a period of $n_c$.

In conclusion, we have studied the Floquet spectrum and the hinge
mode dynamics of driven second-order topological Weyl semimetals
modeled by free fermions hopping on a cubic lattice. Our analysis
reveals specific drive frequencies at which the bulk Floquet modes
become nearly gapless. We also find that the dynamics of the hinge
modes for such a Floquet Hamiltonian depend crucially on the
proximity to these special frequencies; they remain localized close
to their initial position away from these frequencies and propagates
coherently between diagonally analogous hinges close to them. We
suggest that the qualitative difference in such dynamics would be
reflected in LDOS of fermions and shall therefore be measurable via
STM measurements.

\acknowledgements SG acknowledges CSIR NET fellowship award No.
09/080(1133)/2019-EMR-I for support and Roopayan Ghosh for
discussions. KS thanks DST, India for support through SERB project
JCB/2021/000030.

\appendix*
\section{Magnus expansion}

In the appendix, we consider the Floquet Hamiltonian derived by the standard Magnus expansion method. The first-order correction in the Magnus expansion of the Floquet hamiltonian is given by
\begin{equation}
    H_{{\rm mag}}^{(1)}=\frac{1}{T}\int_0^T dt_1 H(t_1),
    \label{Eq:Mag1}
\end{equation}
where $H(t_i)=H_0+H'(t_i)$. Since for both discrete and continuous
driving, we are using protocols which average out to zero over a
complete cycle, therefore $H_{{\rm mag}}^{(1);s}=H_{{\rm mag}}^{(1);
c}=H_{av}=H_0$. We note that this is equivalent to the results
obtained from FPT in the limit $\gamma_1 T \to 0$.

The second-order correction in the Magnus expansion is given by the expression
\begin{equation}
    H_{{\rm mag}}^{(2)}=\frac{1}{2i\hbar T}\int_0^T\int_0^{t_1} dt_1 dt_2
    [H(t_1),H(t_2)].
    \label{eq:Mag2}
\end{equation}
For our chosen continuous drive protocol, $H(t)=H(T-t)$. It can be
shown that for drives satisfying this symmetry condition, $H_{{\rm
mag}}^{(2),c}=0$. Thus, till second order, the spectrum of $H_{{\rm
mag}}^c$ is equivalent to that of $H_{av}$ which doesn't exhibit a
gap closing.

For the square pulse protocol, however, the second-order contribution is non-trivial and can be calculated as follows.
\begin{equation}
     H_{{\rm mag}}^{(2),s}=\frac{1}{2i\hbar T}\int_0^T\int_0^{t_1} dt_1 dt_2 \left( [H'(t_1),H_0] +
     [H_0,H'(t_2)]\right).
\end{equation}
Evaluation of these commutators is straightforward, and one can show that
\begin{align}
    H_{{\rm mag}}^{(2),s}=\frac{\gamma_1 T}{2i\hbar}((a_1-a_2)\Gamma_2\Gamma_4 + (\Gamma_2+\Gamma_4)&(a_3\Gamma_3+a_4\Gamma_1)\nonumber\\&+ ia_5\Gamma_3
    ).
    \label{eq:MagSq2}
\end{align}

The eigenvalues of $H_{{\rm mag}}^{(1),s}+H_{{\rm mag}}^{(2),s}$ are
\begin{widetext}
\begin{align}
E^{{\rm mag},s}_{\pm,\pm}=\pm \frac{1}{2}\Big(4\sum_{i=1}^5a_i^2&+((a_1-a_2)^2+2(a_3^2+a_4^2)+a_5^2)(\gamma_1 T/\hbar)^2\nonumber\\
&\pm 2|a_5|\sqrt{16(a_1^2+a_4^2)+4(2a_1(a_1-a_2)+3a_4^2)(\gamma_1
T/\hbar)^2+((a_1-a_2)^2+2a_4^2)(\gamma_1 T/\hbar)^4}\Big)^{1/2}.
\end{align}
\end{widetext}
However, this neither reproduces the band closing nor even the
substantial reduction in the bandgap at the special points $\sqrt{2}
\gamma_1 T/\hbar =2n\pi$, which we had obtained from first order FPT
and exact numerics respectively.

\end{document}